\documentclass[a4paper,
aps,%
12pt,%
final,%
notitlepage,%
oneside,%
onecolumn,%
nobibnotes,%
nofootinbib,
superscriptaddress,%
showpacs,%
centertags]%
{revtex4}

\def\({\biggl(}
\def\){\biggl)}
\newcommand{\boldnabla}{\bm{\nabla}}
\newcommand{\be}{\begin{equation}}
\newcommand{\ee}{\end{equation}}
\newcommand{\ba}{\begin{eqnarray}}
\newcommand{\ea}{\end{eqnarray}}
\newcommand{\bes}{\begin{equation*}}
\newcommand{\ees}{\end{equation*}}
\newcommand{\bas}{\begin{eqnarray*}}
\newcommand{\eas}{\end{eqnarray*}}

\def\vk{{\bf k}}

\def\vx{{\bf x}}

\newcommand{\bc}{\begin{center}}
\newcommand{\ec}{\end{center}}
\newcommand{\beq}{\begin{equation}}
\newcommand{\eeq}{\end{equation}}

\newcommand{\ket}[1]{\ensuremath{\vert\,#1\,\rangle}}
\newcommand{\braket}[2]{\ensuremath{\langle\,#1\,\vert\,#2\,\rangle}}
\newcommand{\doo}[2]{\ensuremath{{\partial #1\over \partial #2}}}
\newcommand{\fundoo}[2]{\ensuremath{{\delta #1\over \delta #2}}}
\newcommand{\dee}[2]{\ensuremath{{{\rm d} #1\over {\rm d} #2}}}

\def\ann{{\hat{a}}}
\def\Ann{{\hat{A}}}
\def\cre{{\hat{a}^\dagger}}

\def\vk{\bm{k}}

\def\vx{\bm{x}}
\def\vy{\bm{y}}

\def\oH{\hat{H}}

\def\ophi{\hat{\varphi}}
\def\opi{\hat{\pi}}

\newcommand{\matrel}[3]{\ensuremath{\left\langle\,#1\,\left\vert\,#2\,\right\vert\,#3\,\right\rangle}}

\def\mv{{\mathbf v}}

\def\ann{{\hat{a}}}
\def\Ann{{\hat{A}}}
\def\cre{{\hat{a}^+}}

\begin{document}

\selectlanguage{english} 
\title{Field-theoretic technique for irreversible reaction processes}
\author{Michal Hnati\v{c}}
\email{hnatic@saske.sk}
\affiliation{Institute of Experimental Physics, SAS, Ko\v{s}ice, Slovakia}
\affiliation{Faculty of Science, P. J. \v{S}af\'arik University, Ko\v{s}ice, Slovakia}
\author{Juha Honkonen}
\email{juha.honkonen@helsinki.fi}
\affiliation{Department of Military Technology, National Defence University, Helsinki, Finland}
\author{Tom\'a\v{s} Lu\v{c}ivjansk\'y}
\email{lucivjansky@saske.sk}
\affiliation{Institute of Experimental Physics, SAS, Ko\v{s}ice, Slovakia}
\affiliation{Faculty of Science, P. J. \v{S}af\'arik University, Ko\v{s}ice, Slovakia}
\begin{abstract}

\end{abstract}

\maketitle
\section*{INTRODUCTION}
 The irreversible annihilation reaction $A +A \rightarrow\varnothing$, also
 known as the mutually annihilating random walk, is a fundamental model of non-equilibrium physics.
The particles $A$ perform chaotic motion due to diffusion and after the mutual collision they may react with
constant microscopic probability $K_0$ per unit time. Usually it is
assumed that resulting molecule $\varnothing$ is inert,
i.e. chemically inactive and without any backward influence on the motion of
 reacting $A$ particles. Many reactions of this type are observed in diverse
chemical, biological or physical systems. For instance various models
such as formation of domain in some magnetic materials \cite{Derrida95},
annihilation of excitons in crystals \cite{Kroon93} or model for spreading
of opinion of voters in one dimension \cite{Odor04} can be described in terms of annihilation process of this type.

The usual approach to the problems dealing with chemical reactions is based on the use of the kinetic rate equation
\cite{Kampen} for concentration field $n(t,{\mathbf x})$.
It leads to a self-consistent description analogous to the mean-field approximation in the
theory of critical phenomena in the sense that fluctuations in the concentration are neglected.
Equivalently one can assume that the particle
concentration is spatially homogeneous $n=n(t)$. This homogeneity can be thought as a consequence of
either very high mobility of the reactants or of a very small probability
that a reaction actually occurs after mutual collision of reacting particles.
For the annihilation process $A+A\rightarrow\varnothing$
kinetic rate equation can be formulated as
\begin{equation}
   \partial_t n(t) = -K_0 n^2(t),
  \label{eq:doi_mpn}
\end{equation}
which can be easily integrated and the obtained solution is
\be
  n(t) =  \frac{n_0}{1+n_0 K_0 t},
  \label{eq:doi_mpn_sol}
\ee
where $n_0\equiv n(0)$ is initial number of particles.
For long-time ($t\rightarrow\infty$)
 asymptotic decay equation (\ref{eq:doi_mpn}) predicts power law behaviour for concentration
  $n(t)\sim t^{-1}$ without any dependence on the value of space dimension.
  This is a common situation observed in the mean field-like theories.
  In what follows we will refer to the value of time exponent $\alpha$ in
 power law dependence for concentration $n(t)\propto t^{-\alpha}$ as the {\it decay exponent}.
 Note also that the long time behaviour
  does not depend on the initial number $n_0$ of reacting particles.
In the other case, when the particle mobility is sufficiently small, or
equivalently, if the microscopic reaction probability $K_0$ becomes
large enough (so particles react immediately after mutual collision)
there is a possible transition to a new regime. In it is more
probable that the given particle reacts with particles in its neighbourhood
 than with distant ones. This behaviour is known as the diffusion-controlled regime \cite{Kampen,Kang84}.
 To gain physical insight let us consider diffusion process (also known as continuous
 random walk) to be responsible for the motion of particles. A well-known property of diffusion
 \cite{Itzykson_book} is the re-entrancy of the visited sites in low space dimensions.
In particular, for $d=1$ and $d=2$ the probability that the diffusing particle will
 ever return ($t\rightarrow\infty$) to the starting point is equal to $1$.
 Physically it means that the diffusing particle sweeps thoroughly its local neighbourhood and thus
 it is highly probable that it will react with another particle in its vicinity. Hence, it is
 reasonable to expect that after short period of time the system will be in a state, where there is a lot
 of isolated particles, that need effectively longer time to traverse to each other and hence
 to annihilate. This mechanism
 can effectively slow down the time evolution of the process and thus lower the decay exponent to other value than
 $1$ predicted by equation (\ref{eq:doi_mpn_sol}). The approximate value of the decay exponent can
 be guessed according to following scaling argument. The re-entrancy property
 leads to the scaling relation
\be
  V(t) \sim r^d(t)\,\,,
  \label{eq:doi_volum_scale}
\ee
where $\sim$ denotes corresponding scaling relation between
physical quantities. The root mean square distance for the diffusing particle
scales as $r(t)\sim (Dt)^{1/2}$ and therefore the mean particle
number should behave as \be
  n(t)\propto \frac{1}{V(t)} \sim \frac{1}{t^{d/2}} = \frac{1}{t^{1+\Delta}}\,\,,
  \label{eq:doi_mean_part_scale}
\ee
 where the exponent $\Delta$ denotes the deviation from the space dimension $2$ via
relation
\be
  d=2+2\Delta\,\,.
  \label{eq:doi_intro_delta}
\ee
For the space dimension $d=3$ we have $V(t)\sim t$, because now the diffusing particle
effectively explore always new volume and the re-entrancy property can be neglected. Therefore
the same behaviour as the one described by (\ref{eq:doi_mpn}) would be observed.
>From this simple analysis it could be estimated that space dimension $d_c=2$ is the upper
critical dimension for the annihilation process, above which
the mean field approximation is valid. A more rigorous treatment \cite{Lee94} based on renormalization
 group proves this conclusion and
also produces a logarithmic correction for $n(t)$ at the critical dimension,
 which could not be determined by the simple scaling analysis.
In the preceding discussion we have considered only the diffusive motion of reacting particles. However, a typical
 reaction usually occurs in liquid or gaseous environment.
Thermal fluctuations of this environment
or some external advection field such as atmospheric eddies
could have additional influence on motion of the reacting particles.
Therefore, it is interesting to study what effect the
external velocity field can have on the annihilation process.

The most flexible approach to the theoretical analysis of the effects of fluctuations in reaction kinetics
seems to be the second-quantization method due to Doi \cite{Doi76}.
Most of the renormalization-group studies of the effect of random drift
on the annihilation reaction  $\textit{A} +\textit{A} \rightarrow \varnothing$ in the framework
of the Doi approach
have been carried out for the case of a quenched random drift field. Potential random drift with long-range \cite{Park98,Chung}
and short-range correlations \cite{Richardson99} have been studied as well as ''turbulent'' flow (i.e. quenched solenoidal random field) with potential
disorder \cite{Deem98b,Tran}. For a more realistic description of a turbulent flow time-dependent velocity field would be more appropriate.
In Ref. \cite{Tran} dynamic disorder with a given Gaussian distribution has been considered, whereas the most ambitious approach
on the basis of a velocity field generated by the stochastic Navier-Stokes equation has been introduced here by two of the present authors  \cite{Hnatic00}.
>From the point of the Navier-Stokes equation the situation near the critical dimension $d_c=2$ of the pure reaction model is even more intriguing
due to the properties of the Navier-Stokes equation.
It is well-known fact \cite{Frisch} that in the case of space dimension $d=2$, there is inviscid
conservation law of enstrophy absent in the three dimensional case.
Calculations in Ref. \cite{Hnatic00} were performed in the one-loop
approximation. As may readily seen from examination of the Feynman graphs,
in the one-loop approximation there is no influence of the velocity fluctuations on the
renormalization of the interaction vertices. However, the influence of higher
order terms of the perturbation series can have significant effect on the critical properties.

To this end, the most suitable choice for generation of random velocity field is the stochastic Navier-Stokes equation, which
can be used to produce a velocity field
corresponding to thermal fluctuations \cite{Forster} and a turbulent velocity
field with the Kolmogorov scaling behaviour \cite{Adzhemyan}.

   A powerful tool for analyzing asymptotic behaviour of stochastic systems is provided
   by the renormalization-group (RG) method.
It allows to determine long-time -- or infra-red (IR) -- asymptotic
regimes of the system and also it is very a efficient tool for
calculation of various universal physical quantities, e.g. critical
exponents. The aim of this study is to examine the IR behaviour of
the annihilation process under the influence of advecting velocity
fluctuations and to determine its stability. Using the mapping
procedure based on the Doi formalism \cite{Doi76} an effective
field-theoretic model for the annihilation process will be described
in detail in the Sec. \ref{sec:doi_quant_method}. Consequently, the
RG method is applied to this model and within the two-parameter
expansion the renormalization constants and fixed points of the
renormalization group are determined in the two-loop approximation.
The non-linear integro-differential equation, which includes first
non-trivial corrections to the (\ref{eq:doi_mpn}), is obtained for
the mean particle number and it is
shown how the information about IR asymptotics can be extracted from it.

Another aspect of the annihilation problem is connected with its theoretical description in the
form of a functional integral with a given action. This action
 resembles actions for field-theoretic models of critical dynamics obtained
from the Langevin equation \cite{Vas_turbo} within the Martin-Siggia-Rose approach \cite{Martin73}.

In the Langevin equation the
random field compensates for dissipative and reactive losses, hence bringing about a steady state of dynamics of the system.
In reaction kinetics the random sources and sinks, in fact, reflect the real physical situation,
in which during the chemical reaction of follow-up species, particles can appear or disappear due to
uncontrolled random interaction with a particle bath, e.g. due to active chemical radicals.
The interpretation of random fields in the Langevin equation as physical sources and sinks is rather
problematic. Therefore, we propose to analyze the alternative approach provided by the master equation with
terms corresponding to interactions with the bath.

In the present paper we follow the ideas of \cite{Kampen} and
describe the random sources and sinks in terms of new birth and death reactions in the master equation for the
single-species annihilation reaction.
The simplest choice does not conserve the particle number and we have no possibility
to compare it with the standard Langevin approach. A slightly more involved set of birth-death reactions
allows to conserve the particle number and the result may be compared with that of the standard multiplicative noise
in the Langevin equation.

The aim of this paper is to give a detailed account of how the field-theoretic approach
can be applied to the analysis of the large-scale asymptotic behaviour of annihilation process $A+A\rightarrow\varnothing$.
We give an elaborate derivation of the action functional for this reaction and show how the velocity
fluctuations and effects of sinks and sources can be described. For the analysis of the large scale behaviour
methods of the renormalization group are employed.

This paper is organized as follows. In Sec. \ref{sec:doi_quant_method} a detailed
description of the Doi approach for construction of the field-theoretic model for the
annihilation process $A+A\rightarrow\varnothing$ is summarized.
In Sec. \ref{sec:anni1} it is shown how both the Kolmogorov scaling and thermal fluctuations can be included into the model
and the ultraviolet (UV) renormalization of the model and elaborated algorithm for
the calculation of the renormalization constants is described. Fixed points
of the RG are classified together with their stability regions and possible scaling regimes are presented
 and the integro-differential equation for the mean particle number is derived and
 analysis of its solution is given. Sec. \ref{sec:random_sources} is devoted to the introduction of
  random sources and sinks into the field-theoretic model of reaction-diffusion processes.
  We recall the basic features of the Doi formalism and
construct the basic field-theoretic dynamic action functional together with scaling analysis of the dynamic actions
is performed.  
{\section{FIELD-THEORETIC REPRESENTATION OF THE MASTER EQUATION} \label{sec:doi_quant_method} }
{\subsection{INTRODUCTION} \label{sec:doi_intro}}
In our work we are mainly interested in the annihilation process
  $A+A \overset{K_0}{\longrightarrow}  \varnothing$, therefore in this section we describe
  a derivation of the field-theoretic model for such process, which allows
to take into account spatial inhomogenities and randomness in individual reaction
events. Probably the most fundamental description of reaction processes is based on the use
of master equation \cite{Kampen}. We summarize general principles
of how such master equation can be mapped onto a suitable Fock space, which makes it possible
to use very powerful methods of quantum field theory. The resulting action functional
can be treated systematically by such methods as Feynman diagrammatic technique, renormalization group
and operator product expansion \cite{Zinn}. Here we focus mainly on the field-theoretic description of
diffusion and reaction process, but it is possible \cite{Tau05} to generalize this approach to include
effects as multiple reaction schemes, disorder effects, influence of spatial boundaries etc.

Let us start with the particles on a regular, infinite, hypercubic
lattice with the lattice spacing $a$ in $d$-dimensional space. The sites of lattice
can be labeled by natural numbers ($i=1,2,\ldots$).
The $A$ particles are performing continuous random walk on this
 lattice (random hopping betweeen adjoint sites)
  with the diffusion constant $D/a^2$ (the factor $a^2$ will be
 eliminated in the continuum limit).
It is assumed that reaction process can happen only for particles that occupy the same
lattice site with the probability rate $K_0$. A complete
 (microscopic) description of such a stochastic problem can be given in terms of evolution
equations for probabilities $P(t;\{n\})$, where $\{n\}$ is given microstate
$\{n_1,n_2,\ldots\}$ characterized by $n_1$ particles at site $1$ , $n_2$ at site $2$
 and so on.
These equations (known as master equations) express the balance between incoming and outcoming
probabilities \cite{Kampen} for the state $\{n\}$ and in a compact notation they can be
written as
\be
 \frac{dP(t;\{n \})}{dt} = \sum_{\{m\} } R_{m\rightarrow n} P(t;\{n\}) -
  \sum_{\{ m\}} R_{n\rightarrow m} P(t;\{n\}),
 \label{eq:doi_master}
\ee
where $R_{m\rightarrow n}$ is the transition probability rate from the state $m$ to the state $n$.
According to the work of Doi \cite{Doi76} (see also \cite{Peliti85}) such a system
 of coupled differential equations (\ref{eq:doi_master}) can be rewritten in terms of creation
and annihilation operators well known from quantum mechanics.
If there is no site occupation restriction, bosonic operators for
each lattice site $i$ can be introduced with the following commutation relations
\be
   [\ann_i,\ann_j^\dagger]=\delta_{ij},\quad [\ann_i,\ann_j] = [\ann_i^\dagger,\ann_j^\dagger]=0.
\label{eq:doi_boson_oper}
\ee
The ground state $| 0 \rangle$ is defined as
\be
  \ann_i | 0 \rangle = 0 \quad \mbox{for all sites}\mbox{ } i,
\label{eq:doi_gs}
\ee
which corresponds to the empty lattice (without any $A$ particle).
>From the bosonic commutation relations
(\ref{eq:doi_boson_oper}) important relations follow:
\be
  \ann_i^n \ann_i^\dagger = n\ann_i^{n-1}+\ann_i^\dagger \ann^n,\quad
   \ann_i \ann_i^{n\dagger}=n\ann_i^{(n-1)\dagger} + \ann^{n\dagger}\ann_i\,\,.
  \label{eq:doi_boson_pomocka}
\ee
 The state
$|\{n\}\rangle$ with the given lattice configuration $\{ n\}=\{n_1,n_2,\ldots\}$
is introduced with a normalization different from that used in the
second quantization method in quantum field theory
\be
  |\{ n \} \rangle = \ann_1^{\dagger n_1} \ann_2^{\dagger n_2} \ldots|0\rangle.
  \label{eq:doi_def_state}
\ee
 Using relations
(\ref{eq:doi_boson_pomocka}) it can be directly shown that
\ba
   & & \ann_i|\{n\}\rangle = n_i |\{n_1,n_2,\ldots,n_i-1,\ldots\}\rangle,\nonumber\\
   & & \ann_i^{\dagger}|\{n\}\rangle = |\{n_1,n_2,\ldots,n_i+1,\ldots\}\rangle, \nonumber\\
   &&  \ann_i^\dagger \ann_i |\{n\}\rangle = n_i |\{n\}\rangle,
   \label{eq:doi_states}
\ea
 where the last relation legitimizes the identification
 \be
    \hat{n}_i = \ann_i^\dagger \ann_i
    \label{eq:defi_oper_num}
 \ee
for the number operator at site $i$.
The scalar product between two states $|\{n\}\rangle$ and $|\{m\}\rangle$ can be obtained
\be
  \langle \{n\} |\{ m \}\rangle = \prod_{i=1} \delta_{n_i,m_i} n_i!,
  \label{eq:doi_scalar_product}
\ee
where $\delta_{i,j}$ stands for Kronecker symbol and $n!=1\times 2\times\ldots n$ is the factorial function.
The complete information about the stochastic system is
embodied in the probabilities $P(t;\{n\})$ and in the Doi formalism it is
 incorporated into the state vector $|\Phi\rangle$, which is defined as follows
\be
  | \Phi(t) \rangle \equiv \sum_{\{n\}} P(t;\{n\}) |\{n\}\rangle= \sum_{\{n\}} P(t;\{n\})\mbox{ }
  \ann_1^{\dagger n_1} \ann_2^{\dagger n_2}\ldots |0\rangle,
\label{eq:doi_state}
\ee
where the sum runs over all possible lattice occupations.
Now the task is to rewrite master equation (\ref{eq:doi_master})
into the Schr\"odinger-like form for the state vector $|\Phi\rangle$
\be
  \frac{d}{dt} |\Phi(t) \rangle = - \hat{H} |\Phi(t) \rangle,
  \label{eq:doi_schr}
\ee
with some "Hamiltonian" $\hat{H},$ whose exact form depends on the system
under consideration. Then the equation
(\ref{eq:doi_schr}) can be formally integrated to obtain $|\Phi(t)\rangle ={\mathrm e}^{-\oH t}|\Phi(0)\rangle$.
The initial state $|\Phi(0)\rangle$ has to be specified for the full description.
In the case of chemical reactions initial distribution of particles $P(0;\{n\})$ is usually prescribed
and initial state follows from the definition (\ref{eq:doi_state}). From the technical point of view
the most convenient choice for the singe-species annihilation reaction is the Poisson distribution.
Unlike the bimolecular reaction \cite{Lee95}, the long-time behaviour is independent of the
concrete form of initial conditions.

Let us derive the diffusion part $\oH_D$ of the Hamiltonian $H$, which corresponds to the
diffusive (random walk) movement of $A$ particles. First consider a
two-site system with $n_1$ particles at site $1$ and $n_2$ particles at site $2$ with
the one-directional hopping process $1\rightarrow 2$ at the rate $D_0/a^2$.
For such a process the master equation (\ref{eq:doi_master}) is
   \be \frac{dP(n_1,n_2)}{dt} = \frac{D_0}{a^2}(n_1+1)P(n_1+1,n_2-1)-\frac{D_0}{a^2}n_1P(n_1,n_2)\,,
      \label{eq:doi_uni_diff}
   \ee
where $n_1+1$ and $n_1$ are combinatorial factors resulting from the fact that $A$ particles
jump independently of each other. Multiplying both sides of the equation (\ref{eq:doi_uni_diff}) by
the term $ a_{1}^{\dagger n_1} a_{2}^{\dagger n_2} $ and performing
 sum $\sum_{\{n_1,n_2\}} $ over all possible occupations of sites we arrive at
 \be
  \frac{d|\Phi\rangle}{d t} =  \frac{D_0}{\ann^2}(\ann_2^\dagger \ann_1-\ann_1^\dagger \ann_1) |\Phi\rangle.
 \ee
 This result is easily generalized to the two-directional case $1\leftrightarrow 2$
described by the following master equation
\be \frac{d|\Phi\rangle}{d t} =  -\frac{D_0}{a^2}(\ann_2^\dagger-\ann_1^\dagger)(\ann_2-\ann_1) |\Phi\rangle
\ee
and hence the diffusion part $H_D$ between two given sites can be written as
\be
  \oH_D = \frac{D_0}{a^2}(\ann_2^\dagger-\ann_1^\dagger)(\ann_2-\ann_1).
  \label{eq:doi_diffus_part}
\ee
Now consider the annihilation process at given site $1$ and derive
the corresponding part $H_R$ of the Hamiltonian. Because
 any two particles can react together, we can write
\be
\label{eq:master reaction}
\frac{dP(n)}{dt}=K_0 (n+2)(n+1)P(n+2)-K_0 n(n-1)P(n),
\ee
where again the combinatorial factors are taken into account.
After short
 algebraic manipulations and the use of the relations (\ref{eq:doi_boson_oper}) and (\ref{eq:doi_states}),
 relations (\ref{eq:master reaction}) can be rewritten in the Doi formalism as
\be \frac{d|\phi\rangle}{d t} =  K_0(\ann^2-\ann^{\dagger 2}\ann^2) |\phi\rangle\,,
\label{eq:doi_simple_annihil}
\ee
from which we deduce the reaction part of the Hamiltonian
\be
\oH_R=-K_0(\ann^2-\ann^{\dagger 2}\ann^2)\,\,.
  \label{eq:doi_react_part}
\ee
Results ($\ref{eq:doi_diffus_part}$) and ($\ref{eq:doi_react_part}$) are easily generalized
to include all the lattice sites. Consequently, the total Hamiltonian
that accounts for diffusion and annihilation process on the hypercubic lattice
has the following form
\be \oH_D+\oH_R=\frac{D_0}{a^2}\sum_{<ij>}(\ann_i^\dagger-\ann_j^\dagger)(\ann_i-\ann_j)-
K_0\sum_i (\ann_i^2-\ann_i^{\dagger 2} \ann_i^2)\,\,,
\label{eq:doi_hamil_ann}
\ee
where the first sum runs over the neighbouring sites $i$ and $j$.
Non-Hermitian Hamiltonians such as one given in (\ref{eq:doi_hamil_ann})
are often observed in the case of systems out of equilibrium\cite{Odor04}, which
cannot be obtained as dynamical counterparts of some static models.
Non-hermiticity also means that the reaction rates in (\ref{eq:doi_master})
do not satisfy the detailed balance condition and thus the equilibrium state cannot be
characterized by the Gibbs distribution. The Doi formalism also exhibits other
differences from the usual quantum mechanics. They are caused by the fact
 that physical observables cannot be given as bilinear products
$\langle \Phi| A |\Phi\rangle$, since according to (\ref{eq:doi_state}) this would imply expressions bilinear
in probability $P(t;\{n\})$.

Let us now take a closer look at derivation of the ensemble average value for an observable quantity
 $ A$ within the Doi approach. It is physically reasonable to assume
 that $A$ can be expressed as a function of the occupation numbers $A=A(\{n\})$.
Examples of such quantities interesting for the case of chemical reactions are
\begin{enumerate}[(a)]
\item mean particle number (concentration)
\be
  n \longleftrightarrow  \sum_i \ann_i^\dagger \ann_i
  \label{eq:doi_mean_part_num}
\ee
\item two-point correlation function (between sites $i$ and $j$)
\be
  C(i,j) \longleftrightarrow \ann_i^\dagger \ann_i \ann_j^\dagger \ann_j
  \label{eq:doi_two_point_correl}
\ee
\end{enumerate}
The ensemble average of $A$ is then clearly given by the expression
\be
  \langle A(t) \rangle  =  \sum_{\{ n\}} P(t;\{n\}) A(\{n\})
  \label{eq:doi_ens_aver}
\ee
and from the technical point of view it would be  very
convenient to have a projection state $\langle\mathcal{P}|$ such that following identity is valid
\be
   \langle A(t) \rangle
          = \sum_{\{ n\}} P(t;\{n\}) \langle \mathcal{P}| \Ann(\{\ann^\dagger\ann\})
           \ann_1^{\dagger n_1} \ann_2^{\dagger n_2}\ldots |0\rangle 
         =  \langle \mathcal{P}| \hat{A}|\Phi(t)\rangle.
\label{eq:doi_mean_val}
\ee
Substitution of the formal solution of the "Schr\"odinger" equation (\ref{eq:doi_schr}) leads to yet another form
of the last expression on the right side of (\ref{eq:doi_mean_val})
\be
  \langle A(t) \rangle  = \langle \mathcal{P}| \hat{A}\exp(-\oH t)|\Phi(0)\rangle.
  \label{eq:doi_mean_val2}
\ee
Here the operator $\Ann(\{\ann^\dagger\ann\})$ is obtained from the classical function $A(\{n\})$ with the substitution
 $n_i\rightarrow \ann_i^\dagger \ann_i$ at every site $i$, what is justified by (\ref{eq:defi_oper_num}).
  Hence by comparing (\ref{eq:doi_ens_aver}) and (\ref{eq:doi_mean_val})
  it is easy to guess that the projection state $\langle\mathcal{P}|$ should satisfy relation
\be
  \langle \mathcal{P}| \hat{A}(\{\ann^\dagger \ann\})
   \ann_1^{\dagger n_1} \ann_2^{\dagger n_2}\ldots |0\rangle =
  A(\{n\}) \langle\mathcal{P}| \ann_1^{\dagger n_1} \ann_2^{\dagger n_2}\ldots |0\rangle =
  A(\{n\})\,\,.
\ee
It implies that the following two conditions
\be
\langle \mathcal{P}| \ann_i^\dagger =\langle \mathcal{P}|  \quad\mbox{for every site}\mbox{ }i, \quad
\langle\mathcal{P}|0\rangle=1
\label{eq:doi_proper_project}
\ee
 have to be valid for the operator $\mathcal{P}$.
 By direct use of relations (\ref{eq:doi_boson_oper}) we can conclude that the following
 choice of the projection state
 \be
   \langle \mathcal{P}| = \langle 0 | \exp\biggl(\sum_i \ann_i\biggl)
   \label{eq:doi_proj_state}
 \ee
serves our purpose.
Note that if the operator $\Ann$ is written in the normal-ordered form (all creation operators are commuted to the left), it can then
be written in terms of the annihilation operators $a_i$ only using properties (\ref{eq:doi_proper_project}) of the projection
state $\langle \mathcal{P}|$, i.e.
\be
\label{eq:normal form}
  \langle \mathcal{P}|\hat{A}(\{\ann^\dagger \ann\})=\langle \mathcal{P}|N\left[\hat{A}_N(\{\ann^\dagger, \ann\})\right]=
\langle \mathcal{P}|\hat{A}_N(\{1, \ann\})\,.
\ee
The normal form $\hat{A}_N$ of the operator $\Ann$ is defined by relation $\Ann=N\left[\hat{A}_N(\{\ann^\dagger, \ann\})\right]$,
where $N$ denotes the normal product (creation operators are put to the left of annihilation operators).
The $\ann$-dependent operator $\hat{A}_N(\{1, \ann\})$
corresponding to (\ref{eq:doi_mean_part_num}) is then simply $\ann_i$, while the correlation function
(\ref{eq:doi_two_point_correl}) corresponds to operator $\ann_i\ann_j +\delta_{ij}\ann_j$.
For technical reasons it is reasonable to commute the factor ${\mathrm e}^{\sum_i \ann_i}$ to the right
in equation (\ref{eq:doi_mean_val}). In order to do this we employ
the following formula
\be
{\mathrm e}^{\ann_i} \ann_i^\dagger=(\ann_i^\dagger+1){\mathrm e}^{\ann_i},
\label{eq:doi_useful_eq}
\ee
which is derived from relations (\ref{eq:doi_boson_pomocka}) and thus equation (\ref{eq:doi_mean_val2})
can be cast into the form
\be
 \langle A(t) \rangle = \biggl\langle 0\biggl| \hat{A}_N(\{1,\ann\}) \exp(-\oH(\{\ann^\dagger+1,\ann\})t)  \biggl|
   {\mathrm e}^{\sum_i \ann_i}\Phi(0)\biggl\rangle ,
  \label{eq:doi_shifted_mean_val}
\ee
where the substitution $\ann_i^\dagger\rightarrow \ann_i^\dagger+1$ was performed both
in the expression for the operator $\hat{A}$ (with the subsequent
substitution $\ann_i^\dagger\rightarrow 0 $) and also in the original Hamiltonian $\oH(\{a^\dagger,a\})$.
For further use let us write expression for the mean particle number (\ref{eq:doi_mean_part_num})
\be
  n(t) = \biggl\langle 0\biggl| \sum_i \ann_i \exp(-\oH(\{\ann^\dagger+1,\ann\})t)  \biggl| {\mathrm e}^{\sum_i \ann_i}\Phi(0)\biggl\rangle.
  \label{eq:doi_aver_num_dis}
\ee
In the case of annihilation process described by the equation (\ref{eq:doi_hamil_ann}) thus we obtain
\be
  \oH_D+\oH_R=\frac{D_0}{a^2}\sum_{<ij>}(\ann_i^\dagger-\ann_j^\dagger)(\ann_i-\ann_j)+K_0\sum_i (2\ann_i^\dagger \ann_i^2+\ann_i^{\dagger 2} \ann_i^2)\,.
\label{eq:doi_shift_hamil_ann}
\ee
As was mentioned above, the convenient choice for the initial condition is the Poisson distribution, which for a given site $i$ corresponds to
\be
  p(n_i) = {\mathrm e}^{-n_0} \frac{n_0^{n_i}}{n_i!},
\label{eq:doi_poisson}
\ee
where $n_0$ stands for the mean particle number.
Using definition (\ref{eq:doi_state}) we arrive at the initial state vector in the form
\ba
  |\Phi(0)\rangle & = & \sum_{\{ n_1\}} {\mathrm e}^{-n_0}
  \frac{n_0^{n_1}}{n_1!} \ann_1^{\dagger n_1}
  \sum_{\{ n_2\}}{\mathrm e}^{-n_0}
  \frac{n_0^{n_2}}{n_2!} \ann_2^{\dagger n_2}\ldots |0\rangle =
   \prod_i {\mathrm e}^{-n_0}
   {\mathrm e}^{n_0\sum_i \ann_i^\dagger} |0\rangle.
\label{eq:doi_init_state}
\ea
We see that after the substitution $\ann^\dagger\rightarrow \ann^\dagger+1$ the term ${\mathrm e}^{-n_0}$
drops out.
{\subsection{CONTINUUM LIMIT}  \label{sec:doi_con_lim} }
In the field of critical phenomena emphasis often lies in the analysis and determination
of possible behaviour of the studied system. It turns out
that near a second-order phase transition \cite{Cardy_book} a whole set of different models
behave in the same way. Despite the fact that they describe different physical systems, they can
exhibit the same behaviour of so called universal quantities. They are model-independent, but
can depend on universal parameters such as space dimension, number of components
of the order parameter or symmetries of the system. Examples of such universal quantities are
 critical exponents \cite{Fisher74}, describing singular behaviour of various functions.

Now we summarize the main points of the derivation of the continuum limit for the Hamiltonian (\ref{eq:doi_hamil_ann}), which
allows us to study universal properties of the annihilation process $A+A\rightarrow\varnothing$ around its
critical dimension.
In fact, here we just briefly outline a few important steps following \cite{Tau05} of the introduction of the functional integral of the continuum limit
by the popular interpolation procedure. The alternative operator approach will be sketched in Sec. \ref{sec:anni1_model}.
The main task consists of evaluation of matrix elements for the evolution operator $\exp(-\oH(\{a^\dagger,a\})t)$. In order
 to do this we apply the Trotter formula \cite{Schulman_book}, according to which the exponential ${\mathrm e}^{-\oH t}$
 can be written as the infinite product
\be
  \exp(-\oH t) = \lim_{\Delta t\rightarrow 0}(1-\oH\Delta t)^{t/\Delta t} = (1-\oH\Delta t)(1-\oH\Delta t)\ldots
  \label{eq:doi_evol_oper}
\ee
Here we assume that $N\Delta t=t$ and at the end of our derivation we let the number of time slices $N\rightarrow\infty$
(or equivalently $\Delta t\rightarrow 0$).
Now into each time slice complete set of coherent states is inserted, which results
into mapping of operators $\ann_i^\dagger,\ann_i$ onto complex numbers. Coherent states are explicitly defined
 as \cite{Schulman_book}
\be
  |\psi \rangle = \exp\biggl( -\frac{1}{2}|\psi|^2 + \psi \ann^\dagger \biggl)|0\rangle
  \label{eq:doi_def_coh_state}
\ee
and they form the eigenstate basis of the annihilation operator
\be
  \ann|\psi \rangle = \psi|\psi\rangle,\quad \langle \psi| \ann^\dagger= \psi^* \langle\psi|\,,
  \label{eq:doi_eigen_coh}
\ee
where the star stands for complex conjugation.
The important property of coherent states is their overlap function between different eigenstates
\be
 \langle \psi_1|\psi_2\rangle = \exp\biggl(-\frac{1}{2}|\psi_1|^2 - \frac{1}{2}|\psi_2|^2 + \psi_1^*\psi_2 \biggl).
 \label{eq:doi_coh_overlap}
\ee
For a single site we can write an identity resolution in the form
\ba
  1=\sum_{n} \frac{1}{n!} |n\rangle \langle n| =  \sum_{m,n} \frac{1}{n!}|n\rangle \langle m |\delta_{mn} =
    \int \frac{d\psi^*d\psi}{\pi} |\psi\rangle\langle\psi|,
  \label{eq:doi_coh_unit}
\ea
where the orthogonality relation
\be
\delta_{mn} = \frac{1}{\pi m!} \int d\psi^* d\psi\mbox{ } \exp(-|\psi|^2) \psi^{*m}\psi^n
\ee
has been used (note the apperance of the weight function ${\mathrm e}^{-|\psi|^2}$, the measure adopted
here is $ d\psi^* d\psi= {\rm Re} \psi\,{\rm Im}\psi$).
Generalization of (\ref{eq:doi_coh_unit}) to the  whole lattice is straightforward
\be
  1 = \int \prod_i \frac{d\psi_{(i,j)}^* d\psi_{(i,j)}}{\pi} \biggl|\{ \psi \}_j
  \biggl\rangle \biggl\langle \{\psi \}_j\biggl|,
  \label{eq:doi_gen_coh_unit}
\ee
where now $\{\psi\}_j=(\psi_{(1,j)},\psi_{(2,j)},\ldots)$ denotes the set of all eigenvalues corresponding to
the annihilation operators $\ann_i$ at each lattice site at the time instant $j\Delta t\mbox{ } (j=0,\ldots,N)$.
Inserting (\ref{eq:doi_gen_coh_unit}) into each time slice in (\ref{eq:doi_evol_oper}) we get
\be
  \exp(-\oH t) = \frac{1}{\mathcal{N}}  \lim_{\Delta t\rightarrow 0}\int
  [d\psi^*][d\psi] \biggl|\{\psi \}_N \biggl\rangle
  \biggl(
     \prod_{j=1}^N \biggl\langle \{ \psi \}_j \biggl| \exp(-\oH(\{\ann^\dagger,\ann\})
     \Delta t) \biggl|\{ \psi \}_{j-1} \biggl\rangle
  \biggl)
  \biggl\langle \{ \psi\}_0 \biggl|,
  \label{eq:doi_deriv_coh}
\ee
where $\mathcal{N}$ is the normalization constant and we have introduced the notation
\be
  [d\psi^*][d\psi] \equiv \prod_{i,j} d\psi_{(i,j)}^* d\psi_{(i,j)}
  \label{eq:doi_func_measure}
\ee
for the functional measure. Note that
if we deal with normal ordered Hamiltonian (which we shall assume and which explicitly is the case  for (\ref{eq:doi_shift_hamil_ann})),
then
using relations (\ref{eq:doi_eigen_coh}) we can immediately write
\ba
\biggl\langle \{ \psi \}_j \biggl| \exp(-H(\{a^\dagger,a\})\Delta t) \biggl|\{ \psi \}_{j-1}  \biggl\rangle =
\biggl\langle \{ \psi \}_j \biggl| \{ \psi \}_{j-1}\biggl\rangle
 \exp(-H(\{\psi^* \}_j,\{\psi \}_{j-1} )\Delta t),
\label{eq:doi_prepis_elementu}
\ea
where $H(\{\psi^* \}_j,\{\psi \}_{j-1} )$ is obtained by the replacement of operators by their eigenvalues
 $\ann_i\rightarrow\psi_i,\ann_i^\dagger\rightarrow \psi_i^*$.
 The remaining term in (\ref{eq:doi_prepis_elementu}) more precisely stands for the expression
 \be
  \biggl\langle \{ \psi \}_j \biggl| \{ \psi \}_{j-1}\biggl\rangle = \prod_i \biggl\langle \psi_{(i,j)} \biggl|
   \psi_{(i,j-1)}\biggl\rangle,
  \label{eq:doi_debilny_prod}
 \ee
which, with the of the overlap relation (\ref{eq:doi_coh_overlap}), can be rewritten as
\be
  \biggl\langle \psi_{(i,j)} \biggl| \psi_{(i,j-1)} \biggl\rangle=
  \exp\biggl(-\psi^*_{(i,j)}[\psi_{(i,j)} -\psi_{(i,j-1)} ]\biggl) \exp\biggl(
  \frac{1}{2}|\psi_{(i,j)}|^2-\frac{1}{2}|\psi_{(i,j-1)}|^2  \biggl).
  \label{eq:doi_prepis_debility}
\ee
The whole scalar product from (\ref{eq:doi_deriv_coh}) can be expressed to the first order in the time increment as
\be
   \prod_j \biggl\langle \psi_{(i,j)} \biggl| \psi_{(i,j-1)} \biggl\rangle=
   \exp\biggl(-\sum_j \psi^*_{(i,j)} \frac{d\psi_{(i,j)}}{dt} \Delta t +O(\Delta t)   \biggl)
   \exp\biggl(\frac{1}{2}|\psi_{(i,N)}|^2-\frac{1}{2}|\psi_{(i,0)}|^2 \biggl),
\ee
so that in the continuum time limit $\Delta t\rightarrow 0$ we obtain the functional integral representation
for the evolution operator in the form
\ba
  \exp(-\oH t) & = & \int\frac{[d\psi^*][d\psi]}{\mathcal{N}}
  \biggl| \{\psi\}_N \biggl\rangle \times \nonumber\\
  & &
  \exp\biggl(\frac{1}{2}|\psi_i(t)|^2-\frac{1}{2}|\psi_i(0)|^2-
  \int_0^t dt[\psi_i^*\partial_t \psi_i + H(\{\psi^*,\psi\}) ] \biggl)
  \biggl\langle\{\psi\}_0 \biggl|.
  \label{eq:doi_evol_func}
\ea
In calculation of expectation values such as (\ref{eq:doi_mean_part_num}) and (\ref{eq:doi_two_point_correl})
we have to act on the expression (\ref{eq:doi_evol_func}) from the left
by the projection state $\langle \mathcal{P}|$ and from the right by the initial state $|\Phi(0)\rangle$.
It can be done in the following manner: first, we note that the following relation holds
\be
  \biggl\langle \{\psi\}_0\biggl| \Phi(0) \biggl\rangle \prod_i \exp\biggl(-\frac{1}{2}|\psi_i(0)|^2 \biggl)
  = \exp\biggl(\sum_i[n_0\psi_i^*(0) -n_0 - |\psi_i(0)^2|] \biggl),
  \label{eq:doi_uprava1}
\ee
where we have used the initial state in the form (\ref{eq:doi_init_state}) and relations
(\ref{eq:doi_def_coh_state},\ref{eq:doi_eigen_coh}). Using equation (\ref{eq:doi_useful_eq})
we proceed as follows
\ba
 \biggl\langle\mathcal{P}\biggl|\{\psi\}_N\biggl\rangle \prod_i
 \exp\biggl(\frac{1}{2}|\psi_i(t)|^2 \biggl) & = &
  \biggl\langle 0\biggl| {\mathrm e}^{\sum_i \ann_i} {\mathrm e}^{\sum_i(-\frac{1}{2}|\psi_i(t)|^2
  + \psi_i(t)\ann^\dagger_i  )}\biggl|0\biggl\rangle
   \prod_i
   \exp\biggl(\frac{1}{2}|\psi_i(t)|^2 \biggl)\nonumber\\
  & = & \biggl\langle 0\biggl| \prod_i \exp(\ann_i) \exp(\psi_i(t)\ann_i^\dagger) \biggl|0\biggl\rangle \nonumber\\
  & = & \biggl\langle 0 \biggl| \prod_i \sum_{k=0} \frac{\psi_i^k(t)}{k!} \exp(\ann_i) \ann_i^{\dagger k}
  \biggl|0\biggl\rangle \nonumber\\
  & = & \biggl\langle 0 \biggl| \prod_i \exp\biggl(\psi_i(t)(\ann_i^\dagger+1)\biggl) \exp(\ann_i)
  \biggl|0\biggl\rangle \nonumber\\
  & = & \exp\biggl(\sum_i \psi_i(t)\biggl).
 \label{eq:doi_uprava2}
\ea
Putting together terms from (\ref{eq:doi_evol_func}-\ref{eq:doi_uprava2}) and
inserting them into expression (\ref{eq:doi_mean_val2}) for the expectation value of the quantity $A$ we arrive at
the important expression
\be
  \langle A(t) \rangle= \mathcal{N}^{-1} \int [d\psi^*][d\psi] A_N(\{1,\psi\}) \exp[S(\{\psi^*,\psi\})]\,,
  \label{eq:doi_mean_value_func}
\ee
where the functional $A_N(\{1,\psi\})$ is obtained from the normal form $\hat{A}_N\left(\{1,\ann\}\right)$ of the
operator $\hat{A}$ by replacing the operators $\ann_i$ by their eigenvalues $\psi_i$
and the
action functional $S$ is given as
\be
  S(\{\psi^*,\psi\}) = \sum_i\biggl(
     \psi_i(t) + n_0 \psi_i^*(0) -n_0 - |\psi_i(0)|^2 - \int_0^t dt
     [\psi_i^*\partial_t \psi_i + H(\{\psi^*\},\{\psi\})]
  \biggl).
  \label{eq:doi_lattice_action}
\ee
The normalization constant $\mathcal{N}$ is now fixed by the condition
\be
  \mathcal{N} = \int [d\psi^*][d\psi] \exp[S(\{\psi^*,\psi\})]
  \label{eq:doi_normalization}
\ee
Before taking the  continuum limit in space let us note
that the initial term $-|\psi_i(0)|^2$ in (\ref{eq:doi_lattice_action})
can actually be dropped in calculations within perturbation theory. In fact, we will define the
functional integral in (\ref{eq:doi_mean_value_func}) in terms of perturbation expansion and use the bilinear part of the dynamic action
to generate propagators, since otherwise the convergence of the functional integral is somewhat problematic \cite{Tau05}.
Thus, we arrive at perturbation theory with a retarded propagator, which we choose such that its value at coinciding time arguments
is zero by definition. In the traditional field-theoretic parlance this is tantamount to defining the time-ordered product at coinciding time
arguments as the normal-ordered product. Since we are dealing with expectation values of functions of the annihilation operators only, a short
reflection of the perturbation expansion reveals that all graphs containing vertices brought about by the initial term $-|\psi_i(0)|^2$
either are proportional to the equal-time value of the propagator, which we have chosen to vanish, or contain closed loops of propagators and therefore
vanish as well.

Continuum limit then
can be performed in traditional manner according to the substitution
\be
  \sum_i\rightarrow \frac{\int d{\mathbf x}}{a^d},\quad
  \psi_i\rightarrow \psi(t,{\mathbf x}) a^d,\quad
  \psi_i^*(t) \rightarrow \psi^\dagger(t,{\mathbf x}),\quad n_0 \rightarrow n_0 a^d ,
\label{eq:doi_conti_lim}
\ee
that leads to the field-theoretic action for the scalar fields
$\psi^\dagger(\mathbf{x},t)$ and $\psi(\mathbf{x},t)$
\be
  S=-\int_0^t dt\int d{\mathbf x} \biggl\{ \psi^\dagger \partial_{t} \psi -D_{0}\psi^\dagger\nabla^{2}\psi
  -\lambda_0[1-\psi^{\dagger2}]\psi^2 \biggl\} +
  \int d{\mathbf x} \biggl[\psi(t,{\mathbf x}) + n_0 \psi^\dagger(0,{\mathbf x})
   - n_0\biggl].
  \label{eq:doi_beforeS}
\ee
It corresponds to the lattice Hamiltonian (\ref{eq:doi_hamil_ann}).
After the shift $\psi^\dagger\rightarrow\psi^\dagger+1$
we arrive at the desired result, which is the
the field theoretic action $S$ for the annihilation reaction $\textit{A} +\textit{A} \rightarrow \varnothing$
\begin{eqnarray}
   S & = & - \int^{t}_{0} dt \int d{\mathbf x} \biggl\{ \psi^\dagger \partial_{t} \psi -D_{0}\psi^\dagger\nabla^{2}\psi
  +  \lambda_{0} D_{0}[2\psi^\dagger+(\psi^\dagger)^{2}]\psi^{2} +
   n_{0}\int d{\mathbf x} \psi^\dagger({\mathbf x},0)  \biggl\}.\nonumber\\
  \label{eq:doi_S1_anni_final}
\end{eqnarray}
which corresponds to the continuum limit of the action used in (\ref{eq:doi_shift_hamil_ann}). It should be noted that
the shift $\psi^\dagger\rightarrow\psi^\dagger+1$ allows to replace the final term $\int d{\mathbf x}\psi(t,{\mathbf x})$
in (\ref{eq:doi_beforeS}) by the initial term $\int d{\mathbf x}\psi(0,{\mathbf x})$ and the contribution of the latter to
the perturbation expansion vanishes by the same token as the contribution of the quadratic initial term. Therefore, we have
not included this linear initial term in the dynamic action either.

The action functional (\ref{eq:doi_S1_anni_final}) will be used
in calculation of physical quantities such as the mean value (\ref{eq:doi_shifted_mean_val}) that can be expressed as
the functional integral
\begin{equation}
 \langle A(t) \rangle = \int \mathcal{D}\psi^\dagger \mathcal{D}\psi
 A_N\left\{ 1,\psi(t)\right\} \rm{e}^{S}.
  \label{eq:doi_expect_val}
\end{equation}
Here $\mathcal{D}\psi^\dagger \mathcal{D}\psi $ is the continuum functional measure.
The continuum limit of the expression (\ref{eq:doi_aver_num_dis}) for mean particle number  becomes
\be
 n(t,{\bf x})=\biggl\langle0\biggl|
 \psi({\mathbf x}) \exp\biggl(-H\{ \psi^\dagger(t)+1,\psi(t)\}  t\biggl)\exp
 \biggl(n_0\int d{\mathbf x}\mbox{ }\psi^\dagger\biggl)
 \biggl|0\biggl\rangle.
  \label{eq:doi_aver_number}
\ee
{\subsection{LANGEVIN EQUATION AND FOKKER-PLANCK EQUATION} \label{sec:doi_lang_eq} }
Many equations describing
evolution of physical, chemical, biological and social processes
are written as mean-field equations for averages of
quantities, which intrinsically are random processes to some extent.
To take fluctuations around the averages into account, a straightforward
way to proceed is to introduce a source of randomness directly in the
mean-field equation. Then the quantities solved from the mean-field
equations become {\em stochastic processes} depending on coordinate variables,
i.e. {\em stochastic fields}.

The paradigmatic example of this procedure is the Langevin equation for
random walk, which describes the position
$\bm{ r}$ of a test particle subject to
random force $\bm{\eta}$
$$
\frac{d\bm{ r}}{dt}=\bm{\eta}
$$
Here, the random force is of zero mean and uncorrelated in time (white noise), i.e.
\[
\langle \eta_i(t)\eta_j(t')\rangle=D\delta_{ij}\delta(t-t')\,.
\]
Strictly speaking, this standard physical formulation is mathematically inconsistent, which
gives rise to inevitable ambiguities in the case of multiplicative noise (when the noise
term is multiplied by a function of the random position).

Consider, for instance, the average distance between two points of the path of the random walk, i.e.
($d$ is the dimension of space)
-- a hint to the property that the Brownian path, although continuous, is not differentiable anywhere

Increments
of the Brownian path
\[
\bm{W}(t)-\bm{W}(t_0)=\int_{t_0}^{t}\!dt\,\bm{\eta}(t)
\]
constitute an extremely important random process, the
{\em Wiener process}, whose conditional probability density is the Gaussian
\[
p\left(\bm{W},t\vert \bm{W}_0,t_0\right)={1\over [4\pi(t-t_0)]^{d/2}}\,e^{-(\bm{W}-\bm{W}_0)^2/2(t-t_0)}\,.
\]
In particular,
The Wiener process is the basis of the mathematically consistent definition of the Langevin equation (the stochastic differential
equation, SDE).

Critical dynamics.
In the Landau theory of phase transitions
the dynamics of the order parameter $\varphi$ near equilibrium are described
by the kinetic equation [time-dependent Ginzburg-Landau (TDGL) equation]
\be
\label{LandauKinetic}
\doo{\varphi}{t}=-\Gamma\left(-\nabla^2\varphi+a\varphi+{\lambda\over
6}\varphi^3\right)\,.
\ee
In linear response theory, dynamics of fluctuations near equilibrium \cite{LLStatPhys} are often described
by kinetic equations similar to (\ref{LandauKinetic}), but with a random noise term added to the right-hand side.

In a more generic setup, standard models of critical dynamics are based on nonlinear Langevin equations
\be
\label{nonlin_Langevin}
\doo{\varphi}{t}=-\Gamma\fundoo{H}{\varphi}+f:=V(\varphi)+f\,,
\ee
where $H$ is the effective equilibrium Hamiltonian. For the random source a suitable Gaussian
distribution is assumed in which the correlation function is determined through the connection
to the static equilibrium (fluctuation-dissipation theorem).
For instance,
model A for the non-conserved order parameter \cite{Hohenberg77} is described by the SDE obtained from
\ref{LandauKinetic} by the addition of a white-noise field to the right-hand side.

In reaction kinetics and population dynamics
the simplest kinetic description of the  dynamics of the average particle numbers is
given by the {\em rate equation}.
The rate equation is a deterministic differential equation for average particle numbers in
a homogeneous system, therefore
it does not take into account boundary conditions,
spatial inhomogeneities and randomness in the individual reaction
events. Spatial dependence is often accounted for by a diffusion term, which
gives rise to models of {\em diffusion-limited reactions} (DLR).

As a simple example, consider the
coagulation reaction $A+A\to A$. The diffusion-limited rate equation for the
concentration $\varphi$ of the compound $A$ is
$$
\doo{\varphi}{t}=D\nabla^2\varphi-k\varphi^2\,,
$$
where $k$ is the {\em rate constant}\,.

The most straightforward way to take into account various effects of
randomness is to add a random source and sink term to the rate equation:
\be
\label{AAtoAnaive}
\doo{\varphi}{t}=D\nabla^2\varphi-k\varphi^2+f\,.
\ee
This is a nonlinear Langevin equation for the field $\varphi$. Physically, in the case of concentration
$\varphi\ge 0$.

There is an important difference between the reaction models and the critical dynamics:
in the latter, deviations of the fluctuating order parameter from the (usually zero)
mean may physically be of any sign (or direction). In particular, deviations from the equilibrium
value are always allowed.
In the reaction there is often an absorbing steady state, which does not permit fluctuations
therefrom: once the system arrives at the absorbing state, it stays there forever.
In particular, if the empty state is an absorbing state of the reaction, the the random source
should be introduced multiplied by a factor vanishing in the limit $\varphi\to 0$ to prevent the system returning
from the absorbing state by the noise. The simplest choice
yields
$$
\doo{\varphi}{t}=D\nabla^2\varphi-k\varphi^2+f\varphi
$$
instead of (\ref{AAtoAnaive}). This is an equation with a {\em multiplicative noise}.
{\subsection{Multiplicative noise.} \label{sec:doi_multi}}
Consider the Langevin equation with the multiplicative noise of generic form
\be
\label{Langevin}
\doo{\varphi}{t}=V(\varphi)+fb(\varphi):=-K\varphi+U(\varphi)+fb(\varphi)\,,
\ee
where
$f$ is (usually) a Gaussian random field with zero mean and the
white-in-time correlation function
\be
\label{correlator}
\langle
f(t,\bm{ x})f(t',\bm{
x}')\rangle=\overline{D}(x-x')=\delta(t-t')D(\bm{ x}-\bm{ x}')\,,
\ee
where the shorthand notation $x=(t, \bm{ x})$ has been used.
In (\ref{Langevin}), $b(\varphi)$ is a functional of $\varphi$ and
$U(\varphi)$ is a nonlinear functional
of $\varphi$. Both functionals are time-local, i.e. depend only on the current times instant of the
SDE.

The Langevin equation with white-in-time noise $f$ is mathematically
inconsistent, because the time integral of the noise $\int f dt$ is
a Wiener process which not differentiable anywhere as a function of time.

This problem may be approached by starting with
the set of correlation functions consisting of a $\delta$ sequence
in time, i.e.
\be
\label{deltaCorrAdditive}
\langle f(t,\bm{
x})f(t',\bm{ x}')\rangle=\overline{D}(t,\bm{ x};t',\bm{ x}')
\xrightarrow[t' \to t]{\hbox{}}\delta(t-t')D(\bm{ x},\bm{ x}')
\ee
and passing to the white-noise limit at a later stage.
>From the mathematical point of view, this treatment gives rise
to the solution of the stochastic differential equation
(\ref{Langevin}) in the Stratonovich sense
\cite{Gardiner97}. Physically, this is often the most natural way to
approach the white-noise case. However, technically the Stratonovich interpretation
gives rise to a rather complicated treatment and the SDE is most often used
in the Ito interpretation in mathematical analyses.

{\subsection{Fokker-Planck equation.} \label{sec:fokker planck}}

Recall that the point of introducing of the SDE with white noise is to avoid dealing with the limit (\ref{deltaCorrAdditive})
explicitly. To this end,
instead of using the mathematically problematic, although physically
transparent, Langevin equation the stochastic problem (\ref{Langevin}), (\ref{correlator}) may be
equivalently stated in terms of the Fokker-Planck equation (FPE), which is an equation for
both the conditional probability density
$
p\left(\varphi,t\vert\varphi_0,t_0\right)
$
and the probability density
$
p\left(\varphi,t\right)
$
of the variable $\varphi$. Recall that both the master equation and the Fokker-Planck equation are special cases
of the generic (forward) Kolmogorov equation and thus the two problems discussed here are closely related.
The simple way to demonstrate this equivalence uses rules of Ito calculus \cite{Gardiner97}, which is beyond the
scope of the present treatment.
Therefore, only the correspondence between the quantities
specifying the stochastic problem in both approaches will be quoted here. The main advantage of the
Fokker-Planck equation is that the equation itself is completely well-defined partial differential
(or functional-differential for field variables) equation. The ambiguity of the Langevin problem
shows in that the FPE is different for different interpretations of the SDE.

For simplicity of notation, consider zero-dimensional field theory.
The Fokker-Planck equation for the conditional probability density
$
p\left(\varphi,t\vert\varphi_0,t_0\right)
$
in the case of the Ito equation is
\begin{multline}
\label{FokkerPlanckIto0}
\doo{}{t}p\left(\varphi,t\vert\varphi_0,t_0\right)
=-\doo{}{\varphi}\left\{\left[-K\varphi+U(\varphi)\right]p\left(\varphi,t\vert\varphi_0,t_0\right)\right\}\\
+{1\over 2}\doo{^2}{\varphi^2}\left[b(\varphi)Db(\varphi)p\left(\varphi,t\vert\varphi_0,t_0\right)\right]\,.
\end{multline}
If the SDE (\ref{Langevin}) is interpreted in the Stratonovich sense, the FPE is
\begin{multline}
\label{FokkerPlanckStrato0}
\doo{}{t}p\left(\varphi,t\vert\varphi_0,t_0\right)
=-\doo{}{\varphi}\left\{\left[-K\varphi+U(\varphi)\right]p\left(\varphi,t\vert\varphi_0,t_0\right)\right\}\\
+{1\over 2}\doo{}{\varphi}\left\{b(\varphi)\doo{}{\varphi}\left[Db(\varphi)p\left(\varphi,t\vert\varphi_0,t_0\right)\right]\right\}\,.
\end{multline}
The conditional probability density
$
p\left(\varphi,t\vert\varphi_0,t_0\right)
$
is the fundamental solution of the FPE (\ref{FokkerPlanckIto0}) or (\ref{FokkerPlanckStrato0}), i.e.
\beq
p\left(\varphi,t_0\vert\varphi_0,t_0\right)=\delta\left(\varphi-\varphi_0\right)\,.\nonumber
\eeq
Contractions are not quite obvious, when the random variable has several components. For instance, the Fokker-Planck equation in the Ito form
becomes
\begin{multline}
\doo{}{t}p\left(\varphi,t\vert\varphi_0,t_0\right)
=-\doo{}{\varphi_i}\left\{\left[-K_{ij}\varphi_j+U_i(\varphi)\right]p\left(\varphi,t\vert\varphi_0,t_0\right)\right\}\\
+{1\over 2}\doo{^2}{\varphi_i\partial\varphi_j}\left[b_{ik}(\varphi)D_{kl}b_{jl}(\varphi)p\left(\varphi,t\vert\varphi_0,t_0\right)\right]\nonumber
\end{multline}
for a multi-component variable $\varphi_i$.

The Fokker-Planck equation may be regarded as the Schr\"odinger equation with imaginary time. Using this analogy, the solution of the
FPE as well as calculation of expectation values may be represented in a way analogous to quantum field theory \cite{Leschke77}.
Construction with the FPE as the starting
point gives rise to the famous Martin-Siggia-Rose solution of the SDE \cite{Martin73}, but avoids ambiguities
inherent in the SDE (they have been fixed by the choice of the FPE).

Consider, for definiteness, the Fokker-Planck equation (\ref{FokkerPlanckIto0}) corresponding to the Ito
interpretation of the Langevin equation (\ref{Langevin}).
Introduce -- in analogy with Dirac's notation in quantum mechanics -- the state vector $\ket{p_t}$ according to
the following representation of the PDF
\[
p(\varphi,t)=\braket{\varphi}{p_t}\,,
\]
which is the solution of the FPE (\ref{FokkerPlanckIto0}) with the initial condition $p(\varphi,0)=p_0(\varphi)$.
To construct the evolution operator for the state vector, introduce momentum and coordinate operators
in the manner of quantum mechanics by relations
\[
\hat{\pi}f(\varphi)=-\doo{}{\varphi}f(\varphi)\,,\quad \hat{\varphi}f(\varphi)=\varphi f(\varphi)\,,\quad \left[\hat{\varphi},\hat{\pi}\right]=1\,.
\]
In these terms, the FPE for the PDF gives rise to the evolution equation for the state vector in the form
\[
\doo{}{t}\ket{p_t}=-\hat{H}\ket{p_t}\,,
\]
where the "Hamilton" operator for the FPE corresponding to the Ito interpretation of the SDE assumes, according to (\ref{FokkerPlanckIto0}), the form
\beq
\label{LioFPEIto}
\hat{H}=-
\opi\left[-K\ophi+U(\ophi)\right]
-{1\over
2}\opi^2b(\ophi)Db(\ophi)\,.
\eeq
Note that, contrary to quantum mechanics, there is no ordering ambiguity in the construction
of the Hamilton operator here.
In this notation, the conditional PDF may be expressed as the matrix element
\beq
\label{CondPDFMatrEl}
p\left(\varphi,t\vert\varphi_0,t_0\right)=\matrel{\varphi}{e^{-\hat{H}(t-t_0)}}{\varphi_0}\,,
\eeq
and the functional-integral representation may be constructed in the same fashion as in the master-equation case.

{\section{FIELD-THEORETIC STUDY OF  REACTION PROCESS $A+A\rightarrow\varnothing $} \label{sec:anni1} }
{\subsection{FIELD-THEORETIC MODEL OF ANNIHILATION PROCESS} \label{sec:anni1_model}}
Let us study anomalous kinetics of the general type of the irreversible single-species
annihilation reaction
\begin{equation}
  A+A \xrightarrow{K_0} \varnothing,
  \label{eq:anni1_rea}
\end{equation}
with the unrenormalized (mean field) rate constant $K_0$.
The first step of the Doi approach \cite{Doi76} (see also \cite{Peliti85}) consists of the introduction of
the creation and annihilation
operators $\psi^\dagger$ and $\psi$ and the vacuum state $|0\rangle$
satisfying the usual bosonic commutation relations
\begin{eqnarray}
 & & [\psi(\mathbf{x}),\psi^\dagger({\mathbf x'})]=\delta(\mathbf{x}-{\mathbf x'}),\nonumber\\*
 & &[\psi(\mathbf{x}),\psi({\mathbf x'})]=[\psi^\dagger(\mathbf{x}),\psi^\dagger(\mathbf{x'})] = 0, \nonumber\\*
 & & \psi(\mathbf{x})|0\rangle   = 0, \langle 0| \psi^\dagger(\mathbf{x}) = 0, \langle 0|0 \rangle = 1.
  \label{eq:anni1_comm_rel}
\end{eqnarray}
Let $P(\{n_i\},t)$ be the joint probability density function (PDF) for observing $n_i$ particles at positions
$\mathbf{x}_i$. The information about the macroscopic state of the classical many-particle system may be
 transferred into the state vector $ | \Phi(t)\rangle$ defined as the sum over all occupation numbers
\begin{equation}
 | \Phi(t)\rangle = \sum_{ \{n_{i}\} } P(\{ n_i \},t) | \{ n_i \} \rangle,
 \label{eq:anni1_sta_vec}
\end{equation}
where the basis vectors are defined as
\begin{equation}
 | \{ n_i \} \rangle = \prod_{i} [\psi^\dagger(\mathbf{x}_i) ]^{n_i} |0\rangle.
 \label{eq:anni1_blabol}
\end{equation}
The whole set of coupled partial differential equations for the PDFs may be
 rewritten in the compact form  of a master equation \cite{Lee94,Doi76}
\begin{equation}
 \frac{\partial}{\partial t}| \Phi(t)\rangle = -\hat{H} | \Phi(t) \rangle,
 \label{eq:anni1_master}
\end{equation}
where $\hat{H} = \hat{H}_A+\hat{H}_D+\hat{H}_R$ and for the annihilation process $A+A \rightarrow \varnothing$
under consideration
\begin{eqnarray}
   & &\hat{H}_A = \int d{\mathbf x} \psi^\dagger \nabla [{\mathbf v}({\mathbf x},t) \psi({\mathbf x})],\nonumber\\*
   & &\hat{H}_D = - D_{0} \int d{\mathbf x} \psi^\dagger \nabla^{2} \psi({\mathbf x}),\nonumber\\*
   & &\hat{H}_R = \lambda_{0}D_0 \int d{\mathbf x} (\psi^\dagger)^{2} \psi^{2},
   \label{eq:anni1_hamil_react}
\end{eqnarray}
corresponding to the advection, diffusion and reaction part \cite{Hnatic00}.
Due to dimensional reasons we have extracted
the diffusion constant $D_0$ from the rate constant $K_0=\lambda_0 D_0$.
The mean of a physical quantity $A(t)$ may be expressed \cite{Hnatic00} -- with the use of the notation of Sec. \ref{sec:doi_quant_method}
 -- as the vacuum expectation value
\begin{equation}
  \langle A(t) \rangle  =  \langle 0| T\bigl[\hat{A}_N\left(\{1,\psi(t) \}\right)
  {\rm e}^{-\int^{\infty}_{0} \hat{H}'_{I} dt + n_{0} \int d{\mathbf x} \psi^\dagger({\mathbf x},0)}  \bigr]|0\rangle.
  \label{eq:anni1_aver2}
\end{equation}
Here, the interaction operator is defined as $\hat{H}'_I=\hat{H}'-\hat{H}'_0$
and the substitution $\psi^\dagger\rightarrow\psi^\dagger+1: \hat{H}'\equiv \hat{H}(\psi^\dagger+1,\psi)$
is understood. The field operators 
 (\ref{eq:anni1_comm_rel}) have been replaced by the time-dependent operators
 of the interaction representation
\begin{equation*}
  \psi^\dagger(t,{\mathbf x})={\mathrm e}^{\hat{H}_0't}\psi^\dagger{\mathbf x}{\mathrm e}^{-\hat{H}_0't},\quad
   \psi(t,{\mathbf x})={\mathrm e}^{\hat{H}_0't} \psi({\mathbf x}){\mathrm e}^{-\hat{H}_0't}.
\end{equation*}
In this formulation - assuming the Poisson distribution
as the initial condition - the average number density can be computed via the expression
\begin{equation}
   n(t,{\mathbf x}) = \langle 0| \psi({\mathbf x}){\mathrm e}^{-\hat{H}' t}
   {\mathrm e}^{n_0\int d{\mathbf x}\psi^\dagger}|0\rangle.
   \label{eq:anni1_aver_number}
\end{equation}
The expectation value of the time-ordered product in (\ref{eq:anni1_aver2}) can
be cast \cite{Vasiliev} into the form of a
functional integral over scalar fields
$\psi^\dagger(\mathbf{x},t)$ and $\psi(\mathbf{x},t)$:
\begin{equation}
 \langle A(t) \rangle = \int \mathcal{D}\psi^\dagger \mathcal{D}\psi A_N\left(\{1,\psi(t)\}\right) \rm{e}^{S_1},
  \label{eq:anni1_expect_val}
\end{equation}
where the action $S_{1}$ for the annihilation reaction $\textit{A} +\textit{A} \rightarrow \varnothing$  is
\begin{eqnarray}
  & & S_{1}  =  - \int^{\infty}_{0} dt \int d{\mathbf x} \{ \psi^\dagger \partial_{t} \psi +
  \psi^\dagger\nabla({\mathbf v}\psi) -D_{0}\psi^\dagger\nabla^{2}\psi
  + \nonumber\\*
  & & \lambda_{0} D_{0}[2\psi^\dagger+(\psi^\dagger)^{2}]\psi^{2} +
  n_{0}\int d{\mathbf x} \psi^\dagger({\mathbf x},0)  \}.
  \label{eq:anni1_S1}
\end{eqnarray}
In order to analyze the effect of velocity fluctuations on the reaction process
we average the expectation value (\ref{eq:anni1_expect_val}) over
the random velocity field $\mathbf{v}$. The most realistic description of the velocity field ${\bf v}(x)$ is
based on the use of the stochastic Navier-Stokes equation.
Due to the incompressibility conditions $\nabla.\mathbf{v}=0$ and $\nabla.\mathbf{f}^v=0$ imposed
on the velocity field $\mathbf{v}$ and the random-force field $\mathbf{f}^v$ it is possible
to eliminate pressure from the Navier-Stokes equation and hence it is sufficient to consider only the transverse components
\begin{equation}
  \partial_{t}{\mathbf v} + P({\mathbf v}.\nabla){\mathbf v}-\nu_{0}\nabla^{2}{\mathbf v}
  ={\mathbf f}^{v}.
  \label{eq:anni1_NS}
\end{equation}
 Here, $\nu_0$ is the molecular kinematic viscosity,
$P_{ij}({\bf k}) = \delta_{ij}-k_ik_j/k^2$ is the transverse
projection operator and $k=|{\bf k}|$ is the norm of the wave vector ${\mathbf k}$.
Here and below we use the subscript $"0"$ for all "bare" parameters
to distinguish them from their renormalized counterparts, which will appear during
the renormalization procedure.

The large-scale random force per unit mass ${\mathbf f}$ is assumed to be a Gaussian random variable with zero mean and the following correlation function
\begin{eqnarray}
 & & \langle f_{m}({\mathbf x}_1,t_1) f_{n}({\mathbf x}_2,t_2) \rangle =  \delta (t_1-t_2) \times\nonumber\\*
 &  & \int \frac{d{\mathbf k}}{(2\pi)^d} P_{mn}({\mathbf k}) d_f(k)
  \rm{e}^{i{\mathbf k}.({\mathbf x}_{1}-{\mathbf x}_{2}  )},
  \label{eq:anni1_cor_ran_for}
\end{eqnarray}
where the kernel function is chosen in the form
\begin{equation}
 d_f(k) =  g_{10} \nu_{0}^{3}  k^{4-d-2\epsilon}  + g_{20}\nu_{0}^{3}k^{2}.
  \label{eq:anni1_kernel}
\end{equation}
The nonlocal term is often used to generate the turbulent velocity field with Kolmogorov's scaling
\cite{Dominicis79,Adzhemyan,Vas_turbo}. This case
is achieved by setting $\epsilon=2$. The local term $g_{20}\nu_0^3 k^2$ has been added not only because of renormalization reasons but has also
an important physical meaning. Such a term in the force correlation function describes generation of thermal fluctuations
of the velocity field near equilibrium \cite{Forster} and thus can mimic the usual environment in which
chemical reactions take place.

Averaging ($\ref{eq:doi_expect_val}$) over the random velocity field ${\mathbf v}$ is done with the
"weight" functional $\mathcal{W}_{2} = \rm{e}^{S_{2}}$,
where $S_{2}$ is the effective action for the advecting velocity field
\begin{eqnarray}
   S_{2} & = & \frac{1}{2} \int dt d{\mathbf x} d {\mathbf {x}'}\mbox{ } \tilde{\mathbf v}
  ({\mathbf x},t) . \tilde{ {\mathbf v} } ( {\mathbf x}',t)
   d_f (| {\mathbf x} -{\mathbf x}' |)  +
  \int dt d{\mathbf x} \mbox{ }
   \tilde{{\mathbf v}}.[-\partial_{t} {\mathbf v}-( {\mathbf v}.\boldnabla)
  {\mathbf v}+\nu_{0}\nabla^{2}{\mathbf v}].\nonumber\\
  \label{eq:anni1_S2}
\end{eqnarray}
With the use of the complete weight functional
\begin{equation}
\mathcal{W} = \rm{e}^{S_{1} + S_{2}}
\label{eq:anni1_total_func}
\end{equation}
the expectation value of any desirable physical quantity is possible to be
calculated.

Actions  (\ref{eq:anni1_S1}) and (\ref{eq:anni1_S2}) for the studied model are written in the form
convenient for the use of the standard Feynman diagrammatic technique. There is, however, one delicate point
which should be noted. In the pure reaction case we were dealing with the Cauchy problem and, consequently, all time integrals
in the dynamic action were written over the positive time axis. When the random drift is included, it is rather natural and
technically much simpler to regard the drift as a stationary random process given on the whole time axis. Then the integration over the
fields $\psi$ and $\psi^\dagger$ should include negative time arguments as well for consistency. In the pure reaction part this does not
make any difference, because the retarded propagators render the time integrations at interaction vertices insensitive to the lower limit of
integration. In the velocity part this leads to immense technical simplification, because the translation invariance with respect to time
is preserved in construction of the perturbation expansion with the only exception of the initial conditions for the fields $\psi$ and $\psi^\dagger$.
In the case of Poisson initial distribution the initial condition is expressed as a linear term in the action and, in particular, does not affect
renormalization of the model.

\begin{figure}[t!]
  \setcaptionmargin{5mm}
  \onelinecaptionstrue
  \includegraphics{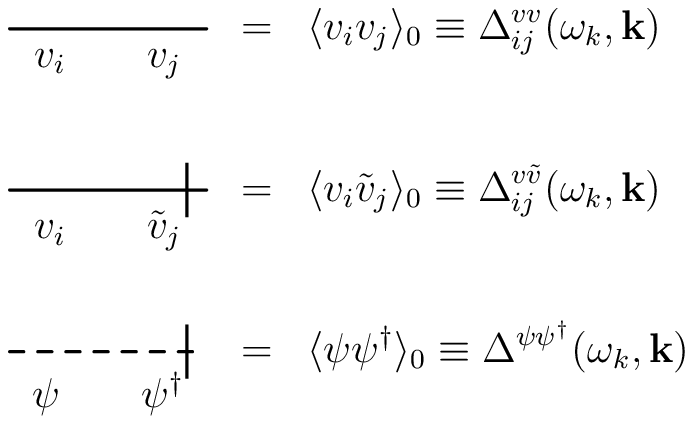}
  \captionstyle{normal}
  \caption{The propagators of the model}
  \label{fig:anni1_propag}
\end{figure}

\begin{figure}[t!]
  \setcaptionmargin{5mm}
  \onelinecaptionstrue
  \includegraphics{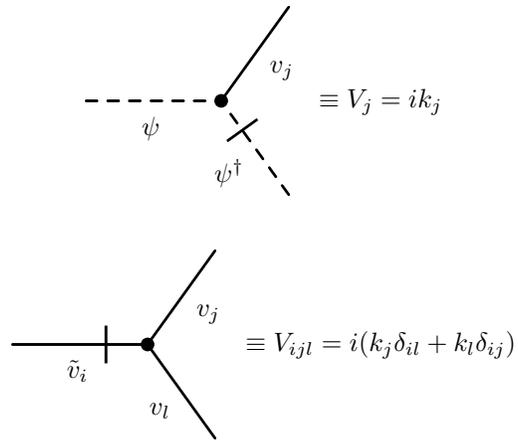}
  \captionstyle{normal}
  \caption{Interaction vertices describing velocity fluctuation and advection
 with corresponding vertex factors}
  \label{fig:anni1_ver_vel}
\end{figure}

\begin{figure}[t!]
  \setcaptionmargin{5mm}
  \onelinecaptionstrue
  \includegraphics{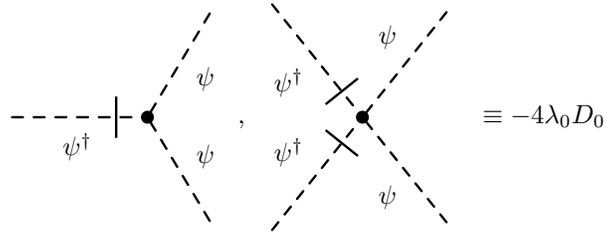}
  \captionstyle{normal}
  \caption{Interaction vertices responsible for density fluctuations with corresponding vertex factor}
  \label{fig:anni1_ver_react}
\end{figure}

Thus, henceforth we assume all time integrals in the dynamic action to be taken over the whole time axis and may use the
standard Fourier representation to express the Feynman rules for the model.
The inverse matrix of the quadratic part of the actions determines the form of the bare propagators.
 It is easily seen that the studied model
 contains three different types of propagators in Figure \ref{fig:anni1_propag}.
In the momentum-frequency representation they are given as
\be
 \Delta^{\psi\psi^\dagger}(\omega_k,{\mathbf k})  =   \frac{1}{-i\omega_k+D_0k^2}
   \label{eq:anni1_prop_freq}
\ee
and in the momentum-time representation as
\be
 \Delta^{\psi\psi^\dagger}(t,{\mathbf k})  =   \theta(t)\exp(-D_0k^2 t).
   \label{eq:anni1_prop_time}
\ee
The vertex factor
\begin{equation}
  V_m(x_1,x_2,\ldots,x_m;\Phi) = \frac{\delta^m V(\Phi)}{\delta\Phi(x_1)\delta\Phi(x_2)\ldots\delta\Phi(x_m)}
  \label{eq:ver_factor}
\end{equation}
 is associated to each interaction vertex of a Feynman graph. Here, $\Phi$ could be any member from the set of all fields $\{\psi^\dagger,\psi,\tilde{v},v \}$.
The interaction vertices from action (\ref{eq:anni1_S2}) describe
interactions between and it may be rewritten in a technically more convenient
form
\begin{equation}
-\int dtd{\mathbf x}\mbox{ } \tilde{\mv}(\mv\partial){\mv} =
  - \int dtd{\mathbf x}\mbox{ }\tilde{v}_i v_k\partial_k v_i = \nonumber
\int dtd{\mathbf x}\mbox{ }(\partial_k \tilde{v}_i) v_k v_i,
  \label{eq:anni1_int_ver}
\end{equation}
where the incompressibility condition $\partial_i v_i=0$ and partial integration method have been used.
We assume that the velocity fields fall off rapidly for $|\mathbf x|\rightarrow\infty$.
Rewriting this functional in the symmetric form $v_i V_{ijl}v_j v_l/2$, it is
easy to find the explicit form for the corresponding
vertex factor in the momentum space
\begin{equation}
  V_{ijl}   = i(k_j\delta_{il}+k_l\delta_{ij}).
  \label{eq:anni1_int_ver_ns}
\end{equation}
Here, the momentum ${\mathbf k}$ is flowing into the vertex through the field $\tilde{v}$.
 The advecting term from the action (\ref{eq:anni1_S1}) can be similarly presented as
\begin{eqnarray}
 & &  -\int dtd{\mathbf x}\mbox{ }\psi^\dagger \boldnabla (\mv \psi)  =
   \int dtd{\mathbf x}\mbox{ } \psi^\dagger \partial_i(v_i\psi)=
    -\int dtd{\mathbf x}\mbox{ }\psi^\dagger v_i\partial_i\psi=
\int dtd{\mathbf x}\mbox{ }(\partial_i\psi^\dagger) v_i\psi.\nonumber\\
  \label{eq:anni1_int_ver_react}
\end{eqnarray}
Rewriting this expression in the form $\psi^\dagger V_j v_j\psi $ we immediately
obtain  the vertex factor in the momentum space
\begin{equation}
  V_j=ik_j,
  \label{eq:anni1_int_ver_adv}
\end{equation}
where the momentum ${\mathbf k}$ represents the momentum flowing into the vertex through the field $\psi^\dagger$.
The vertices $\tilde{v}vv$ and $\psi^\dagger\psi v$ are depicted graphically in Figure \ref{fig:anni1_ver_vel}.
The two reaction vertices derived from the functional (\ref{eq:anni1_S1})
according to the definition (\ref{eq:ver_factor}) are depicted in Figure \ref{fig:anni1_ver_react}
and physically describe the density fluctuations of the reactant particles.

It should be stressed that in our model there is no influence of the reactants on the velocity field itself.
Therefore, the model given by actions (\ref{eq:anni1_S1}) and (\ref{eq:anni1_S2}) may be
characterized as a model for the advection of the "passive" chemically active admixture.


%
%
{\subsection{UV RENORMALIZATION} \label{sec:anni1_uv_renorm}}
The functional formulation provides a theoretical framework
suitable for applying methods of quantum field theory.
 Using RG methods it is possible to determine the IR
asymptotic (large spatial and time scales) behaviour of the correlation functions. First of all
a proper renormalization procedure is needed for the elimination of ultraviolet (UV) divergences.
There are various renormalization prescriptions applicable for such a task, each with its own
advantages. To most popular belong the Pauli-Villars, lattice
and dimensional regularization \cite{Zinn}. In what follows we will employ the
modified minimal subtraction ($\overline{\text{MS}}$) scheme. Strictly speaking, in the analytic
renormalization there is no consistent MS scheme. What we mean here, is the
ray scheme \cite{Adzhemyan05}, in
which the two regularizing parameters $\epsilon$, $\Delta$ ($\epsilon$ has been introduced in (\ref{eq:anni1_kernel})
and $2\Delta=d-2$ was introduced in (\ref{eq:doi_intro_delta})) are taken proportional
to each other: $\Delta=\xi \epsilon$, where the coefficient $\xi$ is arbitrary but fixed. In this case,
only one independent small parameter, say, $\epsilon$ remains and the notion of minimal subtraction becomes meaningful.
UV divergences manifest themselves
in the form of poles in the small expansion parameter and the minimal subtraction scheme
is characterized by discarding all finite parts of the Feynman graphs in the calculation of the renormalization
constants.
In the modified scheme, as usual, certain geometric factors are not expanded in $\epsilon$, however.
This is the content of the $\overline{\text{MS}}$ scheme used in our analysis.

In order to apply the dimensional regularization for the evaluation 
of renormalization constants, an analysis of possible superficial
divergences has to be performed. 
For the power counting in the actions (\ref{eq:anni1_S1}) and
(\ref{eq:anni1_S2}) we use the scheme \cite{Adzhemyan}, in which to each
quantity $Q$ two canonical dimensions are assigned, one with respect
to the wave number $d_Q^k$ and the other to the frequency
$d_Q^\omega$. The normalization for these dimensions is
\be
d^\omega_\omega=-d_t^\omega=1,\quad d_k^k=-d_x^k=1,\quad d_k^\omega=d^k_\omega=0.
\label{eq:anni1_def_power}
\ee
 The canonical (engineering) dimensions
for fields and parameters of the model are derived from the condition for action
to be a scale-invariant quantity, i.e. to have a zero canonical dimension.

The quadratic part of the action (\ref{eq:anni1_S1}) determines only the
canonical dimension of the quadratic product $\psi^\dagger\psi$. In
order to keep both terms in the nonlinear part of the action
\begin{equation}
\lambda_0 D_0 \int
dtd\mathbf{x}[2\psi^\dagger+(\psi^\dagger)^2]\psi^2,
\label{eq:anni1_quad_part}
\end{equation}
the field $\psi^\dagger$ must be dimensionless. If
the field $\psi^\dagger$ has a positive canonical dimension, which
is the case for $d>2$, then
the quartic term should be discarded as irrelevant by the power
counting. The action with the cubic term only, however, does not
generate any loop integrals
corresponding to the density fluctuations and thus is uninteresting for the analysis of fluctuation effects in the
space dimension $d=2$.

Using the normalization choice (\ref{eq:anni1_def_power})
 we are able to obtain the canonical dimensions for all the fields and parameters in the
  $d$-dimensional space. The results
are summarized in Table \ref{tab:anni1_canon_dims}.


Here, $d_Q=d^k_Q+2d_Q^\omega$ is the total canonical dimension and it is determined from the condition that the parabolic
differential operator of the diffusion and Navier-Stokes equation scale uniformly under the simultaneous momentum and frequency dilatation
$k\rightarrow \mu k, \omega\rightarrow \mu^2\omega$.\\
The model is logarithmic when all coupling constants $g_{10},g_{20},\lambda_0$ vanish simultaneously. From
 Table \ref{tab:anni1_canon_dims} it follows that this situation occurs for the choice $\epsilon=\Delta=0$. The UV divergences
 have the form of poles in various
linear combinations of $\epsilon$ and $\Delta$.
 The total canonical dimension of an arbitrary one-particle irreducible Green (1PI)
function $\Gamma=\langle\Phi\ldots \Phi  \rangle_{\rm 1-ir} $ is given by the relation
$  d_\Gamma= d + 2 - N_\Phi d_\Phi,$
where $N_\Phi=\{N_{\psi^\dagger},N_\psi,N_v,N_{\tilde{v}} \}$ are the numbers of corresponding external fields.
The statistical averaging $\langle\ldots\rangle$ means averaging over all possible realizations
of fields $\tilde{\mv},\mv,\psi^\dagger,\psi$ satisfying appropriate boundary conditions
 with the use of the complete weight functional (\ref{eq:anni1_total_func}).
Superficial UV divergences may be present
only in those $\Gamma$ functions for which $d_\Gamma$ is a non-negative integer.
The superficial degree of divergence for a 1PI Green function $\Gamma$ is
\begin{equation}
d_\Gamma=4-N_v-N_{\tilde{v}}-2N_\psi.
\end{equation}
However, the real degree of divergence $\delta_\Gamma$ is smaller, because of the
structure of the interaction vertex (\ref{eq:anni1_int_ver_ns}),which allows for factoring out
 the differential operator $\partial$ to each external line $\tilde{v}$. The real divergence exponent $\delta_\Gamma$
 may then be expressed as
\begin{equation}
  \delta_\Gamma\equiv d_\Gamma-N_{\tilde{v}} = 4-N_v-2N_{\tilde{v}}-2N_\psi
  \label{eq:anni1_real_div_exp}
\end{equation}
 Although the canonical dimension for
the field $\psi^\dagger$ is zero, there is no proliferation of superficial divergent
graphs with arbitrary number of external $\psi^\dagger$ legs. This is due to the fact that $n_{\psi^\dagger}\leq n_\psi$, which
may be established by a straightforward analysis of the graphs \cite{Lee94}.
Brief analysis shows that the UV divergences are expected only for the 1PI Green functions listed in the Table \ref{tab:anni1_canon_green}.


This theoretical analysis leads to the following renormalization of parameters $g_0, D_0$ and $u_0$
\begin{eqnarray}
& &  g_1 = g_{10}\mu^{-2\epsilon} Z_1^3,\quad g_2=g_{20}\mu^{2\Delta}Z_1^3 Z_3^{-1},\quad u=u_0 Z_1 Z_2^{-1}, \nonumber\\*
 & & \lambda = \lambda_0 \mu^{2\Delta} Z_2 Z_4^{-1},\quad \nu=\nu_0 Z_1^{-1},\quad D=D_0Z_2^{-1},
 \label{eq:anni1_ren_rel}
\end{eqnarray}
where $\mu$ is the reference mass scale in the MS scheme~\cite{Zinn}
and we have introduced the inverse Prandtl number $u=D/\nu$ for convenience.
It represents the ratio between diffusion and viscosity forces in a liquid.
In terms of introduced renormalized parameters the total renormalized action for the annihilation reaction in a fluctuating
velocity field is
\begin{eqnarray}
 S_R &=& \int d{\mathbf x} dt \biggl\{
  \psi^\dagger\partial_t\psi +\psi^\dagger\boldnabla({\mathbf v}\psi)-u\nu Z_2\nabla^2\psi+
  \lambda u \nu \mu^{-2\Delta} Z_4[2\psi^\dagger+(\psi^\dagger)^2]\psi^2
  +\nonumber\\& & n_0\int d{\mathbf x}    \psi^\dagger({\mathbf x},0)-
  \frac{1}{2} \tilde{{\mathbf v}}[g_1\nu^3\mu^{2\epsilon}(-\nabla^2)^{1-\Delta-\epsilon} -
  g_2\nu^3\mu^{-2\Delta}Z_3\nabla^2]\tilde{{\mathbf v}}+\nonumber\\*
  & & \tilde{{\mathbf v}}.[\partial_t{\mathbf v}+({\mathbf v}.\boldnabla){\mathbf v}-
  \nu Z_1\nabla^2{\mathbf v}] \biggl\}.
  \label{eq:anni1_total_act}
\end{eqnarray}
The renormalization constants $Z_i,i=1,2,3,4$ are to be calculated perturbatively through
the calculation of the UV divergent parts of the 1PI functions
$\Gamma_{\psi^\dagger\psi},\Gamma_{\psi^\dagger\psi^2},\Gamma_{(\psi^\dagger)^2\psi^2}$,
$\Gamma_{\tilde{v}v}$ and $\Gamma_{\tilde{v}\tilde{v}}$. Interaction terms
corresponding to these functions have to be
added to the original action $S=S_1+S_2$ with the aim to ensure UV finiteness of all Green
functions generated by the renormalized action $S_R$. At this stage the main goal is to
calculate the renormalization constants $Z_i,i=1,2,3,4$.\\
 The singularities in various Green functions will be realized in
the form of poles in $\epsilon$ and $\Delta$ and their linear combinations such as $2\epsilon+\Delta$ or
$\epsilon-\Delta$. Recall that for the consistency of the $\overline{\rm MS}$ scheme it is necessary that the
ratio
\be \xi=\frac{\Delta}{\epsilon}
  \label{eq:anni1_def_xi}
\ee
is a finite real number.
It should be noted that the graphs corresponding to
 $\Gamma_{\psi^\dagger\psi^2}$ and $\Gamma_{(\psi^\dagger)^2\psi^2}$
differ only by one external vertex and thus give rise to equal renormalization of the rate constant $\lambda_0 D_0$.
Therefore, in what follows, we will always consider the function $\Gamma_{\psi^\dagger\psi^2}$.
In order to calculate the renormalization constants $Z_2$ and $Z_4$ we proceed according to the general scheme
suggested in \cite{Adzhemyan05}. We require the fulfillment of UV finiteness (i.e.finite limit
when $\epsilon,\Delta\rightarrow 0$ ) of the 1PI functions $\Gamma_{\psi^\dagger\psi}|_{\omega=0}$
and $\Gamma_{\psi^{\dagger}\psi^2}|_{\omega=0}$. Because
the divergent part of the Feynman graphs should not depend on the value of $\omega$, we have adopted
the simplest choice $\omega=0$. It is convenient to introduce the dimensionless
expansion variables of the perturbation theory as
\begin{equation}
  \alpha_{10} \equiv \frac{g_{10}\overline{S_d}}{p^{2\epsilon}}\,,\quad
   \alpha_{20}\equiv\frac{g_{20}\overline{S_d}}{p^{-2\Delta}}\,,\quad
      \alpha_{30} \equiv \frac{\lambda_0\overline{S_d}}{p^{-2\Delta}},
  \label{eq:anni1_exp_par1}
\end{equation}
where $S_d$ is the surface area of the unit sphere in $d-$dimensional space, $p$ is the total
momentum flowing into the Feynman diagram and $\overline{S_d}=S_d/(2\pi)^d$.
For brevity, in the following we use the abbreviation $\overline{g}_0\equiv g_0\overline{S_d}$ for
the parameters $\{g_{10},g_{20},\lambda_0 \}$ or their renormalized counterparts, respectively.
Next we demonstrate the perturbation series for the 1PI Green functions to the
second order approximation.
The perturbative expansion for $\Gamma_{\psi^\dagger \psi}$ may be written as
\begin{equation}
 \Gamma_{\psi^{\dagger} \psi}|_{\omega=0} = D_0 p^2
 \biggl[-1+\sum_{\begin{subarray}{l}
                    n_1,n_2\ge 0,\\ n_1+n_2\ge 1
                 \end{subarray}}^{n_1+n_2=2}
 \alpha_{10}^{n_1}
\alpha_{20}^{n_2}\gamma_{\psi^{\dagger}\psi}^{(n_1,n_2)}(d,u_0) \biggl],
\label{eq:anni1_GF_prop}
\end{equation}
where $\gamma_{\psi^\dagger\psi}$ are dimensionless coefficients which contain poles in $\epsilon$ and $\Delta$. Explicit
dependence on the space dimension $d$ and inverse Prandtl number $u_0$ is emphasized. It is important
to note that there are no terms in this sequence proportional to the expansion parameter $\alpha_{30}$. In terms
of the renormalized parameters perturbative expansion for the Green function is
(\ref{eq:anni1_GF_prop})
\begin{equation}
   \frac{\Gamma_{\psi^{\dagger}\psi}|_{\omega=0}}{Dp^2} = Z_2 \biggl[-1+\sum_{\begin{subarray}{l}
                    n_1,n_2\ge 0,\\ n_1+n_2\ge 1
                 \end{subarray}}^{n_1+n_2=2}
   \alpha_{1}^{n_1} \alpha_{2}^{n_2}  \gamma_{\psi^{\dagger}\psi}^{(n_1,n_2)}(d,u) \biggl],
  \label{eq:anni1_GF_renprop}
\end{equation}
with the renormalized parameters $\alpha_{1}=  \overline{g_1} s^{2\epsilon} Z_1^{-3}$
and $\alpha_{2}= \overline{g_2} s^{-2\Delta} Z_3 Z_1^{-3}$ in
accordance with the relations (\ref{eq:anni1_ren_rel}) and (\ref{eq:anni1_exp_par1}), where $s\equiv \mu/p$.
Here we would like to stress, that in order to get the correct expansion in $\epsilon$ and $\Delta$, one
has to make replacement
\begin{equation}
d\rightarrow 2+2\Delta, \qquad u_0\rightarrow Z_1^{-1}Z_2 u,
\label{eq:anni1_change}
\end{equation}
 in the arguments of $\gamma_{\psi^{\dagger}\psi}^{(n_1,n_2)}$.
In the same way, the perturbation expansion series for the Green function $\Gamma_{\psi^\dagger \psi^2}$ is
\begin{eqnarray}
\Gamma_{\psi^\dagger \psi^2} |_{\omega =0}& = & -4 D_0 \lambda_0\biggl[1+\sum_{\begin{subarray}{l}
                    n_1,n_2,n_3\ge 0,\\ n_1+n_2+n_3\ge 1
                 \end{subarray}}^{n_1+n_2+n_3=2}
\alpha_{10}^{n_1}\alpha_{20}^{n_2} \alpha_{30}^{n_3}
\gamma_{\psi^\dagger \psi^2}^{(n_1,n_2,n_3)} (d,u_0) \biggl],
\label{eq:anni1_GF_ver}
\end{eqnarray}
where $\gamma_{\psi^\dagger \psi^2}$ are dimensionless coefficients resulting from calculation of Feynman graphs.
Again by replacing the bare parameters with the renormalized counterparts the following series is obtained
\begin{eqnarray}
 \frac{ \Gamma_{\psi^{\dagger} \psi^2} |_{\omega =0}}{4\lambda D \mu^{-2\Delta}} & = &
  -Z_4 \biggl[1+\sum_{\begin{subarray}{l}
                    n_1,n_2,n_3\ge 0,\\ n_1+n_2+n_3\ge 1
                 \end{subarray}}^{n_1+n_2+n_3=2}
   \alpha_{1}^{n_1} \alpha_{2}^{n_2} \alpha_{3}^{n_3}
   \gamma_{\psi^{\dagger}\psi^2}^{(n_1,n_2,n_3)} (d,u) \biggl],
\label{eq:anni1_GF_renver}
\end{eqnarray}
where the dimensionless parameter $\alpha_{3}= \overline{\lambda} s^{-2\Delta} Z_2^{-1} Z_4$ is introduced
and the change (\ref{eq:anni1_change}) is understood. The perturbation series for the Green
function $\Gamma_{(\psi^{\dagger})^2 \psi^2} $ has the same form, so
we do not present it.

Denoting by $Z^{(n)}$ the contribution of the order $g^n,g=\{g_1,g_2,\lambda \}$ the first order
of renormalization constants $Z_2$ and $Z_4$ may be calculated via equations
\ba
 Z_2^{(1)}  &=& \mathcal{L}[ \overline{g_1} s^{2\epsilon}\gamma_{\psi^{\dagger}\psi}^{(1,0)}
  +\overline{g_2} s^{-2\Delta}\gamma_{\psi^{\dagger}\psi}^{(0,1)} ],
  \label{eq:anni1_Z2_ord1}
\ea
\ba
 Z_4^{(1)} &=& - \mathcal{L} [\overline{g_1}s^{2\epsilon}\gamma^{(1,0,0)}_{\psi^\dagger \psi^2}+
 \overline{g_2}s^{-2\Delta}\gamma^{(0,1,0)}_{\psi^\dagger \psi^2}+ \overline{\lambda} s^{-2\Delta}
 \gamma^{(0,0,1)}_{\psi^\dagger \psi^2}],
 \label{eq:anni1_Z4_ord1}
\end{eqnarray}
 where $\mathcal{L}$ stands for the operation of extraction of the UV-divergent part (poles in $\epsilon$ and $\Delta$ or
their linear combination).  In the $\overline{\rm MS}$ scheme  finite terms are discarded, so we do not need to take care of them.
At the second order the term for $Z_2$ can be schematically written as
\ba
   Z_2^{(2)} & = &  \mathcal{L}\biggl[
  -\frac{ \overline{g_1} s^{2\epsilon} }{1+u}\biggl(u Z_2^{(1)}+(u+2)Z_1^{(1)}\biggl)
 \gamma_{\psi^{\dagger}\psi}^{(1,0)} -
 \overline{g_2} s^{-2\Delta} \biggl(\frac{u}{1+u} Z_2^{(1)}+\frac{u+2}{1+u}Z_1^{(1)} -\nonumber\\*
 & & Z_3^{(1)} \biggl)
\gamma_{\psi^{\dagger}\psi}^{(0,1)}+ \overline{g_1}^2 s^{4\epsilon}
 \gamma_{\psi^{\dagger}\psi}^{(2,0)}+ \overline{g_1g_2} s^{2\epsilon-2\Delta}\gamma_{\psi^{\dagger}\psi}^{(1,1)}
 +\overline{g_2}^2 s^{-4\Delta}\gamma_{\psi^{\dagger}\psi}^{(0,2)}\biggl]. \nonumber\\*
 \label{eq:anni1_Z2_ord2}
\ea



\begin{figure}[t!]
  \setcaptionmargin{5mm}
  \onelinecaptionstrue
  \includegraphics{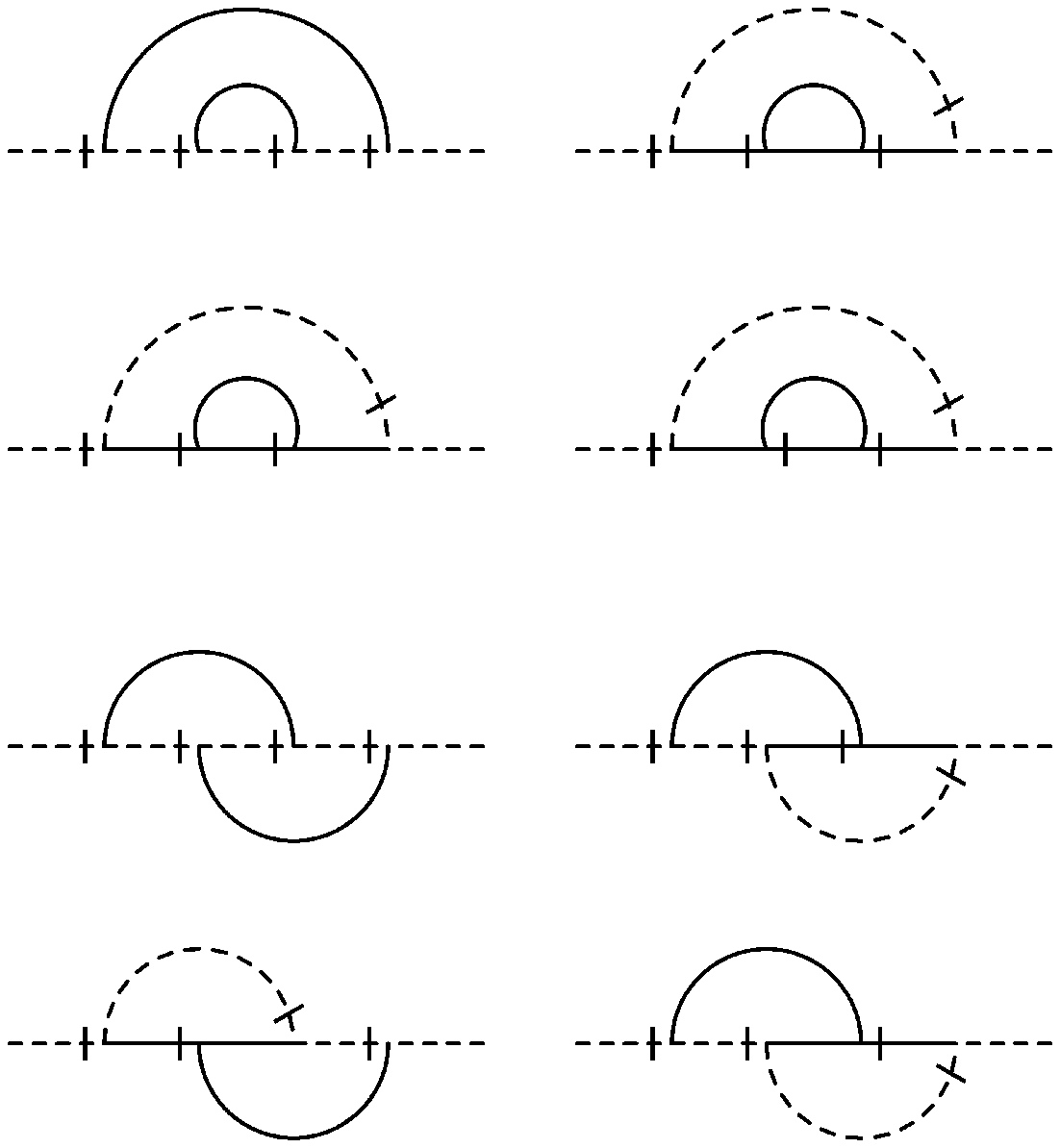}
  \captionstyle{normal}
  \caption{Two-loop graphs for the perturbation expansion of $\Gamma_{\psi^\dagger\psi}$}
  \label{fig:anni1_loop_prop}
\end{figure}

 The two-loop graphs that contribute to the calculation of $Z_2$ are represented by the
graphs depicted in Figure \ref{fig:anni1_loop_prop}.
For the renormalization constants $Z_4$ we have the expression
\begin{eqnarray}
 Z_4^{(2)} &=& - \mathcal{L}[
   \overline{g_1}\overline{\lambda} s^{2(\epsilon-\Delta)} \gamma^{(1,0,1)}_{(\psi^{+})^2 \psi^2}
 +\overline{g_2}\overline{\lambda} s^{-4\Delta}\gamma^{(0,1,1)}_{\psi^\dagger \psi^2}
 + \overline{\lambda}^2 s^{-4\Delta}\gamma^{(0,0,2)}_{\psi^\dagger \psi^2} +
 \overline{g_1} s^{2\epsilon} \gamma^{(1,0,0)}_{\psi^\dagger \psi^2} (-3Z_1^{(1)})
 +\nonumber\\*
 & & \overline{g_2}s^{-2\Delta} \gamma^{(0,1,0)}_{\psi^\dagger \psi^2}(Z_3^{(1)}-3Z_1^{(1)})
+\overline{\lambda}s^{-2\Delta}\gamma^{(0,0,1)}_{\psi^\dagger \psi^2}(2Z_4^{(1)}-Z_2^{(1)})].
 \label{eq:anni1_Z4_ord2}
\end{eqnarray}



\begin{figure}[t!]
  \setcaptionmargin{5mm}
  \onelinecaptionstrue
  \includegraphics{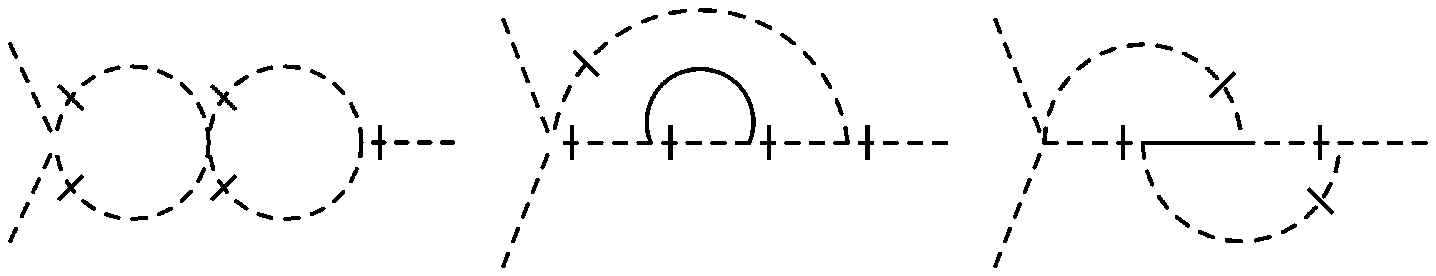}
  \captionstyle{normal}
  \caption{Two-loop graphs for the perturbation expansion of $\Gamma_{\psi^\dagger\psi^2}$}
  \label{fig:anni1_loop_vert}
\end{figure}

The two-loop graphs that contribute to the calculation of $Z_4$ are represented by the
graphs depicted in Figure \ref{fig:anni1_loop_vert}.
>From these expressions the renormalization constants $Z_2$ and $Z_4$ can be calculated in the form
\begin{eqnarray}
  Z_2 & = &  1  -\frac{\overline{g_1}}{8u(1+u)\epsilon}+\frac{\overline{g_2}}{8u(1+u)\Delta}
 +\frac{A_{11}\overline{g_1}^2}{\epsilon^2}
+ \frac{A_{22}\overline{g_2}^2}{\Delta^2}
+\frac{A_{12}\overline{g_1g_2}}{\epsilon\Delta}
+\nonumber\\*
& &
\frac{B_{11}\overline{g_1}^2}{\epsilon}
+\frac{B_{22}\overline{g_2}^2}{\Delta}
+\frac{B_{12}\overline{g_1g_2}}{\epsilon-\Delta}
,
\label{eq:anni1_Z2}
\end{eqnarray}
\begin{eqnarray}
  Z_4 & = & 1 - \frac{\overline{\lambda}}{2\Delta} - \frac{1}{16u(1+u)}
\frac{\overline{g_1\lambda}}{(\epsilon-\Delta)\Delta} +
\frac{1}{32u(1+u)} \frac{\overline{g_2\lambda}}{\Delta^2}
  +\frac{\overline{\lambda}^2}{4\Delta^2}-\nonumber\\*
  & & \biggl(\frac{\overline{g_1\lambda}}{\epsilon-\Delta}-
   \frac{\overline{g_2\lambda}}{\Delta}\biggl)C(u,\xi).
   \label{eq:anni1_Z4}
\end{eqnarray}
The lengthy expressions for the coefficient functions $A_{ij}(\xi,u)$,$B_{ij}(\xi,u)$ and $C(u,\xi)$
 can be found in appendix \ref{sec:appendix_renorm}.

In a similar way we obtain renormalization constants $Z_1$ and $Z_3$ \cite{Adzhemyan05} from
 condition of the UV finiteness
for the 1PI Green functions $\Gamma_{\tilde{v}v}|_{\omega=0}$
and $\Gamma_{\tilde{v}\tilde{v}}|_{\omega=0}$. The perturbation series for $\Gamma_{\tilde{v}v}$ can be written as
\begin{eqnarray}
  & & \Gamma_{\tilde{v}v}|_{\omega=0}=\nu_0 p^2 P_{ij}^p\biggl[-1+\sum_{\begin{subarray}{l}
                    n_1,n_2\ge 0,\\ n_1+n_2\ge 1
                 \end{subarray}}^{n_1+n_2=2}
 \alpha_{10}^{n_1}
\alpha_{20}^{n_2}\gamma_{\tilde{v}v}^{(n_1,n_2)}(d) \biggl],
\end{eqnarray}
and for $\Gamma_{\tilde{v}\tilde{v}}$ as
\begin{eqnarray}
  \Gamma_{\tilde{v}\tilde{v}}|_{\omega=0} & = & P_{ij}^p\biggl[g_{10}\nu_0^3p^{2-2\Delta-2\epsilon}+
 g_{20}\nu_0^3p^2 \biggl\{-1+ \sum_{\begin{subarray}{l} n_1\ge 0,n_2\ge -1,\\ n_1+n_2\ge 1
                 \end{subarray}}^{n_1+n_2=2}
 \alpha_{10}^{n_1}
\alpha_{20}^{n_2}\gamma_{\tilde{v}\tilde{v}}^{(n_1,n_2)}(d) \biggl\}
\biggl].
\end{eqnarray}
>From the definition of the projection operator $P_{ij}^p$ it is easy to see, that after
contracting indices $i$ and $j$ we are left with the constant $d-1$. Hence, rewriting
perturbations series for $\Gamma_{\tilde{v}v}$ and $\Gamma_{\tilde{v}\tilde{v}}$ in
the renormalized variables (\ref{eq:anni1_ren_rel}) and contracting indices $i$ and $j$ we get
\begin{eqnarray}
  & & \frac{\Gamma_{\tilde{v}v}|_{\omega=0}}{\nu p^2(d-1)}=
   -Z_1+Z_1\sum_{\begin{subarray}{l}
                    n_1,n_2\ge 0,\\ n_1+n_2\ge 1
                 \end{subarray}}^{n_1+n_2=2}
 \alpha_{1}^{n_1}
\alpha_{2}^{n_2}\gamma_{\tilde{v}v}^{(n_1,n_2)}(d) \biggl], \nonumber\\*
\end{eqnarray}
\begin{eqnarray}
  \frac{\Gamma_{\tilde{v}\tilde{v}}|_{\omega=0}}{(d-1)g_2\nu^3\mu^{-2\Delta}p^2}
& = & \frac{g_1}{g_2}s^{2\epsilon+2\Delta}+Z_3+Z_3\times  \sum_{\begin{subarray}{l} n_1\ge 0,n_2\ge -1,\\ n_1+n_2\ge 1
                 \end{subarray}}^{n_1+n_2=2}
 \alpha_{1}^{n_1}
\alpha_{2}^{n_2}\gamma_{\tilde{v}\tilde{v}}^{(n_1,n_2)}(d)
.\nonumber\\
\end{eqnarray}
By the same algorithm as described above in detail for the calculation $Z_2$ and $Z_4$
explicit expressions for the renormalization constants $Z_1$ and $Z_3$ are obtained. The results for  them
in the $\overline{\text{MS}}$ scheme can be found in \cite{Adzhemyan05}.

{\subsection{IR STABLE FIXED POINTS AND SCALING REGIMES} \label{sec:anni1_fp}}
The coefficient functions of the RG differential operator for the Green functions
\begin{equation}
  D_{{\rm RG}} = \mu\frac{\partial}{\partial\mu}\biggl|_0 = \mu\frac{\partial}{\partial \mu} +\sum_{g_i}
 \beta_{i}\frac{\partial}{\partial g_i}
  -\gamma_1 \nu\frac{\partial}{\partial\nu},
  \label{eq:anni1_RGeq}
\end{equation}
where the bare parameters are denoted with the subscript ``0'', are defined as
\begin{equation}
  \gamma_1=\mu\frac{\partial  \ln Z_1}{\partial\mu}\biggl|_0\,,\qquad
\beta_{i}=\mu\frac{\partial g_i}{\partial\mu}\biggl|_0,
\label{eq:anni1_RGfunctions}
\end{equation}
with the charges $g_i=\{g_1,g_2,u,\lambda \}$.
 From this definition and the renormalization relations (\ref{eq:anni1_ren_rel}) it follows that
\begin{eqnarray}
& &  \beta_{g_1} = g_1(-2\epsilon+3\gamma_1)\,,\qquad \beta_{g_2} = g_2(2\Delta+3\gamma_1-\gamma_3), \nonumber\\*
& & \beta_\lambda = \lambda(2\Delta-\gamma_4+\gamma_2)\,,\quad \beta_u = u(\gamma_1-\gamma_2),
   \label{eq:anni1_exp_beta}
\end{eqnarray}
where the anomalous dimensions $\gamma_\alpha$ ($\alpha=2,3,4$) are defined as
\begin{equation}
    \gamma_\alpha=\mu\frac{\partial  \ln Z_\alpha}{\partial\mu}\biggl|_0.
    \label{eq:anni1_def_gamma}
\end{equation}
We are interested in the IR asymptotics of small momentum ${\bf p}$ and frequencies $\omega$
of the renormalized functions or, equivalently, large relative distances and time differences in
the $(t,{\bf x})$ representation. Such a behaviour is governed by the IR-stable fixed point $g^*=(g_1^*,g_2^*,u^*,\lambda^*)$, which are determined as zeroes of the $\beta$ functions
$\beta(g^*)=0$. The fixed point $g^*$ is IR stable, if real parts of all eigenvalues
of the matrix $\omega_{ij}\equiv \partial \beta_i/\partial g_j|_{g=g^*}$
are strictly positive. From the knowledge of renormalization constants
$Z_2$ and $Z_4$ (\ref{eq:anni1_Z2}),(\ref{eq:anni1_Z4}) and definitions
 (\ref{eq:anni1_RGfunctions}),(\ref{eq:anni1_def_gamma}) it is possible to calculate anomalous dimensions
$\gamma_2$ and $\gamma_4$
\ba
  \gamma_2 & = &   \frac{\overline{g_1}+\overline{g_2}}{4u(1+u)}  -4B_{11}\overline{g_1}^2+
4B_{22}\overline{g_2}^2 - 2 B_{12}\overline{g_1g_2},\\
  \gamma_4 & = &  -\overline{\lambda} + \overline{\lambda}(\overline{g_1}+\overline{g_2})C(u,\xi).
  \label{eq:anni1_gamma24}
\ea
A straightforward calculation shows that higher order poles cancel each other, so
 that the anomalous dimensions $\gamma_2$ and $\gamma_4$ are finite.
  For completeness we quote also anomalous dimensions $\gamma_1$ and
$\gamma_3$ \cite{Adzhemyan05} to the same order
\begin{eqnarray}
     \gamma_1 & = &  \frac{\overline{g_1}+\overline{g_2}}{16} + \frac{(4\xi+3)}{512(2+\xi)}\overline{g_1}^2 +
  \frac{5\xi+3}{512}\overline{g_1g_2} - \frac{R}{256}(\overline{g_1}+\overline{g_2})^2,
  \label{eq:anni1_gamma1}
\end{eqnarray}
\begin{eqnarray}
  \gamma_3 & = &  \frac{(\overline{g_1}+\overline{g_2})^2}{16\overline{g_2}} -
  \frac{\xi(13+19\xi)}{1024(2+\xi)} \frac{\overline{g_1}^3}{\overline{g_2}} +
  \frac{34\xi+19+6\xi^2}{512(2+\xi)}\overline{g_1}^2 -\frac{3\overline{g_1}^2}{512}+
  \frac{13+31\xi}{1024}\overline{g_1g_2} +\nonumber\\*
  & & \frac{1-R}{256}\frac{(\overline{g_1}+\overline{g_2})^3}{\overline{g_2}},
  \label{eq:anni1_gamma3}
\end{eqnarray}
where the value $R=-0.168$ is a result from numerical integration.
Zeroes of the beta functions (\ref{eq:anni1_exp_beta}) determine possible IR behaviour of the model.
There are four IR stable fixed points and
 one IR unstable fixed point. In this section we present them with their regions of stability.\\
({\it i}$\,$) The trivial (Gaussian) fixed point
\begin{equation}
\label{anni1_Gaussian}
\overline{g_1}^*=\overline{g_2}^*=\overline{\lambda}^*=0\,,
\end{equation}
with no restrictions on the inverse Prandtl number $u$. The Gaussian fixed point is stable, when
\begin{equation}
\label{anni1_Gstable}
\epsilon<0\,,\qquad \Delta >0\,.
\end{equation}
and physically corresponds to the case, when the mean-field solution is valid and fluctuation
effects negligible.\\
({\it ii}$\,$)
The short-range (thermal) fixed point
\begin{eqnarray}
\label{anni1_SR}
& & \overline{g_1}^*=0\,,\quad \overline{g_2}^*=-16\Delta+8(1+2R)\Delta^2,\nonumber\\*
& & u^*=\frac{\sqrt{17}-1}{2}-1.12146\Delta,\nonumber\\*
& &\overline{\lambda}^*=-\Delta+\frac{\Delta^2}{2}\,(\xi-2.64375),
\end{eqnarray}
at which local correlations of the random force dominate over the long-range
correlations. This fixed point
has the following basin of attraction
\begin{eqnarray}
\label{anni1_SRstable}
& & \Delta-\frac{2R-1}{2}\Delta^2<0,\quad 2\epsilon+3\Delta-\frac{3\Delta^2}{2} <0,\\*
& & \Delta+\frac{1}{2}\Delta^2<0,\quad \Delta+0.4529\Delta\epsilon<0
\end{eqnarray}
and corresponds to anomalous decay faster than that due to density fluctuations only, but
slower than the mean-field decay.\\
({\it iii}$\,$)
The kinetic~\cite{Hnatic99} fixed point with finite rate coefficient
 \begin{eqnarray}
& & \overline{g_{1}}^{\ast}=
 \frac{32}{9}\,\frac{\epsilon\,(2\epsilon+3\Delta)}
 {\epsilon+\Delta}+g_{12}^*(\xi)\epsilon^2,\nonumber\\*
& &  \overline{g_{2}}^{\ast}=\frac{32}{9}\,
 \frac{\epsilon^2}{\Delta+\epsilon}+g_{22}^*(\xi)\epsilon^2,\nonumber\\*
& & u^*=\frac{\sqrt{17}-1}{2}+u_1^*(\xi)\epsilon, \nonumber\\*
& & \overline{\lambda}^*=-\frac{2}{3}(\epsilon+3\Delta)+\frac{1}{9\pi}(3\Delta+\epsilon)(Q\epsilon-\Delta),
\label{anni1_K1}
 \end{eqnarray}
 Here $Q=1.64375$. The fixed point (\ref{anni1_K1}) is stable, when inequalities
\begin{equation}
\label{anni1_K1stable}
\Omega_{\pm}>0\,,\qquad\epsilon>0\,,\qquad -\frac{2}{3}\epsilon<\Delta<-\frac{1}{3}\epsilon,
\end{equation}
are fulfilled, where
\begin{eqnarray}
  \Omega_{\pm} & = & \Delta+\frac{4}{3}\epsilon\pm\frac{\sqrt{9\Delta^2-12\epsilon\Delta-8\epsilon^2}}{3} +
\frac{2}{9}\biggl(-(3+2R)\epsilon^2  - 3\epsilon\Delta\pm \nonumber\\*
& & \frac{4\epsilon(\epsilon+3\Delta)R-6\epsilon^2-12\epsilon\Delta-9\Delta^2}{\sqrt{9\Delta^2-12\Delta\epsilon-8\epsilon^2}}\epsilon \biggl)
\end{eqnarray}
The decay rate controlled by this fixed point
of the average number density is faster than the decay rate induced by dominant local
force correlations, but still slower than the mean-field decay rate.

({\it iv}$\,$)
The kinetic fixed point with vanishing rate coefficient:
 \begin{eqnarray}
 & & \overline{g_{1}}^{\ast}=
 \frac{32}{9}\,\frac{\epsilon\,(2\epsilon+3\Delta)}
 {\epsilon+\Delta}+g_{12}^*(\xi)\epsilon^2,\nonumber\\*
& &  \overline{g_{2}}^{\ast}=\frac{32}{9}\,
 \frac{\epsilon^2}{\Delta+\epsilon}+g_{22}^*(\xi)\epsilon^2, \nonumber\\*
& & u^*=\frac{\sqrt{17}-1}{2}+u_1^*(\xi)\epsilon\,,\quad  \overline{\lambda}^*=0\,.
\label{anni1_K2}
 \end{eqnarray}
This fixed point is stable, when the long-range correlations of the random
force are dominant
\begin{equation}
\label{anni1_K2stable}
\Omega_{\pm}>0\,,\qquad\epsilon>0\,,\qquad \Delta>-\frac{1}{3}\epsilon\,,
\end{equation}
and corresponds to reaction kinetics with the normal (mean-field like) decay rate.\\
({\it v}$\,$)
Driftless fixed point given by
\begin{equation}
   \overline{g_1}^*=\overline{g_2}^*=0,\quad u^* \mbox{ not fixed},\quad \overline{\lambda}^*=-2\Delta,
\end{equation}
with the following eigenvalues
\begin{equation}
  \Omega_1=-2\epsilon,\quad \Omega_2=-\Omega_4=2\Delta,\quad \Omega_3=0.
\end{equation}



\begin{figure}[t!]
  \setcaptionmargin{5mm}
  \onelinecaptionstrue
  \includegraphics{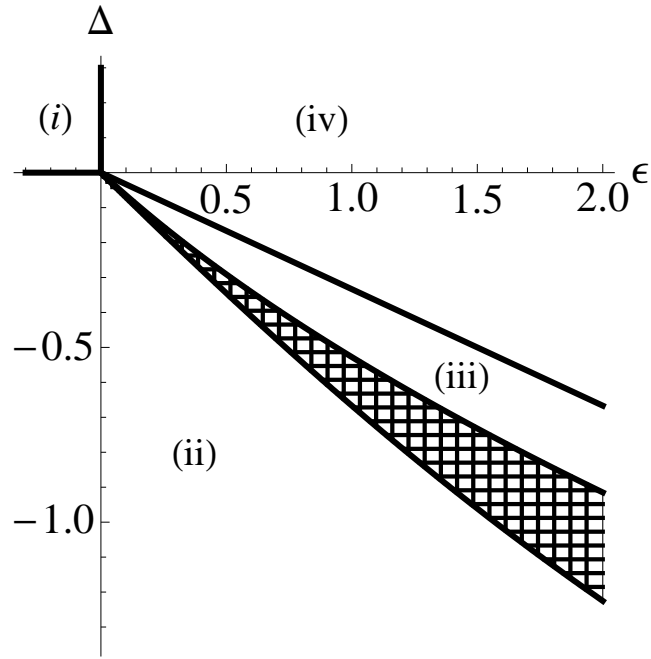}
  \captionstyle{normal}
  \caption{Regions of stability}
  \label{fig:anni1_regions}
\end{figure}

An analysis of
the structure of the fixed points and the basins of attraction leads to the following
physical picture of the effect of the random stirring on the
reaction kinetics. Anomalous behaviour always emerges below two dimensions, when the
{\it local} correlations are dominant in the spectrum of the random forcing
[the short-range fixed point ({\it ii}$\,$)]. However, the random stirring
gives rise to an effective reaction rate faster than the density-fluctuation
induced reaction rate even in this case. The anomaly is present
(but with still faster decay, see the next Section) also, when the long-range part
of the forcing spectrum is effective, but the powerlike falloff of the correlations is
fast [this regime is governed by the kinetic fixed point ({\it iii}$\,$)].
Note that this is different from the case in which the divergenceless
random velocity field is time-independent, in which case there is no fixed
point with $\lambda^*\ne 0$\cite{Deem98b}.
At slower spatial falloff of correlations, however, the anomalous reaction kinetics
is replaced by a mean-field-like behaviour
[this corresponds to the kinetic fixed point ({\it iv}$\,$)]. In particular,
in dimensions $d>1$ this is the situation
for the value $\epsilon=2$ which corresponds to the Kolmogorov spectrum of the velocity field in
fully developed turbulence.
Thus, long-range correlated forcing gives rise to a random velocity field, which
tends to suppress the effect of density fluctuations on the reaction kinetics below two dimensions.
For better illustration, regions of stability for fixed points ({\it i})-({\it iv}) are
depicted in Fig.\ref{fig:anni1_regions}. Wee see
that in contrast to the one-loop approximation \cite{Hnatic00}, overlap (dashed region)
between regions of stability of fixed points ({\it ii}) and ({\it iii}) is observed.
It is a common situation in the perturbative RG approach that higher order terms
lead to either gap or overlap between neighbouring stability regions. The physical realization
of the large-scale behaviour then depends on the initial state of the system.
{\subsection{LONG-TIME ASYMPTOTICS OF NUMBER DENSITY} \label{sec:anni1_longtime}}
Since the renormalization and calculation of the fixed points of the RG are
carried out at two-loop level, we are able to find the first two terms of the
$\epsilon$, $\Delta$ expansion of the average number density, which corresponds to
solving the stationarity equations at the one-loop level.
The simplest way to find the average number density is to calculate it from the stationarity
condition of the functional Legendre transform~\cite{Vasiliev}
(which is often called the effective action) of the generating functional
 obtained by replacing the unrenormalized action by the renormalized
one in the weight functional. This is a convenient way to avoid any
summing procedures used~\cite{Lee94} to take into account the
higher-order terms in the initial number density $n_0$.
We are interested in the solution for the number density, therefore we put
the expectation values of the fields ${\mathbf v}$ and $\tilde{ {\mathbf v} }$ equal to zero
at the outset (but retain, of course, the propagator and the correlation function).
Therefore,
at the second-order approximation 
the effective renormalized action for this model is
\be
\label{anni1_GammaAA1loop}
\Gamma_R  =  S_1+\frac{1}{4}
\raisebox{-4.8ex}{ \epsfysize=2truecm
\epsffile{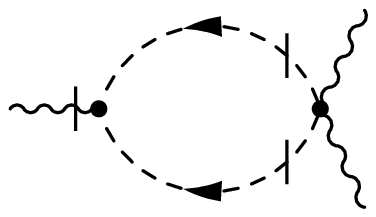}}
+\frac{1}{8}
\raisebox{-4.8ex}{ \epsfysize=2truecm
\epsffile{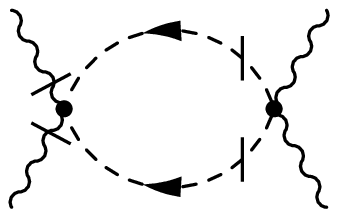}}+
\raisebox{-4.8ex}{ \epsfysize=1.8truecm
\epsffile{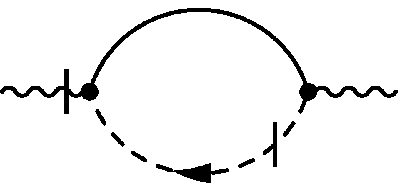}} +\ldots\,,
\ee
where $S_1$ is the action (\ref{eq:anni1_S1}) (within our convention $S_2=0$ in the effective action)
and graphs are shown together with their symmetry coefficients.
The slashed wavy line corresponds to the field $\psi^\dagger$ and the single wavy line
to the field $\psi$.
 The stationarity equations for the variational functional
\begin{equation}
  \frac{\delta \Gamma_R}{\delta \psi^\dagger} = \frac{\delta \Gamma_R}{\delta \psi} = 0
  \label{eq:anni1_var_func}
\end{equation}
  give
rise to the equations
\ba
\label{anni1_EqAA1}
\partial_t \psi & = & u\nu Z_2 \nabla^2 \psi-2\lambda u\nu \mu^{-2\Delta}Z_4 \left(1+\psi^\dagger\right)\psi^2 \nonumber\\
& & +4u^2\nu^2\lambda^2\mu^{-4\Delta}\int\limits_0^\infty\!dt' \int \!dy \,( \Delta^{\psi\psi^\dagger})^2(t-t',\vx-\vy) \psi^2(t',\vy)
\nonumber\\
& & +4u^2\nu^2\lambda^2\mu^{-4\Delta}\psi^\dagger(t,\vx)
\int\limits_0^\infty\!dt'\int\!d\bm{y}\,( \Delta^{\psi\psi^\dagger})^2(t-t',\vx-\vy)\psi^2(t',\vy) \nonumber\\
& & +\doo{}{x_i}\int\limits_0^\infty\!dt' \int \!dy \, \Delta_{ij}^{vv}(t-t',\vx-\vy)\doo{}{x_j}\Delta^{\psi\psi^\dagger}(t-t',\vx-\vy) \psi(t',\vy)+\ldots,
\ea

\ba
-\partial_t \psi^\dagger & = & u\nu Z_2\nabla^2\psi^\dagger-2\lambda u\nu\mu^{-2\Delta}Z_4\left[2\psi^\dagger +\left(\psi^\dagger\right)^2\right]\psi
\nonumber\\*
& & +8u^2\nu^2\lambda^2\mu^{-4\Delta}\int\limits_0^\infty\!dt'\int\!d\bm{y}( \Delta^{\psi\psi^\dagger})^2(t'-t,\vy-\vx)
\psi^\dagger (t',\vy)\psi(t,\vx)\nonumber\\
& & +4u^2\nu^2\lambda^2\mu^{-4\Delta}\int\limits_0^\infty\!dt'\int\!d\bm{y}( \Delta^{\psi\psi^\dagger})^2(t'-t,\vy-\vx)
\times\left[\psi^\dagger (t',\vy)\right]^2 \psi(t,\vx)+\nonumber\\
& & \int\limits_0^\infty\!dt' \int \!dy \, \Delta_{ji}^{vv}(t'-t,\vy-\vx)
\doo{}{x_i}\Delta^{\psi\psi^\dagger}(t'-t,\vy-\vx) \doo{}{y_j}\psi^\dagger(t',\vy)
+\ldots
\label{anni1_EqAA2}
\ea
In (\ref{anni1_EqAA1}) and (\ref{anni1_EqAA2}), in the integral terms it is sufficient to put all renormalization constants equal to unity.
Substituting the solution $\psi^\dagger=0$ of (\ref{anni1_EqAA2}) into (\ref{anni1_EqAA1}) we arrive at the
fluctuation-amended rate equation in the form
\ba
\label{anni1_RateEqAA1loop}
\partial_t \psi & = & u\nu Z_2 \nabla^2 \psi-2\lambda u\nu \mu^{-2\Delta}Z_4 \psi^2\\*
& &
+4u^2\nu^2\mu^{-4\Delta}\lambda^2\int\limits_0^\infty\!dt' \int \!dy \,( \Delta^{\psi\psi^\dagger})^2(t-t',\vx-\vy) \psi^2(t',\vy)
\nonumber\\*
& & +\doo{}{x_i}\int\limits_0^\infty\!dt' \int \!dy \, \Delta_{ij}^{vv}(t-t',\vx-\vy)
\doo{}{x_j}\Delta^{\psi\psi^\dagger}(t-t',\vx-\vy) \psi(t',\vy)
+\ldots,
\ea
This is a nonlinear partial integro-differential equation, whose explicit solution is not known.
It is readily seen that for a homogeneous solution the term resulting from the third graph in
 (\ref{anni1_GammaAA1loop}) vanishes and
hence the influence of the velocity field on the homogeneous annihilation process would be only through the
renormalization of the coefficients $\lambda$ and $D$. However, in case of a nonuniform density field $\psi$
the effect of velocity fluctuations is explicit in (\ref{anni1_RateEqAA1loop}). Such a solution can be most probably
found only numerically.

To arrive at an analytic solution, we restrict ourselves to the homogeneous
 density $n(t)=\langle\psi(t)\rangle$, which
can be identified with the expression (\ref{eq:doi_aver_number}).
In this case the last term in (\ref{anni1_RateEqAA1loop}) vanishes together with the Laplace operator term and the remaining coordinate integral
may be calculated explicitly .
The propagator is the diffusion kernel of the renormalized model (we consider first the system in the general space dimension $d$)
\be
 \Delta^{\psi\psi^\dagger}(t-t',\vx)
 =\frac{\theta(t-t')}{\left[4\pi u\nu(t-t')\right]^{d/2}}\exp\left[-{x^2\over 4u\nu(t-t')}\right].
\label{eq:anni1_diffkernel}
\ee
As noted above, for calculation of the one-loop contribution it is sufficient to put the renormalization constant $Z_2=1$ in the propagator
$\Delta^{\psi\psi^\dagger}$.
Therefore,
evaluation of the Gaussian coordinate integral in (\ref{anni1_RateEqAA1loop}) yields
\begin{equation}
\label{anni1_integral}
\int\!d\bm{y}\,( \Delta^{\psi\psi^\dagger})^2(t-t',\vx-\vy)={\theta(t-t')\over \left[8\pi u\nu(t-t')\right]^{d/2}}
\end{equation}
and we arrive at the ordinary integro-differential equation
\be
\label{anni1_RateEqAA1loopUniform}
\dee{ n(t)}{t}= -2\lambda u\nu \mu^{-2\Delta}Z_4  n^2(t)
+
4\lambda^2u^2\nu^2\mu^{-4\Delta}\int\limits_0^t\!dt'\,{n^2(t')\over \left[8\pi u\nu(t-t')\right]^{d/2}}\,.
\ee
Spatial fluctuations in the particle density show in the integral term and affect rather heavily
even the homogeneous solution. In particular, the integral in (\ref{anni1_RateEqAA1loopUniform}) diverges at
the upper limit in space dimensions $d\ge 2$. This is a consequence of the UV divergences in the model
above the critical dimension $d_c=2$ and near the critical dimension is remedied by the UV renormalization of the
model. To see this, subtract and add the term $n^2(t)$ in the integrand to obtain
\ba
\label{anni1_RateEqAA1loopUniform2}
\dee{ n(t)}{t} & = & -2\lambda u\nu \mu^{-2\Delta}Z_4  n^2(t)
+4\lambda^2u^2\nu^2\mu^{-4\Delta}n^2(t)\int\limits_0^t\!{dt'\over \left[8\pi u\nu(t-t')\right]^{d/2}}\nonumber\\*
& & + 4\lambda^2u^2\nu^2\mu^{-4\Delta}\int\limits_0^t\!dt'\,{n^2(t')-n^2(t)\over \left[8\pi u\nu(t-t')\right]^{d/2}}\,.
\ea
The last integral here is now convergent at least near two dimensions, provided the solution $n(t)$ is
a continuous function. This is definitely the case for the iterative solution constructed below.
The divergence in the first integral in (\ref{anni1_RateEqAA1loopUniform2}) may be explicitly calculated
below two dimensions and
is canceled -- in the leading order in the parameter $\Delta=(d-2)/2$ -- by the one-loop term of the renormalization constant $Z_4$
(\ref{eq:anni1_Z4}). Expanding the right-hand side of (\ref{anni1_RateEqAA1loopUniform2}) in the parameter $\Delta=(d-2)/2$ to the next-to-leading
order we arrive at the equation
\ba
\label{anni1_RateEqAA1loopUniform3}
\dee{ n(t)}{t} & = & -2\lambda  u\nu\mu^{-2\Delta}   n^2(t)+
2\lambda u\nu\mu^{-2\Delta}    n^2(t)\left\{{\lambda\over 4\pi}\,\left[\gamma+\ln\left(2u\nu\mu^2 t\right)\right]\right\}
\nonumber\\*
& & + {\lambda^2u\nu\mu^{-2\Delta}\over 2\pi}\int\limits_0^t\!dt'\,{n^2(t')-n^2(t)\over t-t'}\,.
\ea
without divergences near two dimensions. Here, the factor $\mu^{-2\Delta}$ has been retained intact in order not
to spoil the consistency of scaling dimensions in different terms of the equation. In (\ref{anni1_RateEqAA1loopUniform3}), $\gamma=0.57721$ is Euler's constant and
we have considered the coupling constant $\lambda$ and the parameter
$\Delta=(d-2)/2$ to be small parameters of the same order taking into account the magnitudes of the parameters in the basins of attraction
of the fixed points of the RG. The leading-order approximation for $n(t)$ is given by the first term on the
right-hand side of (\ref{anni1_RateEqAA1loopUniform3}) and it is readily seen that after substitution of this expression the integral
term in (\ref{anni1_RateEqAA1loopUniform3}) is of the order of $\lambda^3$ and thus negligible in the present next-to-leading-order calculation. In this
approximation, Equation (\ref{anni1_RateEqAA1loopUniform3}) yields
\be
\label{anni1_NLsolution}
n(t)=
{n_0\over 1+2{\lambda} u\nu t\left\{1+{\displaystyle{\lambda }\over \displaystyle 4\pi}\left[1-\gamma
-\ln\left(2u\nu\mu^2 t\right)\right]\right\}\mu^{-2\Delta}n_0}\,,
\ee
where $n_0$ is the initial number density.

Since the fields $\Phi=\{v,\tilde{v},\psi,\psi^\dagger\}$ are not renormalized, the renormalized connected
Green functions $W_R$ differ from the unrenormalized $W=\langle\Phi\ldots\Phi\rangle$
\cite{Vas_turbo} only by the choice of parameters and thus one may write
\begin{equation}
  W_R( g,\nu,\mu,\ldots) = W(g_0,\nu_0,\ldots),
  \label{eq:diff_GF}
\end{equation}
where $g_0=\{g_{10},g_{20},u_0,\lambda_0\}$ is the full set of the bare parameters and dots denotes
all variables unaffected by the renormalization procedure. The independence of renormalization mass
parameter $\mu$ is expressed by the equation $\mu\partial_\mu W_R= 0$.
Using this equation the RG equation for the mean particle number $n(t)$ is readily obtained:
\begin{equation}
\left(\mu{\partial\over\partial\mu}+\sum\limits_g\beta_g{\partial\over\partial g}
-\gamma_1\nu{\partial\over\partial\nu}\right)n(t,\mu,\nu,n_0,g)=0.
 \label{eq:anni1_RG}
\end{equation}
We are interested in long-time behaviour of the system ($t\rightarrow\infty$), therefore
we trade the
renormalization mass for the time variable. Canonical scale invariance yields relations
\cite{turbo}
\begin{equation}
  \left(\mu\frac{\partial}{\partial \mu}-2\nu\frac{\partial}{\partial \nu}+
   dn_0\frac{\partial}{\partial n_0}-d \right) n(t,\mu,\nu,n_0,g)=0,
   \label{eq:anni1_scale_mom}
\end{equation}
\begin{equation}
  \left( -t\frac{\partial }{\partial t}+\nu\frac{\partial}{\partial \nu}\right)   n(t,\mu,\nu,n_0,g)=0,
   \label{eq:anni1_scale_freq}
\end{equation}
where the first equation expresses scale invariance with respect to wave number and the second equation
with respect to time.
Eliminating partial derivatives with respect to the renormalization mass $\mu$ and viscosity $\nu$ we
obtain the Callan-Symanzik equation for the mean particle number
\begin{equation}
  \biggl[(2-\gamma_1)t\frac{\partial}{\partial t}+\sum_{g} \beta_g \frac{\partial}{\partial g}-
 dn_0 \frac{\partial}{\partial n_0}+d \biggr] n\left(t,\mu,\nu,n_0,g\right)=0
  \label{eq:anni1_Callan}
\end{equation}
To separate information given by the RG, consider the dimensionless normalized mean particle number
\begin{equation}
\label{anni1_normaln}
{n\over n_0}=\Phi\left(\nu\mu^2 t,\lambda u\,{n_0\over \mu^d},g\right)\,.
\end{equation}
For the asymptotic analysis, it is convenient to express the particle density in the combination used here.
Solution of (\ref{eq:anni1_Callan}) by the method of characteristics yields
\begin{equation}
\Phi\left(\nu\mu^2 t,\lambda u\,{n_0\over \mu^d},g\right)=\Phi\left({\nu}\mu^2 \tau,\overline{\lambda}\overline{u}\,{\overline{n}_0\over \mu^d},\overline{g}\right)
 \label{eq:anni1_solNLRG}
\end{equation}
where $\tau$ is the time scale.
In Equation (\ref{eq:anni1_solNLRG}), $\overline{g}$ and $\overline{n}_0$
are the first integrals of the system of differential equations
\begin{equation}
  t\frac{d}{dt}\overline{g}=-\frac{\beta_g (\overline{g})}{2-\gamma_1(\overline{g})}\,,\quad
  t\frac{d}{dt}\overline{n}_0=d\frac{\overline{n}_0}{2-\gamma_1(\overline{g})}.
  \label{eq:anni1_first_int}
\end{equation}
Here $\overline{g}=\{\overline{g_1},\overline{g_2},\overline{u},\overline{\lambda} \}$ with
initial conditions $\overline{g}|_{t=\tau}=g$ and $\overline{n}_0|_{t=\tau}=n_0$.
In particular,
\begin{equation}
\label{eq:anni1_initialdensity}
\overline{\lambda}\overline{u}\overline{n}_0=\lambda u\,n_0\,\left({t\over \tau}\right)\exp\left[{\displaystyle  \int_\tau^t{\displaystyle \gamma_4 ds\over\displaystyle  (2-\gamma_1)s}}\right]\,.
\end{equation}
The asymptotic expression of the integral on the right-hand side of
(\ref{eq:anni1_initialdensity}) in the
vicinity of the IR-stable fixed point $g^*$ is of the form
\be
\label{eq:anni1_asyamplitude}
{ \int_\tau^t{\displaystyle \gamma_4 ds\over\displaystyle  (2-\gamma_1)s}}
\operatornamewithlimits{\sim}_{t\to\infty}{\gamma_4^*\over 2-\gamma_1^*}\ln\left({t\over\tau}\right)+
{2\over 2-\gamma_1^*}\int\limits_\tau^\infty\!{ (\gamma_4-\gamma_4^*)ds\over  (2-\gamma_1)s}
={\gamma_4^*\over 2-\gamma_1^*}\,\ln\left({t\over \tau}\right)+\tilde{c}_4(\tau)\,,
\ee
corrections to which vanish in the limit $t\to\infty$. In (\ref{eq:anni1_asyamplitude}) and henceforth, the
notation $\gamma_1^*=\gamma_1\left(g^*\right)$ has been used.
>From the point of view of the long-time asymptotic
behaviour the next-to-leading term in (\ref{eq:anni1_asyamplitude}) is an inessential constant.
In the vicinity of the fixed point
\be
\label{eq:anni1_initialdensityasy}
\overline{\lambda}\overline{u}\,{\overline{n}_0\over \mu^d}
\sim \lambda u\,{n_0\over \mu^d} \left({t\over \tau}\right)^{\displaystyle 1+{\displaystyle \gamma_4^*\over \displaystyle 2-\gamma_1^*}}\widetilde{C}_n
\equiv \lambda u\,{n_0\over \mu^d} \left({t\over \tau}\right)^\alpha \widetilde{C}_n
\equiv \overline{y}\,\widetilde{C}_n\,,
\ee
where a shorthand notation $\overline{y}$ has been introduced for the long-time scaling of the normalized number density as well as
the dimensional normalization constant
\[
\widetilde{C}_n=e^{\tilde{c}_4(\tau)}\,.
\]
and the decay exponent
\begin{equation}
\label{eq:anni1_defalpha}
\alpha=\displaystyle 1+{\displaystyle \gamma_4^*\over \displaystyle 2-\gamma_1^*}
\end{equation}
The asymptotic behaviour of the normalized particle density is described by the scaling function $f(x,y)$
\be
\Phi\left(\nu\mu^2 t,\lambda u\,{n_0\over \mu^d},g\right)
\sim\Phi\left({\nu}\mu^2 \tau,\widetilde{C}_n\overline{y},{g}^*\right)\equiv
f\left({\nu}\mu^2 \tau,\widetilde{C}_n\overline{y}\right)\,.
 \label{eq:anni1_scalingf}
\ee
The free parameters in the variables of the scaling function $f(x,y)$ correspond to the choice of units of these variables, whereas the
objective information is contained in the form of the scaling
 function~\cite{turbo,Vas_turbo}. Here, it is convenient to use the explicit solution (\ref{anni1_NLsolution})
to obtain the $\varepsilon$, $\Delta$ expansion for the inverse $h(x,y)={1/ f(x,y)}$ of the scaling function. We obtain the generic expression
\be
\label{eq:anni1_ExpansionOfScaling}
h(x,y)={1\over f(x,y)}
=1+2xy\left\{1+{\displaystyle{{\lambda}^* }\over \displaystyle 4\pi}\left[1-\gamma
-\ln\left(2u^*x\right)\right]\right\}\,,
\ee
the substitution in which of the various fixed-point values $\lambda^* $ (at the leading order $\lambda^*\approx 2\pi\overline{\lambda}^*$)
and $u^*$ in the leading approximation yields the corresponding $\varepsilon$, $\Delta$ expansions.

Below, we list the scaling functions  $h(x,y)$ and the dynamic exponents $\alpha$ at the stable fixed points
in the next-to-leading-order approximation.\\
({\it i}$\,$) At the trivial (Gaussian) fixed point (\ref{anni1_Gaussian})
the mean-field behaviour takes place with
\begin{align}
\label{eq:anni1_hGaussian}
h(x,y)&=1+2xy\,,\nonumber\\*
\alpha&=1\,.
\end{align}
({\it ii}$\,$)
The thermal (short-range) fixed point
(\ref{anni1_SR}) leads to scaling function and decay exponent
\begin{align}
\label{eq:anni1_hSR}
h(x,y)&=1+2xy\left\{1-{\displaystyle{\Delta }\over \displaystyle 2}\left[1-\gamma
-\ln \left(\sqrt{17}-1\right)\,x\right]\right\}
\,,\nonumber\\*
\alpha&=1+{\Delta\over 2}+{\Delta^2\over 2}\,.
\end{align}
Here, the last coefficient is actually a result of numerical calculation, which in the standard accuracy of Mathematica
is equal to 0.5. We have not been able to sort out this result analytically, but think that most probably the coefficient of
the $\Delta^2$ term in the decay exponent $\alpha$ in (\ref{eq:anni1_hSR}) really is ${1\over 2}$.\\
({\it iii}$\,$)
The kinetic fixed point with an anomalous reaction rate
(\ref{anni1_K1}) corresponds to
\begin{eqnarray}
\label{eq:anni1_hK1}
h(x,y)&=&1+2xy \biggl\{ 1-{\displaystyle{\epsilon+3\Delta }\over \displaystyle 3}
\left[1-\gamma-\ln \left(\sqrt{17}-1\right)\,x\right]\biggl\}
\,,\nonumber\\*
\alpha&=&1+{3\Delta+\epsilon\over 3-\epsilon}\,,
\end{eqnarray}
with an exact value of the decay exponent. \\
({\it iv}$\,$)
At the kinetic fixed point with mean-field-like reaction rate
(\ref{anni1_K2}) we obtain
\begin{align}
\label{eq:anni1_hK2}
h(x,y)&=1+2xy
\,,\nonumber\\*
\alpha&=1\,.
\end{align}
In the actual asymptotic expression corresponding to (\ref{eq:anni1_scalingf}) the argument $y\to \widetilde{C}_n\overline{y}$
is different from that of the Gaussian fixed point.

To complete the picture, we recapitulate -- with a little bit more detail --
the asymptotic behaviour of the number density in the physical space dimension $d=2$ predicted
within the present approach \cite{Hnatic00} (it turns out that for these conclusions the one-loop
calculation is sufficient).
On the ray $\varepsilon\le 0$, $\Delta=0$ logarithmic corrections to the mean-field decay take place. The integral
determining the asymptotic behaviour of the variable (\ref{eq:anni1_initialdensity}) yields in this case
\begin{equation}
\label{eq:anni1_asyamplitudelog}
{ \int_\tau^t{\displaystyle \gamma_4 ds\over\displaystyle  (2-\gamma_1)s}}
\operatornamewithlimits{\sim}_{t\to\infty}-{1\over 2}\,\ln\ln\left({t\over \tau}\right)+\tilde{c}_4(\tau)\,,
\end{equation}
with corrections vanishing in the limit $t\to\infty$. Therefore, in the vicinity of the fixed point
\begin{equation}
\label{eq:anni1_initialdensityasylog}
\overline{\lambda}\overline{u}\,{\overline{n}_0\over \mu^d}
\sim \lambda u\,{n_0\over \mu^d} \left({t\over \tau}\right)\ln^{-1/2}\left({t\over \tau}\right)\widetilde{C}_n
\equiv \overline{y}\,\widetilde{C}_n\,.
\end{equation}
The scaling function $h$ is of the simple form
\begin{equation}
\label{eq:anni1_hlog}
h(x,y)=1+2xy\nonumber
\end{equation}
and gives rise to asymptotic decay slower than in the mean-field case by a logarithmic factor:
\[
{n}\sim {\ln^{1/2}\left(t/ \tau\right)\over 2\nu\lambda u \widetilde{C}_n t}\,.
\]
It is worth noting that this logarithmic slowing down is weaker than that brought about the density fluctuations only \cite{Peliti85}
and this change is produced even by the ubiquitous thermal fluctuations of the fluid, when the reaction is taking
place in gaseous or liquid media.

On the open ray $\varepsilon> 0$, $\Delta=0$ the kinetic fixed point with mean-field-like reaction rate (\ref{anni1_K2}) is
stable and the asymptotic behaviour is given by (\ref{eq:anni1_hK2}) regardless of the value of the falloff exponent of the
random forcing in the Navier-Stokes equation. In particular, only the amplitude factor in
the asymptotic decay rate in two dimensions is affected by the developed turbulent flow with Kolmogorov scaling, which
corresponds to the value $\varepsilon=2$.
This is in accord with the results obtained in the case of quenched solenoidal flow with long-range correlations \cite{Deem98b,Tran} as
well as with the usual picture of having the maximal reaction rate in a well-mixed system.

{\section{ROLE OF RANDOM SOURCES AND SINKS ON REACTION PROCESSES} \label{sec:random_sources}}
{\subsection{MASTER EQUATION FOR RANDOM SOURCES AND SINKS} \label{sec:random_sourcesink}}

We will consider the annihilation reaction $A+A\rightarrow\varnothing$ in
a random drift field in a more general setup than in the previous parts.
For this purpose we introduce random sources and sinks
of the reacting particles in order to maintain a steady state in the system.
In most cases this is carried out by including an additive noise
term in the Langevin equation of the stochastic process
as was done e.g. in (\ref{eq:anni1_NS}) to have steady turbulent state. Since our
analysis is based on the master equation, this is not quite
appropriate here. Unfortunately, there is no unique way to introduce
random sources in the master equation corresponding to the random
noise of the mean-field (Langevin) description. We use the simplest
choice, described in detail in \cite{Kampen}, which is equivalent to adding
processes $A\to X$ and $Y\to A$ .to the whole reaction scheme. Here $X$ and $Y$ stand for
 particle baths of the sink and the source respectively. In a homogeneous system these reactions
 leads to the master equation
\begin{eqnarray}
\dee{P(t,n)}{t}&=&\mu_+V\left[P(t,n-1)-P(t,n)\right]
+\mu_-\left[(n+1)P(t,n+1)-nP(t,n)\right]\ldots
\label{eq:random_MasterSource}
\end{eqnarray}
where $P(t,n)$ is the probability to find $n$ particles at the
time instant $t$ in the system. The ellipsis in (\ref{eq:random_MasterSource})
represents terms describing the annihilation reaction, diffusion
and advection in the system. In (\ref{eq:random_MasterSource}) $\mu_+$ and $\mu_-$ are the
reaction constants of the creation and annihilation reactions, respectively. The transition rate
has been chosen proportional to the particle number $n$, which can be understood as consequence
of independent processes $A\rightarrow X$ and this choice
also preserves the empty state as an absorbing state. In the transition rate for creation process
$V$ is the volume of the (for the time being) homogeneous system and will be important in passing to
the continuum limit of the inhomogeneous system. The master equation (\ref{eq:random_MasterSource})
gives rise to the reaction-rate equation
\beq
\label{eq:random_RateSource}
\dee{\langle n\rangle}{t}=\mu_+V-\mu_-\langle n\rangle+\ldots
\eeq
where $\langle n\rangle$ is the mean particle number.

We recall that the basic idea of the Doi approach \cite{Doi76} is to rewrite the set of master
equations for probability distributions of a stochastic problem in
the form of a single kinetic equation for a state vector
incorporating all probabilistic information about the system
constructed in a suitable Fock space. The kinetic equation is
defined by the Liouville operator acting in the Fock
space and generated by the set of master equations. Although the
basic procedure has been thoroughly exposed in the literature, the
introduction of random sources and sinks of particles in the master
equation has specific features, which should be presented in detail.
Therefore,
let us briefly recall the basic quantities and relations of the Doi
approach. For simplicity, consider probabilities $P(t,n)$ to find
$n$ particles at the time instant $t$ on a fixed lattice site. Then the spatial dependence
may be described by labeling the particle number by the coordinates of the
lattice and introducing necessary sums and products over the lattice sites.
The construction of corresponding Fock space was presented in Section. \ref{sec:doi_quant_method},
 namely equations (\ref{eq:doi_boson_oper}-\ref{eq:doi_state}).
The set of master equations for a birth-death process may also be cast in the
form of a single evolution equation for the state vector
(\ref{eq:doi_state}) without any explicit dependence on the
occupation number
\begin{equation}
\label{L} \dee{\ket{\Phi}}{t}=-\hat{H}(\cre,\,\ann)\ket{\Phi}\,.
\end{equation}
Master equations (\ref{eq:random_MasterSource}) give rise to the following terms in
the Hamilton operator
\be
\label{eq:random_LSource}
\hat{H}_g(\cre,\ann)= -\mu_+V\left(\cre-I\right)-\mu_-\left(I-\cre\right)\ann\,,
\ee
where $I$ is the identity operator. The expectation value of any function $A(n)$ of the
random particle number
\be
\label{eq:random_average}
\langle A(t)\rangle=\sum\limits_{n=0}^\infty A(n)P(t,n)\,,
\ee
may be expressed in the form of the functional integral over the
functions $\tilde{a}(t)$ and $a(t)$
\be
\label{eq:random_functional integral}
\langle A(t)\rangle=\int\!{\cal D}\tilde{a} {\cal D}a\, A_N(1,a(t))
e^{S_1}\,,
\ee
where $A_N(\tilde{a}, a)$ is the normal form \cite{Vasiliev} of the operator $A(\cre\ann)$ and $S_1$ is the
dynamic action
\be
\label{eq:random_action1}
S_1(\tilde{a} ,a)=\int\limits_0^\infty\!\! dt\,\left[-\tilde{a} (t)\partial_t a(t)
+\mu_+V\tilde{a} (t)-\mu_-\tilde{a} (t)a(t)\right]\ldots
\ee
Only the generic time-derivative term and terms brought about by the random source model are expressed
here explicitly, while the ellipsis stands for terms corresponding to other reactions and initial conditions.

Let the transition rates $\mu_\pm$ be the random functions uncorrelated in time with a probability distribution given
in terms of the moments $\langle\mu_\pm^n\rangle=E_{\pm\,,n}$. To keep the problem translation-invariant in time
we assume stationary stochastic processes determined on the whole time axis. Therefore, henceforth all time integrals in the
action functional and, correspondingly, in the perturbation expansion shall be taken over the whole time axis.
At this point we also generalize the treatment to the
case of a spatially inhomogeneous system and introduce a lattice subscript as the spatial argument, i.e.
$a(t)\to a_i(t)$. In this case the volume $V$ becomes the volume element attached to the lattice site.
For simplicity, we replace the time integral with the integral sum $\int_{-\infty}^\infty\! dt\to \sum_\alpha\Delta t$
and assume that the transition rates at each time instant and lattice site $\mu_{\pm\\,\alpha,i}$
are independent random variables. Then the average of the expectation value (\ref{eq:random_functional integral})
over the distribution of random sources reduces to the calculation of the expectation value
\be
\label{eq:random_g-average}
\prod_{\alpha,i}
\biggl\langle \exp\biggl(\mu_{+\,,\alpha,i}V\tilde{a} _{\alpha,i}\Delta t-
\mu_{-\,,\alpha,i}\tilde{a} _{\alpha,i}a_{\alpha,i}\Delta t\biggl)\biggl\rangle\,.
\ee
For each particular time instant and lattice  (we assume that the moments of $\mu_\pm$ are the same for
all $\alpha$ and $i$ and omit labels for brevity) this gives rise to the usual cumulant expansion
\begin{eqnarray}
  \biggl\langle {\mathrm e}^{\mu b\Delta t}\biggl\rangle & = &
  1+b\Delta t E_1+{1\over 2}E_2(b\Delta t)^2+{1\over 6}E_3(b\Delta t)^3+\cdots \nonumber\\
  & = & \exp\biggl(b\Delta t E_1+\frac{E_2-E_1^2}{2}(b\Delta t)^2
  +\frac{E_3-3E_1E_2+E_1^3}{6}(b\Delta t)^3+\cdots\biggl)
  \label{eq:random_expansion-average}
\end{eqnarray}
Here, $b$ stands for either $V\tilde{a} $ or $-\tilde{a} a$. Thus, for instance the average over $\mu_+$ assumes the form
\begin{eqnarray}
\nonumber
\prod_{\alpha,i}
\biggl\langle \exp \biggl(\mu_{+\,,\alpha,i}V\tilde{a} _{\alpha,i}\Delta t\biggl) \biggl\rangle
&=& \exp\biggl(\sum_\alpha\sum_i\left[\Delta t E_{+1}V\tilde{a} _{\alpha,i}
+{1\over 2}\left(E_{+2}-E_{+1}^2\right)(V\tilde{a} _{\alpha,i}\Delta t)^2\right]\biggl)\\
&\times& \exp\biggl(\sum_\alpha\sum_i\left[\frac{E_{+3}-3E_{+1}E_{+2}+E_{+1}^3}{6}
(V\tilde{a} _{\alpha,i}\Delta t)^3+\cdots\right]\biggl) .\nonumber\\
\label{eq:random_+average}
\end{eqnarray}
In the continuum limit the function $\tilde{a} _{\alpha,i}$ is replaced
by the field $\psi^+(t,\vx)$, whereas in the limit $V\to 0$ the expression $a_{\alpha,i}/V$ gives rise to the field
$\psi(t,\vx)$. The sum over $\alpha$ together with $\Delta t$ gives rise
 to the time integral and the
sum over $i$ together with the volume element leads
 to the spatial integral $\sum_iV\to \int d\vx$.
In the first term of the exponential in (\ref{eq:random_+average}) we thus obtain
\be
\sum_\alpha\sum_i\Delta t E_{+1}V\tilde{a} _{\alpha,i}\to E_{+1}\int dt\int d\vx\,\psi^+(t,\vx)\,.
\ee
The continuum limit for the cumulants of second and higher order is not so obvious.
We assume the simplest nontrivial distribution for $\mu_\pm$, in which only the variance term has
a finite limit, when $\Delta t\to 0$ and $V\to 0$, whereas the contributions of higher-order cumulants
vanish, for instance
\ba
\label{eq:random_limit_variance}
\left(E_{+2}-E_{+1}^2\right)V\Delta t \to \sigma_+\,,\quad \Delta t\to 0\,,\ V\to 0\,,\\
\label{eq:random_limit_cumulants}
\left(E_{+3}-3E_{+1}E_{+2}+E_{+1}^3\right)(V\Delta t)^2\to 0\,,\quad \Delta t\to 0\,,\ V\to 0\,.
\ea
Therefore, the contribution of the average over $\mu_+$ to the effective dynamic action assumes the form
\be
\label{eq:random_+}
S_+= \int\! dt\int\! d\vx\,\left\{E_{+1}\psi^+(t,\vx)+{1\over 2}\sigma_+\left[\psi^\dagger(t,\vx)\right]^2\right\}\,.
\ee
For the average over $\mu_-$ a similar argument yields
\be
\label{eq:random_-}
S_-= \int\! dt\int\! d\vx\,\left\{-E_{-1}\psi^\dagger(t,\vx)\psi(t,\vx)
+{1\over 2}\sigma_-\left[\psi^\dagger(t,\vx)\psi(t,\vx)\right]^2\right\}\,.
\ee
These contributions to the effective dynamic action may, of course, be generated by suitably chosen normal
distributions of $\mu_\pm$.

This way of introduction of random sources and sinks has the annoying feature that it does not conserve the
number of particles in the system. For a comparison with the treatment of this problem in the Langevin approach the
random sources and sinks should be introduced in such a way that the particle number is conserved.
The simplest way how to deal with this problem
 is to add to the random source a term proportional to the particle number, i.e. to
use the ''reaction constant'' $\mu_+V+\mu_{1+}n$ instead of $\mu_+V$ in the master equation.
 The source terms on
the right-hand side of the master equation (\ref{eq:random_MasterSource}) in this case assume the form
\begin{eqnarray}
\nonumber
\dee{P(t,n)}{t}&=&\mu_+V\left[P(t,n-1)-P(t,n)\right]\\
&+&\mu_{1+}\left[(n-1)P(t,n-1)-nP(t,n)\right]\ldots
\label{eq:random_MasterSource2}
\end{eqnarray}
The new part of the master equation corresponds to a branching process \cite{Kampen}.

The added term gives rise to the following contribution to the Hamilton operator
\beq
\label{eq:random_LSource2}
\hat{H}_{g2}(\cre,\ann)= -\mu_{1+}\left(\cre-I\right)\cre\ann\,.
\eeq
Performing the steps described above we arrive at the contribution to the dynamic action in
the following form
\beq
\label{eq:random_1+}
S_{1+}= \int\! dt\int\! d\vx\,\left\{E_{1+1}\psi^\dagger\left(\psi^\dagger +1\right)\psi
+{1\over 2}\sigma_{1+}{\psi^\dagger}^2\left(\psi^\dagger +1\right)^2\psi^2\right\}\,.
\eeq
Now it is easy to see, that if we exclude the plain source (i.e. letting $E_{+1}=\sigma_+=0$) and choose
$E_{1+1}=E_{-1}$, the empty state remains absorbing one and the ''mass term'' $\propto \psi^\dagger\psi$
disappears in the dynamic action and we arrive at the dynamic action of random sources and sinks
\be
S_{gc}= \int\! dt\int\! d\vx\,\biggl\{E_{1+1}{\psi^\dagger}^2\psi+
{1\over 2}\sigma_-\left(\psi^\dagger \psi\right)^2
+{1\over 2}\sigma_{1+}{\psi^\dagger}^2\left(\psi^\dagger +1\right)^2\psi^2\biggl\},
\label{eq:random_gc}
\ee
which conserves the average number of particles.

The effects of the high-order terms are drastically different
in the two cases amenable for a scaling analysis with the aid of the renormalization group.
The time derivative term in the dynamic action
\be
S= -\int\! dt\int\! d\vx\,\psi^\dagger(t,\vx)\partial_t\psi(t,\vx)+\ldots
\ee
must be dimensionless in order to have nontrivial dynamics. Therefore
the total scaling dimension of the number-density operator $\psi^\dagger(t,\vx)\psi(t,\vx)$ is equal to
the dimension of space and thus is positive.

First, if the
scaling dimension of the field $\psi^\dagger$ is equal to zero, $d_{\psi^\dagger}=0$, then the dimension of the
field $\psi$ is positive (more precisely $d_{\psi}=d$)
and the operator monomials in the second and third terms in (\ref{eq:random_gc})
have the same scaling dimension. Since they are carrying the factor $\psi^2$, their scaling dimension
is larger than that of ${\psi^\dagger}^2\psi$. Therefore, the second and third terms
in (\ref{eq:random_gc}) are IR irrelevant and should be discarded in the asymptotic analysis.\\
 Second,
if the scaling dimensions of both fields are positive, then in the operator monomials
in the second and third terms in (\ref{eq:random_gc}) there is at least one ''excessive'' field factor
in comparison with the first term, which renders them irrelevant.
Thus, in these cases
the IR relevant dynamic action of random sources and sinks reduces to the single term
\beq
\label{eq:random_gcPrime}
S'_{gc}= \int\! dt\int\! d\vx\,E_{1+1}{\psi^\dagger}^2\psi\,,\quad d_{\psi^\dagger}=0 \ \vee \ d_{\psi^\dagger}>0\,, \ d_{\psi}>0 \,.
\eeq
Third,
if the scaling dimension of the field $\psi$ is zero, the scaling dimension
of the field $\psi^\dagger$ is positive and terms with ''excessive'' powers of $\psi^\dagger$ are
IR irrelevant. So the starting point for the subsequent RG analysis is the source and sink action
in the form
\beq
\label{eq:random_gcPos}
S_{gc}''= \int\! dt\int\! d\vx\,\left\{E_{1+1}{\psi^\dagger}^2\psi+
{1\over 2}\left(\sigma_-+\sigma_{1+}\right)\left(\psi^\dagger \psi\right)^2\right\}\,,
\quad d_{\psi}=0\,.
\eeq
{\subsection{ANNIHILATION PROCESS WITH RANDOM SOURCES AND SINKS} \label{sec:random_AAsourcesink}}

Let us analyze the dynamic action of the diffusion-limited annihilation reaction
$A+A\rightarrow\varnothing$
\ba
S_1&=&-\int\! dt\!\int\! d{\bf x} \,\biggl\{\psi^\dagger \partial_t\psi
-D_0 \psi^\dagger \nabla^2\psi +\lambda_0D_0\left[2\psi^\dagger +
(\psi^\dagger)^2\right]\psi^2
\biggl\}
+n_0\int\! d{\bf x}\, \psi^\dagger ({\bf x},0), \nonumber\\
\label{eq:random_action11}
\ea
from the point of view of scaling behaviour sketched in section \ref{sec:random_sourcesink}.

In the first case with $d_{\psi^+}=0$ the nonlinear terms in action (\ref{eq:random_action11})
are of equal scaling dimension. However, the source-sink part (\ref{eq:random_gcPrime}) is linear
in the field $\psi$ with positive scaling dimension in contrast to the quadratic
 $\psi$ terms of (\ref{eq:random_action11}). Therefore, the IR relevant interaction above two dimensions
is (\ref{eq:random_gcPrime}) and the corresponding dynamic action is
\ba
S_{IR1}&=&-\int\! dt\!\int\! d{\bf x} \,\left\{\psi^\dagger \partial_t\psi
-D_0 \psi^\dagger \nabla^2\psi
-E_{1+1}{\psi^\dagger}^2\psi
\right\}+n_0\int\! d{\bf x}\, \psi^\dagger ({\bf x},0)\,.
\label{eq:random_IRaction1}
\ea
This dynamic action does not bring about any graphs with closed loops of the
density propagator, which implies suppresion of the fluctuation effects.
However, the scaling dimension of the interaction term is negative and may compensate for
the positive dimensions of the irrelevant interaction terms. Therefore, the rest of the interaction
terms are in fact dangerous irrelevant operators and in
this case  definitive conclusion about the
IR relevant action cannot be reached on the basis of the analysis of the scaling dimensions.

In the second case with $d_{\psi^\dagger}>0$ and $d_{\psi}>0$ the fourth-order term in action (\ref{eq:random_action11})
becomes irrelevant. Either of the remaining third-order terms alone does not generate loops, therefore
density fluctuation effects are brought about only, when both fields have the same scaling dimension
$d_{\psi^\dagger}=d_{\psi}=d/2$. In this case the IR relevant dynamic action is
\ba
S_{IR2}&=&-\int\! dt\!\int\! d{\bf x} \,\biggl\{\psi^+\partial_t\psi
-D_0 \psi^+\nabla^2\psi+2\lambda_0D_0\psi^+\psi^2 -E_{1+1}{\psi^+}^2\psi
\biggl\}
+n_0\int\! d{\bf x}\, \psi^+({\bf x},0).\nonumber\\
\label{eq:random_IRaction2}
\ea
Here, the scaling dimension of both interaction terms is $(d/2)-2$ and vanishes at the
critical dimension $d_c=4$, at which the dimensions of all the other interaction terms are positive
and they are unambiguously irrelevant.
Effective action (\ref{eq:random_IRaction2})
is the dynamic action of the Gribov process \cite{Gribov67}, also known as the Reggeon model.
Effects of random drift in this case with the use of the Obukhov-Kraichnan compressible velocity field
have been analyzed in \cite{Antonov10}.

In the third case with $d_{\psi}=0$ the fourth-order term in action (\ref{eq:random_action11})
becomes irrelevant as well due to the positive dimension of the field $\psi^+$. By the same token, however,
both terms of the source-sink action (\ref{eq:random_gcPos}) are also irrelevant and we arrive at the IR relevant
dynamic action
\ba
S_{IR3}&=&-\int\! dt\!\int\! d{\bf x} \,\biggl\{\psi^\dagger \partial_t\psi
-D_0 \psi^\dagger\nabla^2\psi
+2\lambda_0D_0\psi^\dagger\psi^2
\biggl\}+n_0\int\! d{\bf x}\, \psi^+({\bf x},0)\,.
\label{eq:random_IRaction3}
\ea
An argument similar to that used for (\ref{eq:random_IRaction1}) shows, that the scaling analysis with this choice of
field dimensions does not allow to resolve relevance of interaction terms. It should be recalled that
the scaling dimensions of auxiliary quantities and the asymptotic behaviour of individual graphs is actually independent
of the choice of the values of the field dimensions. Therefore, the effective action (\ref{eq:random_IRaction2}) with
unambiguous classification of relevant and irrelevant interaction terms describes the critical scaling behaviour
amenable to the RG analysis.

In summary, if the sources and sinks are chosen such that they conserve the mean number of particles in
the system, the anomalous scaling behaviour in the system is that of the Gribov process.

A different situation arises, if the plain source term is included into the analysis. Then there is a possibility
that the system does not tend to the absorbing empty state but to an active state with a finite concentration
of particles. In this case the starting point is the dynamic action with all the terms quoted above, i.e.
\begin{eqnarray}
\nonumber
S&=&\int\! dt\!\int\! d{\bf x} \,\Biggl\{-\psi^\dagger\partial_t\psi
+D_0 \psi^\dagger\nabla^2\psi
-\lambda_0D_0\left[2\psi^\dagger +
\left(\psi^\dagger \right)^2\right]\psi^2
+E_{+1}\psi^\dagger\\
\nonumber
&+&{1\over 2}\sigma_+\left(\psi^\dagger\right)^2
+E_{1+1}\psi^\dagger \left(\psi^\dagger+1\right)\psi
+{1\over 2}\sigma_{1+}{\psi^+}^2\left(\psi^\dagger +1\right)^2\psi^2
-E_{-1}\psi^\dagger \psi\\
\label{eq:random_actionfull}
&+&{1\over 2}\sigma_-\left(\psi^\dagger \psi\right)^2
\Biggl\}
+n_0\int\! d{\bf x}\, \psi^\dagger({\bf x},0)\,.
\end{eqnarray}
The stationarity equation brought about by this dynamic action for the field $\psi$ is (the stationary
value $\psi^\dagger=0$ as usual)
\beq
\label{eq:random_statfull}
\partial_t\psi
-D_0 \nabla^2\psi=-2\lambda_0D_0\psi^2
+E_{+1}
+E_{1+1}\psi
-E_{-1}\psi\,.
\eeq
However, the action expanded around the stationary value is rather complicated.
To keep expressions simple, continue to consider the case $E_{1+1}=E_{-1}$. Then the
re-expanded action is
\begin{eqnarray}
\nonumber
S&=&\int\!\! dt\!\int\! d{\bf x} \Bigl\{-\psi^\dagger\partial_t\psi
+D_0 \psi^\dagger\nabla^2\psi        -{\sqrt{8}}{\sqrt{ E_{+1} {\lambda_0D_0} }} \psi^\dagger \psi \\
 \nonumber
 &+&\left( \frac{- E_{+1} }{2} +
       \frac{ E_{-1} {\sqrt{ E_{+1} \,{\lambda_0D_0} }}}{{\sqrt{2}}{\lambda_0D_0} } +
       \frac{ E_{+1} \sigma_{1+} }{4{\lambda_0D_0} } +
       \frac{ E_{+1}  \sigma_{-} }{4 {\lambda_0D_0} } + \frac{ \sigma_+}{2} \right)
      { \psi^\dagger }^2\\
\nonumber
&+& \frac{ E_{+1}  \sigma_{1+} { \psi^\dagger }^3}
     {2{\lambda_0D_0} } + \frac{ E_{+1}  \sigma_{1+} { \psi^\dagger }^4}{4{\lambda_0D_0} }
 + \frac{{\sqrt{2}}\,{\sqrt{ E_{+1} \,{\lambda_0D_0} }}\,
         \sigma_{1+} \,{ \psi^\dagger }^3\psi}{{\lambda_0D_0} } \\
     \nonumber
&+&\left(  E_{-1}  - {\sqrt{2}}\,{\sqrt{ E_{+1} \,{\lambda_0D_0} }} +
        \frac{{\sqrt{ E_{+1} \,{\lambda_0D_0} }}\, \sigma_{1+} }{{\sqrt{2}}\,{\lambda_0D_0} } +
        \frac{{\sqrt{ E_{+1} \,{\lambda_0D_0} }}\, \sigma_{-} }{{\sqrt{2}}\,{\lambda_0D_0} } \right){ \psi^\dagger }^2\psi
      \\
\nonumber
&+&
     \frac{{\sqrt{ E_{+1} \,{\lambda_0D_0} }}\, \sigma_{1+} \,{ \psi^\dagger }^4\psi}
      {{\sqrt{2}}\,{\lambda_0D_0} }
        -2\,{\lambda_0D_0} \, \psi^\dagger {\psi }^2  +
     \left( -{\lambda_0D_0}  + \frac{ \sigma_{1+} }{2} + \frac{ \sigma_{-} }{2} \right) \,
      { \psi^\dagger }^2 {\psi }^2\\
&+&  \sigma_{1+} \,{ \psi^\dagger }^3 {\psi }^2+
     \frac{ \sigma_{1+} \,{ \psi^\dagger }^4{\psi }^2}{2}
     \Bigr\}
+n_0\int\! d{\bf x}\, \psi^\dagger ({\bf x},0)\,.
\label{eq:random_actioneffective}
\end{eqnarray}
In the critical limit $E_{+1}\to 0$. Since it is the expectation value of a nonnegative random
quantity $\mu_+$, the variance $\sigma_+$ vanishes as well. In the vicinity of the critical point
we keep only the leading $E_{+1}$ and $\sigma_+$ putting them equal zero in terms, where they are
subleading. This simplifies the action a little bit

\begin{eqnarray}
S&=&\int\! dt\!\int\! d{\bf x}\Bigl\{-\psi^\dagger\partial_t\psi
+D_0 \psi^\dagger \nabla^2\psi        -{\sqrt{8}}{\sqrt{ E_{+1}{\lambda_0D_0} }} \psi^\dagger \psi +
\left(
       \frac{ E_{-1} \,{\sqrt{ E_{+1}  }}}{{\sqrt{2{\lambda_0D_0}}}\, } + \frac{ \sigma_+}{2} \right){ \psi^\dagger }^2
\nonumber\\
 &+&  \frac{ E_{+1} \, \sigma_{1+} \,{ \psi^\dagger }^3}
     {2\,{\lambda_0D_0} } + \frac{ E_{+1} \, \sigma_{1+} \,{ \psi^\dagger }^4}{4\,{\lambda_0D_0} }
+  E_{-1}  \, { \psi^\dagger }^2\psi       +
\frac{{\sqrt{2}}\,{\sqrt{ E_{+1} \,{\lambda_0D_0} }}\,
         \sigma_{1+} \,{ \psi^\dagger }^3\psi}{{\lambda_0D_0} }
\nonumber\\
&+&  \frac{{\sqrt{ E_{+1} \,{\lambda_0D_0} }}\, \sigma_{1+} \,{ \psi^\dagger }^4\psi}
      {{\sqrt{2}}\,{\lambda_0D_0} }
        -2\,{\lambda_0D_0} \, \psi^\dagger{\psi }^2
        +\left( -{\lambda_0D_0}  + \frac{ \sigma_{1+} }{2} + \frac{ \sigma_{-} }{2} \right) \,
      { \psi^\dagger }^2 {\psi }^2  \nonumber\\
 &+&   \sigma_{1+} \,{ \psi^\dagger }^3 {\psi }^2+
     \frac{ \sigma_{1+} \,{ \psi^\dagger }^4{\psi }^2}{2}
     \Bigr\}+n_0\int\! d{\bf x}\, \psi^\dagger({\bf x},0)\,.
\label{eq:random_actioneffective2}
\end{eqnarray}
Dimensional analysis of the canonical dimensions then yields the following cases. In the nonlinear parts
without the critical parameters  $E_{+1}$ and $\sigma_+$  the
previous arguments hold, but in terms having powers of these parameters as coefficients the positive scaling
dimensions of them must be taken into account. The free-field part of the action (\ref{eq:random_actioneffective2})
suggests that the canonical dimension of $E_{+1}$ is four. In fact, the canonical dimension of $\sigma_+$ remains
a free parameter.

Proceeding in the same manner as above, we arrive at the following effective actions for the IR scaling limit.
In the first case with $d_{\psi^\dagger}=0$ the third and fourth powers of ${\psi^\dagger}$ and independent of ${\psi}$
or first order in ${\psi}$
 are irrelevant (due to the coefficients proportional to $E_{+1}$ or its square root) compared with terms $\propto{\psi^\dagger}^2$
in action (\ref{eq:random_actioneffective2}). Nonlinear in ${\psi}$ terms are irrelevant against the linear terms due to the
positive dimension of ${\psi}$.
Therefore, the IR effective action in this case is
\begin{eqnarray}
\nonumber
S&=&\int\! dt\!\int\! d{\bf x} \,\Bigl\{-\psi^\dagger\partial_t\psi
+D_0 \psi^\dagger\nabla^2\psi        -2\,{\sqrt{2}}\,{\sqrt{ E_{+1} \,{\lambda_0D_0} }}\, \psi^\dagger \psi\\
 &+&\left(
       \frac{ E_{-1} \,{\sqrt{ E_{+1}  }}}{{\sqrt{2{\lambda_0D_0}}}\, } + \frac{ \sigma_+}{2} \right)
      \,{ \psi^\dagger }^2
      +  E_{-1}  \,
      { \psi^\dagger }^2\psi
     \Bigr\}
+n_0\int\! d{\bf x}\, \psi^\dagger({\bf x},0)\,.
\label{eq:random_IRactionE+1}
\end{eqnarray}
Again, the interaction term remaining after the formal dimensional analysis does not bring about loops, although here we have a nontrivial
correlation function of the field ${\psi}$. The scaling dimension of this term is negative, however, rendering the irrelevant
terms dangerous and prohibiting any definitive conclusion about the relevance of individual interaction terms.

In the second case with $d_{\psi^\dagger}>0$ and $d_{\psi}>0$ higher powers than the leading corrections
to the free-field action of both fields are irrelevant. This argument leaves us with the dynamic action
\begin{eqnarray}
S&=&\int\! dt\!\int\! d{\bf x} \,\Bigl\{-\psi^\dagger\partial_t\psi
+D_0 \psi^\dagger\nabla^2\psi        -2\,{\sqrt{2}}\,{\sqrt{ E_{+1} \,{\lambda_0D_0} }}\, \psi^\dagger \psi \nonumber\\
 &+&\left(
       \frac{ E_{-1} \,{\sqrt{ E_{+1}  }}}{{\sqrt{2{\lambda_0D_0}}}\, } + \frac{ \sigma_+}{2} \right)
      \,{ \psi^\dagger }^2
      +  E_{-1}  \,
      { \psi^\dagger }^2\psi   -2\,{\lambda_0D_0} \, \psi^\dagger{\psi }^2
     \Bigr\}+  n_0\int\! d{\bf x}\, \psi^\dagger({\bf x},0)\,.
\label{eq:random_IRactionE+2}
\end{eqnarray}
Contrary to the case discussed above, here the interaction term $-2\,{\lambda_0D_0} \, \psi^\dagger{\psi }^2$
generates loops alone due to the presence of the correlation function of the field ${\psi}$. Therefore,
two effective actions with nontrivial fluctuation contributions are possible.

a)
$d_{\psi^\dagger}>d_{\psi}$. To keep the correlation function of the field ${\psi}$ for the loops, the variance
$\sigma_+$ must have a dimension less than that of $\sqrt{E_{+1}}$. This yields the effective action
\begin{eqnarray}
\nonumber
S&=&\int\! dt\!\int\! d{\bf x} \,\Bigl\{-\psi^\dagger\partial_t\psi
+D_0 \psi^\dagger\nabla^2\psi        -2\,{\sqrt{2}}\,{\sqrt{ E_{+1} \,{\lambda_0D_0} }}\, \psi^\dagger \psi\\
 &+&\frac{ \sigma_+}{2}
      \,{ \psi^\dagger }^2
  -2\,{\lambda_0D_0} \, \psi^\dagger{\psi }^2
     \Bigr\}
+n_0\int\! d{\bf x}\, \psi^+({\bf x},0)
\label{eq:random_IRactionE+2a}
\end{eqnarray}
with the critical dimension depending on the scaling dimension of $\sigma_+$ in the spirit of the
description of tricritical scaling behaviour \cite{Vas_turbo}. The model is logarithmic at six dimensions, however,
because apart from the coefficient of the $\propto {\psi^\dagger}^2$ the action is that of critical dynamics of the
$\varphi^3$ model. The upper critical dimension is determined by the scaling behaviour of $\sigma_+$ in the
critical limit $\sigma_+\to 0$, ${E_{+1}}\to 0$.

b) $d_{\psi^\dagger}=d_{\psi}=d/2$. Both third-order terms are relevant and the effective action is
basically (\ref{eq:random_IRactionE+2}). In this case the dimension of
$\sigma_+$ is larger than that of $\sqrt{E_{+1}}$ and for simplicity we omit $\sigma_+$. Thus, the
effective dynamic action may be written as
\begin{eqnarray}
\nonumber
S&=&\int\! dt\!\int\! d{\bf x} \,\Bigl\{-\psi^\dagger\partial_t\psi
+D_0 \psi^\dagger\nabla^2\psi        -2\,{\sqrt{2}}\,{\sqrt{ E_{+1} \,{\lambda_0D_0} }}\, \psi^\dagger \psi\\
 &+&
       \frac{ E_{-1} \,{\sqrt{ E_{+1}  }}}{{\sqrt{2{\lambda_0D_0}}}\, }
      \,{ \psi^\dagger }^2
      +  E_{-1}  \,
      { \psi^\dagger }^2\psi   -2\,{\lambda_0D_0} \, \psi^\dagger{\psi }^2
     \Bigr\}
+n_0\int\! d{\bf x}\, \psi^\dagger({\bf x},0)\,.
\label{eq:random_IRactionE+2b}
\end{eqnarray}
Note that this is a dynamic action describing the Gribov process with a random source independent of
the active agent density. That the rate of change of the density due to the random sink is proportional
to a power of density is a natural assumption. The assumption that the rate of change of the density due
to the random source is proportional to a power of density is not natural. Therefore, the dynamic action
(\ref{eq:random_IRactionE+2b}) possibly predicts a critical behaviour of the Gribov process different from that
discussed in the literature.

In the third case with $d_{\psi^\dagger}>0$ and $d_{\psi}=0$ we arrive at the effective action
 (\ref{eq:random_IRactionE+2a}).

The analysis of scaling dimensions shows that we may actually lift most of the restrictions on the probability
distribution of the transition rates of the type (\ref{eq:random_limit_variance}) and
 (\ref{eq:random_limit_cumulants}).
Indeed, even if the higher order cumulants are finite, the scaling dimensions of corresponding terms
in the dynamic action grow with the order of the cumulant with the exception of the case, when the transition
rate is independent of the agent density.

The scaling-dimension analysis of relevant and irrelevant interaction terms presented above appears somewhat formal.
In particular, the arbitrariness of the scaling dimensions of the fields is an irritating detail.
Therefore, it is instructive to repeat the analysis with the use of standard power counting. Consider first the
case of particle-number conserving sources and sinks. The dynamic action is then (\ref{eq:random_actionfull}) at the critical point, i.e. with
$E_{+1}=\sigma_+=0$ and $E_{1+1}=E_{-1}$:
\begin{multline}
\label{actionfullcritical}
S=\int\! dt\!\int\! d{\bf x} \,\biggl\{-\psi^+\partial_t\psi
+D_0 \psi^+\nabla^2\psi
-\lambda_0D_0\left[2\psi^++
\left(\psi^+\right)^2\right]\psi^2\\
+E_{1+1}{\psi^+}^2\psi
+{1\over 2}\sigma_{1+}{\psi^+}^2\left(\psi^++1\right)^2\psi^2
+{1\over 2}\sigma_-\left(\psi^+\psi\right)^2
\biggr\}
+n_0\int\! d{\bf x}\, \psi^+({\bf x},0)\,.
\end{multline}
The divergence index of a one-irreducible graph is
\beq
\label{Dindex}
\delta=(d+2)L-2I\,,
\eeq
where $d$ is the dimension of space, $L$ the number of loops and $I$ the number of internal lines.
The usual conditions relate the number of loops, lines, vertices,
external field arguments $E_\psi$ and $E_{\psi^+}$  as well as the number of vertices $V_{ij}$, in which the first subscript denotes
the number of fields $\psi$ and the second the number of fields $\psi^+$ in the action (\ref{actionfullcritical}):
\begin{align}
\label{indexEqs}
L&=I+1-\left(V_{12}+V_{21}+V_{22}+V_{23}+V_{24}\right)\,,\nonumber\\
I&=V_{12}+2V_{21}+2V_{22}+2V_{23}+2V_{24}-E_\psi\,,\\
I&=2V_{12}+V_{21}+2V_{22}+3V_{23}+4V_{24}-E_{\psi^+}\,.\nonumber
\end{align}
>From these equations it follows, in particular,
\beq
\label{EDifference}
E_\psi-E_{\psi^+}=-V_{12}+V_{21}-V_{23}-2V_{24}\,.
\eeq
Eliminating the number of lines $I$ and the number of vertices $V_{12}$ we obtain
\beq
\label{Dindex2}
\delta=d+2+(d-4)V_{21}+(d-2)V_{22}+dV_{23}+ (d+2)V_{24}-(d-2)E_\psi-2E_{\psi^+}\,.
\eeq
Multiplying relation (\ref{EDifference}) by an arbitrary coefficient $a$ and combining with expression (\ref{Dindex2})
we arrive at the following representation of the divergence index
\begin{multline}
\label{Dindex3}
\delta=d+2-aV_{12}+(d-4+a)V_{21}+(d-2)V_{22}+(d-a)V_{23}\\
+ (d+2-2a)V_{24}-(d-2+a)E_\psi-(2-a)E_{\psi^+}\,.
\end{multline}
Here, it is immediately seen that the choice of the value of the parameter $a$ is tantamount to choosing the values
of the scaling dimensions of the fields. Since the divergence index $\delta$ is independent of $a$, we conclude that the
choice of the scaling dimension of the fields has actually nothing to do with the UV or IR behaviour of a one-irreducible
graph. There is no mass parameter in the model, therefore a divergence index $\delta$ with at least one negative coefficient of
a vertex number indicates potential IR divergence, while a positive
$\delta$ corresponds to the usual superficial UV divergence.

For further discussion, let us denote the generic vertex as $v_{nm}\psi^n{\psi^+}^m$ and refer to any particular interaction term
by its coefficient function $v_{nm}$.
>From expression (\ref{Dindex2}) it would appear that vertices $v_{23}$ and $v_{24}$ always make a positive contribution to $\delta$ regardless of the
space dimension, therefore they are definitely IR irrelevant. The vertex $v_{22}$ is irrelevant above two dimensions, whereas the vertex $v_{21}$
gives rise to IR divergences below four dimensions. Since there is no known regular way to cope with these directly, we have to rely
on the usual connection between the IR and UV divergences in the logarithmic theory and extrapolate its results below the critical
dimension with the aid of the  $\varepsilon$ expansion. Thus, if the vertex $v_{21}$ is present, the best we can do is to carry out the
RG analysis around the critical dimension $d_c=4$, at which the vertex $v_{22}$ is also irrelevant, and we are left with the Gribov process.
>From the point of view of the previous scaling dimension analysis representation (\ref{Dindex2}) corresponds to
effective action (\ref{eq:random_IRaction2}) at the logarithmic dimension
$d=4$ with the field dimensions $d_\psi=d-2$, $d_{\psi^+}=2$ and dimensionless coupling constant of the vertex $v_{12}$.

On the other hand, with the use of relation (\ref{EDifference}) we may eliminate any one vertex number, except $V_{ 22}$, which then leads
to changes in the coefficients of the remaining vertex numbers (apart from $V_{22}$) and to a different classification of the relevance of a given
vertex, although the index of any graph does not feel the change. For instance, choosing $a=4-d$ in (\ref{Dindex3}) to eliminate $V_{21}$ we
arrive at
\beq
\label{Dindex4}
\delta=d+2+(d-4)V_{12}+(d-2)V_{22}+(2d-4)V_{23}+ (3d-6)V_{24}-2E_\psi-(d-2)E_{\psi^+}\,.
\eeq
Here, $v_{22}$, $v_{23}$ and $v_{24}$ are all irrelevant at $d>2$, whereas $v_{12}$ is relevant at $d<4$. Thus, if $V_{12}>0$, the only critical regime
amenable to an RG analysis is that of the Gribov process.
It is evident from the preceding discussion that eliminating a vertex number $V_{ij}$
from the expression for the divergence index is tantamount to putting the scaling dimension of that vertex equal to zero and thus fixing the
scaling dimensions of the fields correspondingly. The aim of the power-counting analysis is to discard all graphs of a perturbation
expansion of a Green function with the scaling dimension (i.e. the divergence index) larger than that of the leading-order expression.
The divergence index is independent of the choice of the scaling dimension of the fields. In fact, the only tunable parameter it depends on is the
space dimension. Putting the scaling dimension of a vertex equal to zero means that we are looking for the critical behaviour of the model
at a space dimension at which there are no restrictions on the number these vertices in the relevant and marginal graphs.

Since we are analyzing the effect of random sources and sinks on the pair annihilation process, we should keep the $v_{21}$ vertex in the effective IR model
at any rate and therefore, for the purposes of the physical model, put the scaling dimension of this vertex equal to zero. This means that
$d_\psi=2$ and $d_{\psi^+}=d-2$ and the that the critical behaviour of the model is that of the Gribov process, which belongs to a universality class
significantly different from that of the pure annihilation reaction model.

>From the point of view of classification of terms in power counting the two effective actions (\ref{eq:random_IRaction1}) and (\ref{eq:random_IRaction3})
of the previous scaling-dimensions analysis correspond to elimination of $E_{\psi^+}$ and $E_\psi$, respectively, from the expression for the divergence index. This yields
\begin{align}
\label{Dindex6a}
\delta&=d+2-2V_{12}+(d-2)V_{21}+(d-2)V_{22}+(d-2)V_{23}+ (d-2)V_{24}-dE_\psi\\
\label{Dindex6b}
&=d+2+(d-2)V_{12}-2V_{21}+(d-2)V_{22}+(2d-2)V_{23}+ (3d-2)V_{24}-dE_{\psi^+}\,.
\end{align}
Relation (\ref{Dindex6a}) suggests that all other vertices than $v_{12}$ are irrelevant above two dimensions leading to effective action (\ref{eq:random_IRaction1}).
However, all these are ''dangerous'' irrelevant operators in the sense that the negative dimension of the vertex $v_{12}$ may render the dimension of a graph with
irrelevant operators negative anyway. Since all dangerous operators have equal dimensions, they cannot be classified by the degree of ''dangerousness''.
Relation (\ref{Dindex6b}) leads to a similar phenomenon with respect vertices other than $v_{21}$, which has negative
dimension there. However, here irrelevant operators have different dimensions and this might serve as a basis for classification of some operators
more dangerous than others.

Let us recall that all expressions (\ref{Dindex2}) -- (\ref{Dindex6b}) for the divergence index are equal and reflect different choices of the ambiguous
field dimensions. We see that in a multicoupling case it is rather difficult to arrive at the consistent conclusion about the relevance of different vertices
solely on the basis of the analysis of scaling dimensions of fields and vertex operators. In particular, to put either of the field dimensions equal to zero
at the outset appears to be quite misleading. Assuming both field dimensions nonvanishing, put them equal to each other to obtain
\beq
\label{Dindex7}
\delta=d+2+\left({d\over 2}-2\right)V_{12}+\left({d\over 2}-2\right)V_{21}+(d-2)V_{22}+\left({3d\over 2}-2\right)V_{23}+ (2d-2)V_{24}-{d\over 2}\,E_\psi-{d\over 2}\,E_{\psi^+}\,.
\eeq
Here, indeed, all coefficients of vertex numbers depend of the space dimension explicitly and immediately lead to the conclusion, that the
critical scaling behaviour tractable within the RG is that of the Gribov process.

Consider then the action (\ref{eq:random_actioneffective}) for the active state. The divergence index does not give the whole truth in this case due
to the presence of the ''temperature'' parameter $E_{+1}$, but it reflects the contribution of the wave-vector and frequency integral anyway.
Many new types of vertices appear, but the set of conditions imposed on their numbers remains the same. In this case it is convenient to analyze the
IR and UV behaviour separately. The reason is that the temperature parameter is not involved in the power counting of UV divergences in any other way that
new vertices have appeared. Thus, the power counting goes in the same fashion as above.
Denoting new vertices by tilde, we obtain
\begin{align}
\label{indexEqs2a}
L&=I+1-\left(\tilde{V}_{02}+\tilde{V}_{03}+\tilde{V}_{04}+V_{12}+\tilde{V}_{13}+\tilde{V}_{14}+V_{21}+V_{22}+V_{23}+V_{24}\right)\,,\\
\label{indexEqs2b}
I&=V_{12}+\tilde{V}_{13}+\tilde{V}_{14}+2V_{21}+2V_{22}+2V_{23}+2V_{24}-E_\psi\,,\\
\label{indexEqs2c}
I&=2\tilde{V}_{02}+3\tilde{V}_{03}+4\tilde{V}_{04}+2V_{12}+3\tilde{V}_{13}+4\tilde{V}_{14}+V_{21}+2V_{22}+3V_{23}+4V_{24}-E_{\psi^+}\,.
\end{align}
Consequently
\beq
\label{EDifferenceB}
E_\psi-E_{\psi^+}=-2\tilde{V}_{02}-3\tilde{V}_{03}-4\tilde{V}_{04}-V_{12}-2\tilde{V}_{13}-3\tilde{V}_{14}+V_{21}-V_{23}-2V_{24}\,.
\eeq
>From (\ref{indexEqs2a}) and (\ref{indexEqs2b}) it follows that
\begin{multline}
\label{DindexB1}
\delta=(d+2)L-2I=d+2-\left(d+2\right)\tilde{V}_{02}-\left(d+2\right)\tilde{V}_{03}-\left(d+2\right)\tilde{V}_{04}-2V_{12}
-2\tilde{V}_{13}\\
-2\tilde{V}_{14}+\left({d}-2\right)V_{21}
+\left(d-2\right)V_{22}+\left({d}-2\right)V_{23}+\left(d-2\right)V_{24}
-{d}\,E_\psi\,.
\end{multline}
Adding the relation (\ref{EDifferenceB}) multiplied by the coefficient $a$ to expression (\ref{DindexB1}) we arrive at the representation
\begin{multline}
\label{DindexB2}
\delta=d+2+\left(2a-d-2\right)\tilde{V}_{02}+\left(3a-{d}-2\right)\tilde{V}_{03}+\left(4a-{d}-2\right)\tilde{V}_{04}\\
+\left(a-2\right)V_{12}
+\left(2a-2\right)\tilde{V}_{13}
+\left(3a-2\right)\tilde{V}_{14}+\left({d}-2-a\right)V_{21}\\
+\left(d-2\right)V_{22}+\left({d}-2+a\right)V_{23}+\left(d-2+2a\right)V_{24}
-{a}\,E_\psi-\left(d-a\right)\,E_{\psi^+}\,.
\end{multline}
Here, by the choice of the value of the parameter $a$ we could try to find a representation convenient for the classification of the
vertices. However, in this case the vertex with the negative dimension $\psi^+\psi^+$,
may be absorbed in
the pair correlation function of the field $\psi$ included in the elements of the graphical representation.
In calculation of the UV index the field dimensions are then fixed by the condition that the coupling constant of the term $\propto\psi^+\psi^+$
is dimensionless. There is no ambiguity in the expression for the index and denoting the number of correlation functions in
a one-irreducible graph $\tilde{I}$ we arrive at relations
\begin{align}
\label{indexEqs3}
L&=I+\tilde{I}+1-\left(\tilde{V}_{03}+\tilde{V}_{04}+V_{12}+\tilde{V}_{13}+\tilde{V}_{14}+V_{21}+V_{22}+V_{23}+V_{24}\right)\,,\nonumber\\
I+2\tilde{I}&=V_{12}+\tilde{V}_{13}+\tilde{V}_{14}+2V_{21}+2V_{22}+2V_{23}+2V_{24}-E_\psi\,,\\
I&=3\tilde{V}_{03}+4\tilde{V}_{04}+2V_{12}+3\tilde{V}_{13}+4\tilde{V}_{14}+V_{21}+2V_{22}+3V_{23}+4V_{24}-E_{\psi^+}\,.\nonumber
\end{align}
>From here it follows that
\begin{multline}
\label{DindexC}
\delta=(d+2)L-2I-4\tilde{I}=d+2+\left({d\over 2}+1\right)\tilde{V}_{03}+\left({d}+2\right)\tilde{V}_{04}\\
+\left({d\over 2}-1\right)V_{12}
+d\tilde{V}_{13}+\left({3d\over 2}+1\right)\tilde{V}_{14}
+\left({d\over 2}-3\right)V_{21}
+(d-2)V_{22}\\
+\left({3d\over 2}-1\right)V_{23}+2dV_{24}
-\left({d\over 2}-1\right)E_\psi-\left({d\over 2}+1\right)E_{\psi^+}\,.
\end{multline}
The same relation follows from (\ref{DindexB2}), when the parameter $a$ is chosen such that the coefficient in front of $\tilde{V}_{02}$ vanishes.
>From (\ref{DindexC}) we see that when the vertex $v_{21}$ is marginal, all the rest are irrelevant and we arrive at the situation described
by the action (\ref{eq:random_IRactionE+2a}).

This, however, is not the whole story, because
in the case of IR behaviour we are interested in the limit of vanishing temperature parameter. Since coupling constants of new vertices are functions
of $E_{+1}$, they affect the limit of small $E_{+1}$ directly. The dependence of a graph on $E_{+1}$ in the critical limit ($\omega_i\to 0$, $\vk_i\to 0$,
$E_{+1}\to 0$, $\omega_i=O\left(\sqrt{E_{+1}}\right)$, $\vk^2_i=O\left(\sqrt{E_{+1}}\right)$) is readily estimated by a suitable scaling of variables in the loop integrals, if these
integrals have negative dimensions, in which case the upper limit
in integrals of a renormalized graph may be sent to infinity. The result of the scaling of
integration variables is a power of the parameter $E_{+1}$ multiplied by a function of reduced frequencies and wave-vectors
$\omega_i/\sqrt{E_{+1}}$, $\vk_iE_{+1}^{-1/4}$. The wave-number integrals defining this function are UV and IR finite for all values
of the reduced frequencies and wave-vectors and thus possess a finite limit, when the latter vanish. Therefore, any IR-singular behaviour is signalled
by a negative overall power of the parameter $E_{+1}$, which takes into account both the scaling of integration variables (this gives the
divergence index $\delta$) and the additional powers of $E_{+1}$ at the vertex factors of the graph. Thus, we might use the following "IR divergence index"
brought about by the action (\ref{eq:random_actioneffective}) (here, $\sigma_+$ is assumed to be subleading for simplicity)
\begin{multline}
\label{DindexIRB}
\delta_{\rm IR}=\delta+2\tilde{V}_{02}+4\tilde{V}_{03}+4\tilde{V}_{04}+2\tilde{V}_{13}+2\tilde{V}_{14}
=d+2+\left({d\over 2}+2\right)\tilde{V}_{03}+\left({d}+2\right)\tilde{V}_{04}\\
+\left({d\over 2}-2\right)V_{12}
+d\tilde{V}_{13}
+\left({3d\over 2}\right)\tilde{V}_{14}+\left({d\over 2}-2\right)V_{21}
+(d-2)V_{22}\\
+\left({3d\over 2}-2\right)V_{23}+(2d-2)V_{24}
-{d\over 2}\,E_\psi-{d\over 2}\,E_{\psi^+}\,.
\end{multline}
to guide in the classification of the interaction terms. The same form may be obtained the generic expression (\ref{DindexB2}) with the choice of $a$ such that
the coefficient in front of $\tilde{V}_{02}$ in $\delta_{\rm IR}$ vanishes.
In order to keep the IR behaviour tractable, all coefficients of the vertex numbers in (\ref{DindexIRB}) must be nonnegative. The lowest
space dimension conforming to this requirement is the upper critical dimension of the model. We see from (\ref{DindexIRB}) that $d_c=4$
for model (\ref{eq:random_actioneffective}) and the effective action is indeed (\ref{eq:random_IRactionE+2b}).


{\section*{CONCLUSIONS} }


In conclusion, we have analyzed the effect of density and velocity
fluctuations on the reaction kinetics of
the single-species decay $A+A\rightarrow\varnothing$ universality class in the
framework of field-theoretic renormalization group and calculated the
scaling function and the decay exponent of the mean particle density for the four asymptotic patterns
predicted by the RG.

We have calculated the relevant renormalization constants at two-loop level
and found the decay exponent of the mean particle density at this order
of the $\epsilon$, $\Delta$ expansion for four IR stable fixed points of the
RG, whose regions of stability cover the whole parametric space in the vicinity
of the origin in the $\epsilon$, $\Delta$ plane. The decay exponent assumes
the mean-field value in the basins of attraction of the trivial fixed point (\ref{anni1_Gaussian}) and
of the kinetic fixed point (\ref{anni1_K2}) with dominant fluctuations of the random force
of the Navier-Stokes equation. At the kinetic fixed point with finite rate coefficient (\ref{anni1_K1}) the decay
value of the decay exponent is determined exactly by the fixed-point equations. At the thermal (short-range)
fixed point (\ref{anni1_SR}) the decay exponent possesses a non-trivial $\epsilon$, $\Delta$ expansion.
We have calculated three first terms of this expansion.

Using a variational approach, we have inferred a renormalized fluctuation-amended rate
equation with the account of one-loop corrections. This
non-linear integro-differential equation has been solved iteratively in the framework of the
$\epsilon$, $\Delta$ expansion and the scaling function for the mean particle density
has been calculated for the four IR stable regimes.
The scaling function assumes the mean-field form (exactly)
in the basins of attraction of the trivial fixed point and
the kinetic fixed point with dominant fluctuations of the random force.
At the kinetic fixed point with finite rate coefficient  and at the thermal fixed point the scaling function possesses a non-trivial
$\epsilon$, $\Delta$ expansion, which we have calculated at the linear order.
Fluctuations of the random advection field affect heavily the long-time asymptotic behaviour of the
system: the kinetic fixed points are brought about by the velocity fluctuations as well as the
non-trivial series expansion of the decay exponent at the thermal fixed point (without velocity
fluctuations, the decay exponent is fixed to the one-loop value, because there are no high-order
corrections to the rate constant in this case).
Predictions of the renormalization-group analyses for the reaction $A+A\rightarrow \varnothing$ in
quenched random fields have been corroborated by numerical simulations \cite{Chung,Tran}. In the case of
dynamically generated random drift this is seems to be a much more demanding task, but would surely be
highly desirable, since the experimental data for reaction processes is quite scarce.s

We have investigated possible effects of random sources and sinks on the pair annihilation reaction $A+A\rightarrow\varnothing$.
Contrary to the frequently used approach,
in which the sources and sinks are introduced into the Langevin equation, we have included them directly to
the master equation, where their physical sense is clear. We have considered linear in particle number creation and annihilation reactions
with random rate coefficients to model the sources and sinks.
On the basis of the analysis of canonical scaling dimensions we have constructed effective actions, which are the starting point for an RG
analysis of the critical behaviour of the systems under consideration.
In all cases the effect of random sources and sinks to the large-scale, long-time behaviour of the Green functions
is significant and changes the universality class of the model. Instead of the universality class of the pair annihilation  reaction $A+A\rightarrow\varnothing$
, the asymptotic behaviour of the model with random sources and sinks belongs to the universality class of the Gribov process in the critical
case and to a modified Gribov process in the critical limit of the noncritical model.
In the former case it is demonstrated once again that the description of a stochastic process with the use of the
Langevin equation is significantly different from the description in terms of a master equation.
The random noise term in the Langevin equation corresponds rather to the account of effects of genuine
random sources and sinks than to a description of the effect of microscopic degrees of freedom on the mesoscopic process.
Here, the universality classes of the same reaction process are completely different in the case of the master equation
without sources and sinks in comparison to the case of Langevin equation for the same process.

In the noncritical case with a random source independent of the agent density the Gribov process is modified
to account for effects in critical  behaviour, when sources and sinks asymptotically vanish. The analysis of the
dependence of scaling functions on the parameters of the probability distribution of sinks and sources in infrared limit
is called for. This reminds the situation, which
takes place in the theory of phase transitions, where statistical correlations of the order parameter depend on a "mass"
(deviation of temperature from the critical value) and the dependence of the scaling functions on the "`mass" is investigated.

\begin{acknowledgments}
The work was supported by VEGA grant 0173 of Slovak Academy of Sciences, and by Centre of Excellency for
Nanofluid of IEP SAS. This article was also created by implementation
of the Cooperative phenomena and phase transitions
in nanosystems with perspective utilization in nano- and
biotechnology projects No 26220120021 and No 26220120033. Funding for the operational
research and development program was provided by
the European Regional Development Fund.
\end{acknowledgments}


\begin{appendix}
{\section{Explicit form of the renormalization constants } \label{sec:appendix_renorm}}
\begin{eqnarray*}
 & & A_{11} = -\frac{(1+\xi)u^2+(3\xi+2)u+6\xi+1}{512u(1+u)^3\xi}, \quad
 A_{12}= -\frac{(1+\xi)u^2+(1+3\xi)u+6\xi-4}{256u(1+u)^3(1-\xi)},\\*
& &A_{22}=-\frac{u+5}{512u(1+u)^3},\\*
& & B_{11} = B_1(u)+ B_2(u,\xi),\quad B_{12} = 4[B_1(u)+ B_3(u,\xi)],\quad B_{22} = -B_1(u)-B_2(u,-1),
 \end{eqnarray*}
 where the functions $B_1,B_2$ and $B_3$ are given as
 \begin{eqnarray*}
 B_1(u)& = &\frac{1}{1024u^4(1+u)^3(u-1)}\biggl[-12u^3(1+u)\ln\frac{2}{1+u}+32u^4(1+u)^2\ln 2+2u^3\times\nonumber\\
 & & (1+u)^2(u+10)\ln\frac{4}{3}- 32u^3(1+u)^3{\rm arctgh}\frac{1}{2} + 4(2u^6+6u^5+7u^4+10u^3-\nonumber\\
 & & 4u-1) \ln\frac{1+2u}{(1+u)^2}-4u^3(1+u)[4u^2+20u+9]\ln\frac{1+2u}{2+2u}+16u^3(1+u)^3\times\nonumber\\
 & & {\rm arctgh}\frac{u}{u+1}+ 8u^3(1+u)^2(u-1)(\gamma+\psi(3/2))u^2(23u^4+38u^3+17u^2+\nonumber\\
 & & 22u+4)\biggl]- \frac{1}{128\pi u(1+u)^2(u-1)}\int_1^\infty dq\int_{-1}^1 dz F(q,z,u) \\
 B_2(u,\xi)  & = & \xi\frac{(2+4\xi)u^2+(10\xi+8)u+38+22\xi}{1024u(1+u)^3(2+\xi)},\\
 B_3(u,\xi)  & = & -\frac{(8\xi+4)u^2+(14\xi+10)u+22-10\xi}{1024u(1+u)^3}
\end{eqnarray*}
with function $F$ given by the expression
\begin{eqnarray*}
F(x,z,u) & = & (1-z^2)^{1/2}\frac{M(x,z,u)}{N(x,z,u)} \\
 M(x,z,u) & = & (x^6+1)[z^3(24u^3+24u^2+72u+72)-z(8u^3+12u^2+8u+60)]+ \nonumber\\
& &(x^5+x)[-z^4(40u^3+88u^2+120u+264)+ z^2(-4u^3+16u^2+108u+168)+\nonumber\\
& & 4u^3+14u^2+28u+18]+(x^4+x^2)[z^5(16u^3+96u^2+48u+288)+  z^3(12u^4+\nonumber\\
& & 64u^2+ 128u^2+96u+180)-z(4u^4+26u^3+92u^2+174u+312)] + \nonumber \\
 & &x^3[-z^6(32u^2+96)- z^4(8u^4+64u^3+240u^2+144u+600)+ z^2(-8u^4+\nonumber\\
 & & 4u^3+84u^2+108u+452)+2u^4+ 6u^3+26u^2+58u+36],\\
N(x,z,u) &= &(1+x^2-2xz)(1+x^2-xz)((1+u)x^2+2-2xz)(1+u+2x^2-2xz)
\end{eqnarray*}
\begin{eqnarray}
  C(u,\xi) & = &
  - \frac{1}{8u\pi } \int_{-1}^1 dz (1-z^2)^{\frac{1}{2}} G(z,u)
  +  \frac{1}{8 u(1+u)}\biggl(\ln\frac{1+u}{2u}+1+\frac{2+u}{u}\times\nonumber\\
  & & \ln\frac{u+2}{2u+2}+\xi\biggl),
  \label{eq:C}
\end{eqnarray}
where
\begin{eqnarray}
  G(z,u) & = & \frac{4}{(1-u)^2+4uz^2} \biggl\{\frac{u-1}{2}\ln\frac{2u}{1+u}-
  \frac{2(1+u)z}{\sqrt{1-z^2}}\biggl[\frac{\pi}{2}-\arctan\sqrt{\frac{1+z}{1-z}} \biggl]+\nonumber\\
  & & \frac{u(u+3)z}{\sqrt{2u(1+u)-u^2z^2}}\biggl[\pi-\arctan\frac{zu+u+1}{\sqrt{2u(1+u)-u^2z^2}}-\nonumber\\
  & & \arctan\frac{(2+z)u}{\sqrt{2u(1+u)-u^2z^2}} \biggl]\biggl\}
  \label{eq:F}
\end{eqnarray}
{\section{Fixed points} \label{sec:appendix_fp}}
\begin{eqnarray}
  u_1^*(\xi) & = & \frac{8R}{3\sqrt{17}}-\frac{8192}{3\sqrt{17}}B_1(u^*_0)-
  \frac{1}{432\sqrt{17}(1+\xi)^2(2+\xi)}\biggl(
  (21384-648\sqrt{17})\xi^4+ \nonumber\\
 & & (52512-2592\sqrt{17}) \xi^3+(22192-2736\sqrt{17})\xi^2
  +(72\sqrt{17}-29064)\xi+\nonumber\\
  & & 720\sqrt{17}-18768\biggl), \\
 g_{12}^*(\xi) & = & \frac{64(R(2+3\xi)-1)}{27(1+\xi)}-\frac{16(2+3\xi)}{243(1+\xi)^4(2+\xi)}\biggl[
   45\xi^4+213\xi^3+349\xi^2+231\xi+50
   \biggl], \nonumber\\
   & & \\
 g_{22}^*(\xi) & = & \frac{64(1+R)}{27(1+\xi)}-\frac{16(2+3\xi)}{243(1+\xi)^4(2+\xi)}\biggl[
  57\xi^4+171\xi^3+185\xi^2+93\xi+22 \biggl].
\end{eqnarray}
\end{appendix}

\newpage
{\section{REFERENCES} \label{sec:reference}}

\newpage \mbox{ }

\newpage \mbox{ }

\newpage \mbox{ }

\newpage \mbox{ }

\newpage \mbox{ }

\newpage \mbox{ }
\newpage \mbox{ }
\begin{table}
  \setcaptionmargin{0mm}
  \onelinecaptionstrue
  \captionstyle{flushleft}
  \caption{Canonical dimensions for the parameters and the fields of the model}
  \begin{tabular}{|c|c|c|c|c|c|c|c|c|c|c|}
    \hline
    $Q$ & $\psi$ & $\psi^\dagger$ & $v$ & $\tilde{v}$ & $\nu_0$ & $D_0$ & $\lambda_0$ & $g_{10}$ & $g_{20}$ \\ \hline
    $d_Q^k$ & $d$ & $0$ & $-1$ & $d+1$ & $-2$ & $-2$ & $-2\Delta$ & $2\epsilon$ & $-2\Delta$ \\ \hline
    $d_Q^\omega $& $0 $ & $0$ & $1$ & $-1$ & $1$ & $1$ & $0$ & $0$ & $0$\\ \hline
    $d_Q $& $d$ & $0$ & $1$ & $d-1$ & $0$ & $0$  & $-2\Delta$ & $2\epsilon$ & $-2\Delta$\\ \hline
  \end{tabular}
  \label{tab:anni1_canon_dims}
\end{table}

\newpage \mbox{ }
\begin{table}
  \setcaptionmargin{0mm}
  \onelinecaptionstrue
  \captionstyle{flushleft}
  \caption{Canonical dimensions for the (1PI) divergent Green functions of the model}
  \begin{tabular}{|c|c|c|c|c|c|c|c|c|c|c|}
    \hline
    $\Gamma_{1-ir}$ & $\langle\psi_\dagger \psi\rangle $& $\langle\psi_\dagger\psi v\rangle$
                  & $\langle \tilde{v} v\rangle $ & $ \langle \tilde{v} v v\rangle $
                  & $\langle \tilde{v}\tilde{v}\rangle$
                  & $\langle \psi^\dagger \psi^2\rangle $ & $\langle (\psi^\dagger)^2\psi^2 \rangle$
    \\ \hline
    $d_\Gamma$ & $2$ & $1$
         & $2$ & $1$ & $2$
             & $0$ & $0$
    \\ \hline
    $\delta_\Gamma$ & $2$ & $1$
         & $1$ & $0$ & $0$
             & $0$ & $0$
    \\ \hline
  \end{tabular}
  \label{tab:anni1_canon_green}
\end{table}

\end{document}